\documentclass[utf8]{FrontiersinHarvard} 

\usepackage{url,hyperref,lineno,microtype,subcaption}
\usepackage[onehalfspacing]{setspace}
\usepackage{float}

\def\keyFont{\fontsize{8}{11}\helveticabold }
\def\firstAuthorLast{Santos {et~al.}} 
\def\Authors{\^Angela R. G. Santos\,$^{1,*}$, Diego Godoy-Rivera\,$^{2,3}$, Adam J. Finley\,$^4$, Savita Mathur\,$^{2,3}$, Rafael A. Garc\'ia\,$^{4}$, Sylvain N. Breton\,$^5$ and Anne-Marie Broomhall\,$^{6}$}
\def\Address{$^{1}$Instituto de Astrof\'isica e Ci\^encias do Espa\c{c}o, Universidade do Porto, CAUP, Rua das Estrelas, PT4150-762 Porto, Portugal \\
$^{2}$ Instituto de Astrof\'\i sic
a de Canarias (IAC), E-38205 La Laguna, Tenerife, Spain\\
$^3$ Universidad de La Laguna (ULL), Departamento de Astrof\'isica, E-38206 La Laguna, Tenerife, Spain\\
$^4$ Universit\'e Paris-Saclay, Universit\'e Paris Cit\'e, CEA, CNRS, AIM, 91191, Gif-sur-Yvette, France\\
$^5$ INAF – Osservatorio Astrofisico di Catania, Via S. Sofia, 78, 95123 Catania, Italy\\
$^6$ Department of Physics, University of Warwick, Coventry, CV4 7AL, UK}
\def\corrAuthor{\^Angela R. G. Santos, CAUP, Rua das Estrelas, PT4150-762 Porto, Portugal}

\def\corrEmail{Angela.Santos@astro.up.pt}

\newcommand{\kep}{\textit{Kepler}}

\newcommand{\prot}{$P_\text{rot}$}
\newcommand{\sph}{$S_\text{ph}$}

\newcommand{\teff}{$T_\text{eff}$}

\begin{document}
\onecolumn
\firstpage{1}

\title{\textit{Kepler} main-sequence solar-like stars: surface rotation and magnetic-activity evolution}

\author[\firstAuthorLast ]{\Authors} 
\address{} 
\correspondance{} 

\extraAuth{}

\maketitle

\begin{abstract}

\section{}

While the mission's primary goal was focused on exoplanet detection and characterization, \kep\ made and continues to make extraordinary advances in stellar physics. Stellar rotation and magnetic activity are no exceptions. \kep\ allowed for these properties to be determined for tens of thousands of stars from the main sequence up to the red giant branch. From photometry, this can be achieved by investigating the brightness fluctuations due to active regions, which cause surface inhomogeneities, or through asteroseismology as oscillation modes are sensitive to rotation and magnetic fields. This review summarizes the rotation and magnetic activity properties of the single main-sequence solar-like stars within the \kep\ field. We contextualize the \kep\ sample by comparing it to known transitions in the stellar rotation and magnetic-activity evolution, such as the convergence to the rotation sequence (from the saturated to the unsaturated regime of magnetic activity) and the Vaughan-Preston gap. While reviewing the publicly available data, we also uncover one interesting finding related to the intermediate-rotation gap seen in \kep\ and other surveys. We find evidence for this rotation gap in previous ground-based data for the X-ray luminosity. Understanding the complex evolution and interplay between rotation and magnetic activity in solar-like stars is crucial, as it sheds light on fundamental processes governing stellar evolution, including the evolution of our own Sun. 

\tiny
 \keyFont{ \section{Keywords:} Stars: activity, Stars: evolution, Stars: late-type, Stars: low-mass, Stars: magnetic field, Stars: rotation, Stars: solar-type, starspots} 
\end{abstract}

\section{Introduction}

Low-mass stars with convective outer layers, also known as solar-like stars, can sustain internal dynamos and have the potential to harbor magnetic activity. Magnetic fields and magnetic cycles are generated by an interaction between differential rotation and convection \citep[see][for a recent review]{Brun2017}. As the strong magnetic fields emerge at the stellar photosphere, they form active regions, where dark spots appear, usually in pairs or groups of opposite polarity \citep[e.g.][]{Hale1919,Solanki2003,Hathaway2015}. Such active regions can be associated with eruptive events like flares and coronal mass ejections \citep[e.g.][]{Zirin1970,Solanki2003}. As active regions decay their magnetic fields disperse and concentrate at the edges of the convective cells, forming the bright faculae \citep[e.g.][]{vanDriel-Gesztelyi2015}. In the chromosphere, these regions of intermediate magnetic-field strength appear as bright plage. All these phenomena are part of the star's magnetic activity, which varies at different timescales. In particular, the Sun undergoes an 11-year cycle of magnetic activity. Analogously, other stars are known to exhibit magnetic activity cycles \citep[e.g.][]{Baliunas1995,Olah2009,Garcia2010,Karoff2018,BoroSaikia2018}, with cycle periods ranging from a few years to over 20 years. As the dynamo mechanism is powered by the interplay of convection and rotation \citep[e.g.][]{Brun2017}, the cycle and rotation periods are found to be related, with slow-rotating stars having longer cycles than fast-rotating stars. Stars are often seen to group along two branches: active and inactive \citep[e.g.][]{Brandenburg1998,Bohm-Vitense2007}. However, there is still debate whether these are properly determined or even exist \citep[e.g.][]{BoroSaikia2018,Bonanno&Corsaro2022}.

The level of magnetic activity is also intrinsically linked to rotation \citep[e.g.][]{Kraft1967,Walter1981,Pallavicini1981,Noyes1984b,Soderblom1993,Pizzolato2003}. At the beginning of their main-sequence lifetime, stars are relatively fast rotators and exhibit high activity levels \citep[e.g.][]{Skumanich1972,Barnes2003,Brown2021,Fritzewski2021b}. Stars gradually lose angular momentum due to their magnetized winds \citep[e.g.][]{Weber1967,Kraft1967,Skumanich1972, Kawaler1988,Pinsonneault1989,Gallet2013,Matt2015}, in a process known as magnetic braking. The rate at which the stars spin down depends on their rotation, with faster rotators losing angular momentum faster than slower rotators. Eventually, stars will converge into the so-called slow-rotation sequence (see for example Fig.~3 in \citealt{Gallet2013}), and from that point onwards the rotation rate decays proportionally to the square root of the age, known as the Skumanich spin-down law \citep{Skumanich1972}. The spin-down process is also mass-dependent, with lower-mass stars taking longer to converge into the rotation sequence than higher-mass stars, but once reached, lower-mass stars spin down faster than higher-mass stars \citep[e.g.][]{Barnes2003,Barnes2007,vanSaders2013,Matt2015}. The magnetic activity also decays with time: as stars evolve and spin down, they gradually become less active \citep{Wilson1963,Skumanich1972,Soderblom1991}. Therefore, generally, fast-rotating stars have stronger magnetic activity than slow-rotating stars. This activity-rotation relationship can be represented as a function of the Rossby number, Ro \citep{Noyes1984b}. Ro can be defined as the ratio between the star's rotation period and its convective turnover timescale. The latter corresponds to the typical timescale for convective motions in the stars' envelopes and remains mostly constant during the main-sequence lifetime, increasing as the effective temperature decreases \citep[e.g.][]{Lehtinen2021}. Metallicity, however, can complicate this picture. Due to a larger opacity, metal-rich stars have deeper convection zones than their metal-poor counterparts \citep[e.g.][]{vanSaders2012}. A deeper convection zone leads to a more vigorous dynamo, consequently to higher magnetic activity. Observational evidence for this effect has been found in both large samples \citep{See2021,See2023} and in the particular case of the seismic solar-analog HD 173701 (KIC 8006161). This $\sim1\text{M}_\odot$ star exhibits a magnetic cycle with more than twice the amplitude of the solar cycle \citep{Karoff2018}. Strong magnetic activity, in turn, would lead to a more efficient loss of the angular momentum. Thus, metal-rich stars are expected to spin down faster than metal-poor stars \citep{Amard2020} and, so far, this theoretical expectation has been supported by observations \citep[e.g.][]{Amard2020b,Santos2023}.

The Skumanich spin-down law led to the development of gyrochronology \citep[e.g.][]{Barnes2003,Barnes2007,Mamajek2008,Garcia2014,Angus2015,Lu2023}, enabling estimation of stellar ages from surface rotation. Analogously, magnetochronology and magnetogyrochronology relations have also been established \citep[e.g.][]{Mamajek2008,Pace2013,Lorenzo-Oliveira2018,Mathur2023,Ponte2023}. As rotation and magnetic activity measurements are available for large numbers of stars, these techniques are powerful tools for estimating stellar ages. However, the evolution of stellar rotation and magnetic activity is not yet fully understood, as we will discuss below. 

The stars' magnetic fields can be measured through Zeeman Broadening of spectral lines and Zeeman Doppler Imaging \citep[ZDI; e.g.][see also \citealt{Kochukhov2021} for a recent review]{Semel1989,Donati1997,Reiners2022,Vidotto2014,See2019a,EBrown2022}. These techniques shed light on the evolution of stellar magnetic fields and allow the large-scale magnetic field topology to be recovered. The ZDI technique has been used to follow entire stellar magnetic cycles which, in the case of 61 Cyg A, share many similarities to that of the solar cycle \citep{Saikia2018L}. However, due to the challenges of directly measuring magnetic fields, indirect measures are often used. These are magnetic activity proxies and link to phenomena associated with the presence of strong magnetic fields, which can be constrained by spectroscopic and photometric observations. 

In spectroscopy, magnetic activity is often constrained through the analysis of particular spectral lines. Examples of such lines are the Ca \textsc{ii} H \& K lines in the near ultra-violet (NUV) and the Ca \textsc{ii} infrared triplet (IRT). In the presence of active regions, these absorption lines show emission at their cores due to chromospheric heating. By measuring such emission and removing the basal and photospheric contributions, these lines provide proxies for magnetic activity in the chromosphere \citep[e.g.][]{Leighton1959,Wilson1968,Wilson1978,Baliunas1995,Karoff2016,GomesDaSilva2021,Fritzewski2021b}. Since active regions go in and out of view as the star rotates, these activity proxies can show short-term quasi-periodic variations, allowing to constrain rotation periods \citep[e.g.][]{Suarez-Mascareno2017,Lorenzo-Oliveira2019}. Another activity proxy that can be derived from spectra is the spot filling factor \citep[e.g.][]{Gully-Santiago2017,Gosnell2022,Cao2022}, which follows upon the fact that magnetic spots in solar-like stars are cool features in comparison to their surroundings. 
 
In white-light photometry, active regions lead to variations in the stellar brightness, known as rotation modulation. Once again, as active regions co-rotate, in this case with the stellar surface, the periodicity of these brightness variations is related to the surface rotation period at the latitudes where the active regions are formed \citep[e.g.][]{Gaidos2000,Reinhold2013a,McQuillan2014,Garcia2014,Lanzafame2018,Santos2021ApJS,Gordon2021,Lu2022,Distefano2023,Claytor2023}. The amplitude of the brightness variations relates to the surface coverage by active regions \citep[e.g.][]{Garcia2010,Basri2010,Mathur2014b,Salabert2017} and is found to be well correlated with chromospheric activity proxies \citep{Salabert2016a,Ponte2023}. One of the advantages associated with the advent of photometric space missions is that they allow the retrieval of magnetic activity proxies and rotation for stellar samples orders of magnitude larger than those from ground-based surveys.

Stellar flares can also be detected and characterized from white-light photometry \citep[e.g.][]{Davenport2016,Yang2019,Ilin2019,Ilin2021,Gunther2020}, providing a proxy for magnetic activity.
Another widely used magnetic activity proxy is X-ray emission from hot plasma confined in coronal loops \citep[e.g.][]{Schmitt1995,Pizzolato2003,Pillitteri2006,Wright2018}, which is correlated with the stellar wind mass-loss rates \citep{Wood2021}. Other activity indicators used in the literature are the H$\alpha$ emission and NUV excess \citep[e.g.][]{Findeisen2011,Newton2017,Godoy-Rivera2021,Zhong2023}. 

Asteroseismology can also be effective in measuring rotation and magnetic activity in solar-like stars, thanks to the acoustic modes (p modes) being sensitive to both properties. In this case, solar-like acoustic oscillations give information mostly from subphotosphere layers \citep[e.g.][]{Basu2012,Benomar2015}. Rotation can be measured from the splitting of the modes' azimuthal orders if the stellar inclination angle is not too small \citep[][inclination of $90^\circ$ and $0^\circ$ correspond, respectively, to observing the star equator-on and pole-on]{Gizon2003,Ballot2006}. In addition to an average rotation \citep[e.g.][]{Gizon2013,Davies2015,Hall2021}, asteroseismology also allows us to obtain information about surface latitudinal differential rotation \citep{Benomar2018,Bazot2019}. Unfortunately, main-sequence solar-like stars pose a significant challenge in the determination of their radial differential rotation due to uncertainties in observations and stellar models \citep{Benomar2015,Schunker2016,Schunker2016a,Nielsen2017}. Moreover, the number of visible modes is limited, as well as their sensitivity to greater depths in the stellar interiors. Magnetic activity affects different properties of the acoustic modes \citep[e.g.][]{Woodard1985,Elsworth1990,Jimenez-Reyes1998,Jain2009,Tripathy2011,Garcia2010,Broomhall2014,Kiefer2017,Santos2018}. Particularly, the mode frequencies are observed to increase with the magnetic activity level, while mode amplitudes decrease. For low-degree modes, those that are possible to observe for stars other than the Sun, modes of different angular degrees are affected differently by stellar activity \citep[e.g.][]{Jimenez-Reyes1998,Chaplin2004,Broomhall2012,Salabert2015}, depending on the latitudes where active regions emerge (active latitudes). This fact reveals another capability of asteroseismology, in this case, to constrain active latitudes in stars by investigating the magnetic signatures in modes of different angular degrees, as it was done for the well-characterized solar-analog HD~173701 by \citet{Thomas2019}. However, the suppression of mode amplitudes by magnetic activity prevents the detection of acoustic modes in stars with strong magnetic activity \citep{Chaplin2011b,Mathur2019,Gehan2022,Gehan2024}. Therefore, seismic samples are biased towards weakly active slow rotators. 

The \kep\ mission \citep{Borucki2010}, launched by NASA (National Aeronautics and Space Administration), provided one of the most significant contributions to the expansion of stars with known surface rotation and measured activity levels \citep[e.g.][]{Nielsen2013,Reinhold2013a,Reinhold2023,McQuillan2013a,McQuillan2014,Garcia2014,Ceillier2017,Santos2019a,Santos2021ApJS}. Comparatively, {\it Gaia}, launched by ESA (European Space Agency), has yielded a much larger number of rotation measurements \citep{Lanzafame2018,Distefano2023}. Nevertheless, \kep\ and {\it Gaia} yields are complementary \citep[e.g.][]{Lanzafame2019}, with \kep\ probing typically slower rotators than {\it Gaia}, including stars similar to our Sun (for reference, at 5000 K, the 5\textsuperscript{th} and 95\textsuperscript{th} percentiles of \kep\ \prot\ distribution are 8.2 and 38.8 days, while the analogous limits for \textit{Gaia} are 0.4 and 12.6 days). \kep\ revealed two potential deviations to the Skumanich spin-down law, whose origins are still under debate.

Most of the \kep\ main-sequence sample has already converged to the slow-rotation sequence, where the Skumanich spin-down law is generally assumed to be valid. However, \kep\ data suggests the existence of a transition within this regime, with the surface rotation distribution being bimodal \citep[e.g][]{McQuillan2013a,McQuillan2014,Davenport2018,Santos2019a,Santos2021ApJS}, which is particularly evident at low temperatures, resulting on an intermediate-rotation gap ($\sim 15$ days at 4500 K). Spin-down stalling, likely associated with this gap, is evident in stellar clusters with ages around 1 Gyr \citep{Curtis2019}. Recently the intermediate-rotation gap was found in K2 \citep{Howell2014} and ground-based data for partially convective stars \citep{Reinhold2020,Gordon2021,Lu2022}, but is absent in fully convective stars \citep{Lu2022}. Different hypotheses to explain the gap were proposed, with the core-envelope coupling theory gaining traction \citep[e.g][]{McQuillan2014,Angus2020,Spada2020,Gordon2021,Lu2022}. In this scenario, the angular momentum transfer between the stars' fast core and slow envelope would momentarily stall the spin-down. Once the coupling is completed, the Skumanich-like spin-down would resume. 

The second transition \kep\ unveiled concerns relatively old main-sequence stars, around the age of the Sun and older. Given their asteroseismic ages, some of these stars spin faster than expected if their spin-down was consistent with the Skumanich law \citep{Angus2015,vanSaders2016,Hall2021}. This observation led to the formulation of the weakened magnetic braking (WMB), which would take place around the middle of the main-sequence lifetime, at a given critical Ro \citep[e.g.][]{vanSaders2016,Metcalfe2016,Metcalfe2017,Saunders2023}. Despite the observational support for WMB, its physical cause(s) remains unclear. As the efficiency of angular momentum transport is primarily governed by the stellar magnetic field \citep{Reville2015}, one possible explanation is that the stellar dynamo becomes less efficient, leading to a weaker or more complex magnetic field configuration \citep[explored in numerical experiments, e.g.][]{Brun2022}. This hypothesis has been investigated using spectropolarimetric observations of solar-like stars, which recover the large-scale magnetic field strength and topology \citep{See2019b,Metcalfe2021,Metcalfe2022}. Other explanations range from the influence of latitudinal differential rotation \citep{Tokuno2023}, to decreases in the stellar wind mass-loss rates due to closed magnetic fields \citep{Garraffo2016} or less efficient wind heating/acceleration \citep{Shoda2020}. As the Sun lies around this transition, attempts have been made to measure the present-day solar wind braking torque \citep{Finley2019c}. Both observations and numerical models show the braking torque to be a factor of two to three times smaller than required by the Skumanich relation \citep{Matt2015,Finley2018b}, lending support to the weakened braking hypothesis.

Looking to the magnetic-activity evolution, another transition may exist, the so-called Vaughan-Preston (VP) gap \citep[e.g.][]{VaughanPreston1980,Vaughan1980,Henry1996,GomesDaSilva2021}. The VP gap is characterized by a lack of stars with intermediate Ca \textsc{ii} H \& K emission. However, there is extensive debate in the community surrounding its existence. In particular, when exploring larger and more complete samples, the VP gap attenuates and in some cases almost disappears \citep[e.g.][]{BoroSaikia2018,EBrown2022}. Although \citet{EBrown2022} did not find a clear gap, the authors also found evidence supporting a phase of rapid evolution, consistent with the parameter space of the VP gap. Therefore, it is not clear yet whether the VP gap is a result of a transition in the magnetic-activity evolution or a result of observational bias. So far, there is no evidence for it in \kep\ data. 

These recent discrepancies between the observations and the expected behavior reinforce the need for a better understanding of rotation and magnetic-activity evolution. This review places the \kep\ rotational sample in the context of the known transitions during the main sequence and describes them in more detail in the following sections. 

\section{\textit{Kepler} main-sequence solar-like rotation sample}\label{sec:sample}

The \kep\ mission provided exquisite data for stellar physics. In addition to high-precision photometry, \kep\ monitored the same stars in a continuous, long-term manner, spanning up to 4 years of observations. Still, \kep\ data are not free of systematics and instrumental artifacts. Therefore, it is important to correct and calibrate them \citep[e.g.][]{Jenkins2010,Garcia2011,Garcia2014a}, while preserving stellar signals at long timescales, such as the rotation modulation of slow rotators. Once processed, \kep\ data constrained surface rotation periods and photometric magnetic activity for several tens of thousands of solar-like stars from the main-sequence to the red-giant phase \citep[e.g.][]{McQuillan2014,Garcia2014,Ceillier2017,Santos2019a,Santos2021ApJS}. While the focus of this review is the main-sequence (MS) solar-like stars, it is worth noting that magnetic fields and activity are also found in earlier spectral types \citep[e.g.][]{Balona2015,Balona2019,Mathys2017,Henriksen2023}.

We begin with the sample of 55,252 stars with known rotation periods from \citet{Santos2019a,Santos2021ApJS}\footnote{KEPSEISMIC light curves were adopted in these works and are available on MAST (Mikulski Archive for Space Telescopes): DOI: 10.17909/t9-mrpw-gc07; \href{https://archive.stsci.edu/prepds/kepseismic/}{https://archive.stsci.edu/prepds/kepseismic/}.}, which included subgiant stars by design. To select solely the MS stars we adopt the selection criteria based on the color-magnitude diagram (CMD) from {\it Gaia} Data Release 3 \citep{Gaia_DR3}{, as detailed in Appendix A of \citet{Garcia2023}}. The magnitudes were corrected for extinction and the selection criteria also remove potential binaries and outliers that sit above or below the MS in the CMD, as well as targets with large ($\geq1.2$) renormalized united weighted error \citep[RUWE;][]{Gaia_DR3}, \textit{Gaia} radial velocity variables \citep{Katz2023}, stars in the \textit{Gaia} non-single-star sample \citep{Binaries_GaiaDR3}, and eclipsing binaries \citep{Kirk2016}. As we do not yet fully understand the rotational signals from targets flagged as close-in binary candidates in \citet{Santos2019a, Santos2021ApJS}, we keep those that pass the criteria (1,311 targets). This leaves us with a reference \kep\ sample of 34,898 single MS stars with known rotation rates. The selection criteria are relatively stringent to ensure a clean sample. Applying the same criteria to the sample of \citet{McQuillan2014} would reduce it to 21,685 stars. Comparing the respective clean samples, we verify that the latest catalog still pushed the upper edge of the rotation-period distribution towards slower rotators \citep[see Fig.~12 in][]{Santos2021ApJS}.

\begin{figure}[h]
    \centering
    \includegraphics[width=0.95\hsize]{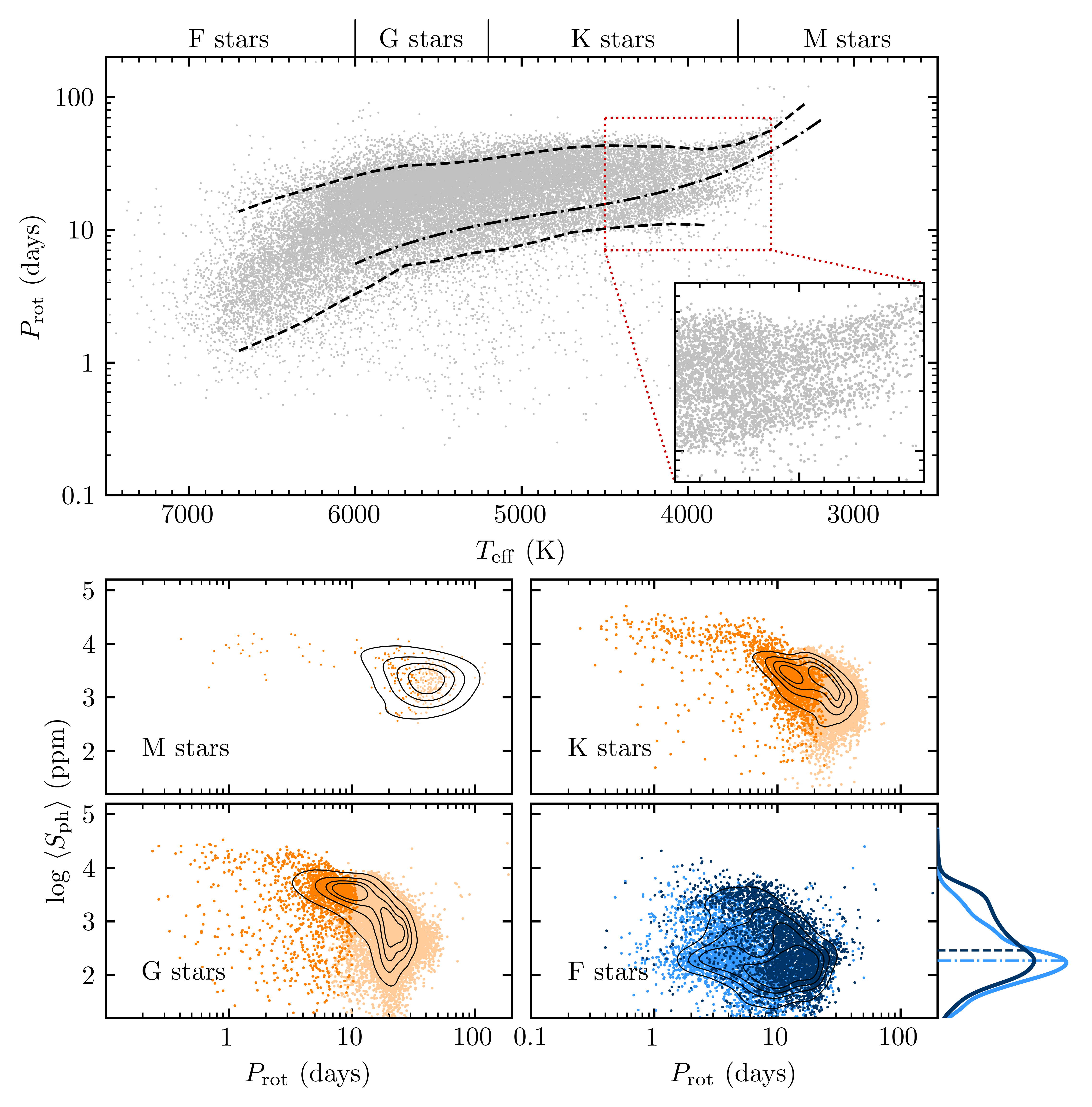}
    \caption{Summary figure of the sample of \kep\ MS solar-like stars with known \prot, illustrating the parameter space probed by \kep. {\it Top:} \prot\ as a function of \teff. The dashed black lines mark the upper and lower edges of the \prot\ distribution. The black dot-dashed line marks the intermediate-\prot\ gap. The subpanel shows a zoom-in into the low-temperature regime. To better see the intermediate-\prot\ gap, we removed the lines. {\it Bottom:} Activity-rotation diagrams for each spectral type. The solid lines mark the density contours. Orange and beige symbols correspond to stars faster and slower than the intermediate-\prot\ gap, respectively. Dark and light blue show F stars cooler and hotter than the Kraft break, respectively. The side panel shows the corresponding \sph\ distributions and median values.}
    \label{fig:summary}
\end{figure}

The top panel of Figure~\ref{fig:summary} shows the \kep\ rotation MS sample. The dashed lines mark the upper and lower edges of the \prot\ distribution, corresponding to the 95\textsuperscript{th} and 5\textsuperscript{th} percentiles, whose origin is discussed in more detail below. In general, hotter stars are fast rotators than cooler stars, which is expected, as for most of the MS, the magnetic braking is more efficient for less massive stars \citep{vanSaders2013,Matt2015}. Another feature that can be seen in the \kep\ sample is the so-called intermediate-\prot\ gap, leading to a bimodal \prot\ distribution. For cooler solar-like stars (K and M), a lower-density region, in between two populations or sequences of stars, can be found. For G dwarfs, the lower density region disappears but the \prot\ distribution is still bimodal \citep[e.g.][]{Davenport2017}. The dot-dashed line indicates the intermediate-\prot\ gap computed for the clean sample as described in \citet{Santos2023}. 

For the \kep\ sample, the magnetic activity is quantified from the rotational modulation in the light curve using the activity proxy \sph, computed in \citet{Santos2019a,Santos2021ApJS} as defined by \citet[][see also \citealt{Garcia2010}]{Mathur2014,Mathur2014b}. The bottom panels of Fig.~\ref{fig:summary} show the activity-rotation diagram for the \kep\ MS sample split by spectral type (\teff\ boundaries at 3700, 5200, and 6000 K). Generally, fast rotators are more magnetically active than slow rotators. The black solid lines show the respective density contours, as reference for the figures below. GKM stars are color-coded according to their location concerning the intermediate-\prot\ gap. The color code for the F stars indicates whether their \teff\ is below or above the Kraft break \citep[6250 K;][]{Kraft1967}. While the activity and rotation of cool F stars are correlated, similarly to the case of the GKM stars, hot F stars tend to be fast rotators with low activity levels \citep[see also Appendix~B in ][]{Santos2021ApJS}. The solid blue lines show the \sph\ distribution for F stars cooler and hotter than the Kraft break, whose median values are indicated by the dashed and dot-dashed lines (287.8 ppm and 185.8 ppm, respectively). This behavior can be explained by the shallow convective envelopes of the latter, which are unable to produce a strong magnetic field yielding inefficient magnetic braking \citep[e.g.][]{vanSaders2013}.

Below we discuss the transitions and detection biases in the \kep\ rotation sample. We discuss other transitions that happen in a regime not probed by \kep\ or, if probed, are not observed in \kep\ data. In the activity-rotation figures, we opt to show \prot\ and not Ro, as \prot\ can be measured directly from the observations. In addition, the comparison between Ro numbers from different studies is not straightforward, as Ro depends on the adopted definition for the convective turnover timescale. Nevertheless, to split the samples into different regimes, when possible, we adopt the Ro number and the respective transitions as determined by the authors in the respective original studies.

\section{From saturated to unsaturated: convergence to the rotation sequence}

The members of young open clusters exhibit a wide range of rotation periods \citep[e.g.][]{Stauffer1987,Soderblom1993,Barnes2003}, consistent with stars transitioning from an ultra-fast rotation (a result of spin-up due to contraction onto the main sequence) to the converged rotation sequence. These two regimes are separated by a lower-density region, i.e. a gap at ultra-fast rotation \citep[few days, depending on \teff;][]{Barnes2003}. The signature of this transition can also be identified in the magnetic activity of low-mass stars, particularly in their X-ray emission \citep[][]{Wright2011}. Two regimes in the coronal magnetic activity are found by \citet[][]{Wright2011}: a \textit{saturated} regime where X-ray emission is almost independent on \prot; and a \textit{unsaturated} regime where X-ray emission strongly depends on rotation, corresponding to the converged rotation sequence. The ultra-fast-\prot\ bimodality found by \citet{Barnes2003} marks the transition from saturated to unsaturated magnetic activity. \citet[][hereafter B2003]{Barnes2003L,Barnes2003} proposed that this transition is related to the core-envelope coupling, where the core and envelope would be coupled for stars in the converged rotation sequence (B2003's \textit{interface} sequence), and decoupled for ultra-fast rotators (B2003's \textit{convective} sequence). However, fully convective M dwarfs also follow the same activity-rotation relation characterized by the saturated and unsaturated regimes \citep{Wright2016,Wright2018}. Alternatively, \citet{Brown2014} attributed the rotation gap to a transition between weak (saturated) to strong (unsaturated) coupling with the stellar wind, which in turn might be related to a transition from complex to simple magnetic field morphology \citep{Reville2015,Garraffo2018}. 

Physically motivated models of angular momentum evolution can describe the observed convergence of rotation, particularly reproducing the mass-dependence of the spin-down \citep[e.g.][]{vanSaders2013,Matt2015}. Lower-mass solar-like stars spend more time in the saturated regime than higher-mass solar-like stars, which converge earlier into the unsaturated regime. Moreover, the efficiency of magnetic braking depends on the rotation period: stars born with fast rotation spin down faster than stars born with slow rotation. Eventually, they converge into the same sequence, the unsaturated regime, where they lose angular momentum following the Skumanich law and, thus, gyrochronology becomes valid. 

Placing the \kep\ sample in context, most of the MS stars have already converged into the unsaturated regime. This is not surprising as the \kep\ sample corresponds to field stars, with a mix of populations of typical ages of several Gyr, and thus stars have had time to spin down. In fact, the transition between the saturated and unsaturated regimes happens at faster rotation rates than the lower edge of the \kep\ \prot\ distribution. Both saturated regime and transition are very sparsely populated in the \kep\ field. Furthermore, this region of the parameter space, particularly $P_\text{rot}<7$  days, is found to be dominated by tidally-synchronized binaries, as determined by \citet{Simonian2019} and \citet{Angus2020}. Close-in binary candidates identified by \citet{Santos2019a,Santos2021ApJS} that survived the selection criteria (Sect.~\ref{sec:sample}) were kept in the \kep\ MS sample, but they are found to occupy the same region of the parameter space as the tidally-synchronized binaries (with some common targets). It is, thus, unclear whether the surviving \kep\ rapidly-rotating targets are young solar-like stars still in the saturated regime or tidally-synchronized binaries, which were not identified as such yet.

\begin{figure}[b]
    \centering
    \includegraphics[width=\hsize]{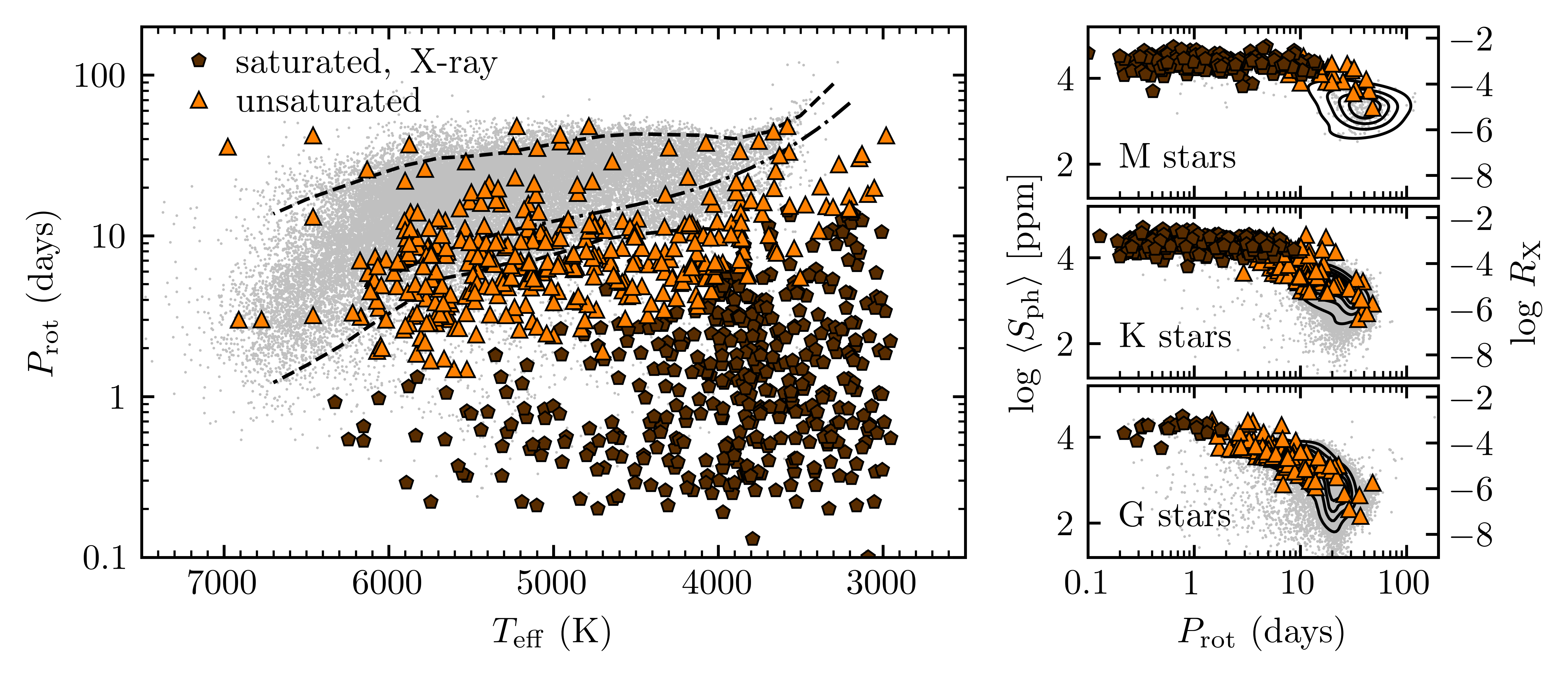}
    \caption{Comparison between the rotation and magnetic activity of the \kep\ sample (gray and black lines) and the X-ray emission sample in \citet[][colored symbols]{Wright2011}: \prot-\teff\ diagram (left) and activity-\prot\ diagram per spectral type (right). The saturated and unsaturated regimes identified by \citet{Wright2011} are shown by the brown hexagons and orange triangles respectively. The solid lines in the activity-\prot\ diagram show the density contours of the \kep\ sample for reference (see Fig.~\ref{fig:summary}). The activity proxy of the \kep\ sample is \sph\ (left y-axis), while $R_\text{X}$ (right y-axis) corresponds to the ratio between the X-ray and the bolometric luminosities for the X-ray emission sample. The right panels employ dual y-axes, avoiding the need to convert an activity proxy into the other. As the \prot\ spans roughly the same interval in both samples, we simply align them.}
    \label{fig:Wright}
\end{figure}

Figure~\ref{fig:Wright} compares the parameter space of rotation and activity for the X-ray emission sample in \citet[][from where we take \teff, \prot, and color index]{Wright2011} and for the \kep\ MS sample. The left panel show the \prot-\teff\ diagram, while the right panels show the activity-\prot\ diagram, where F stars were ignored as they are only a few. The color index is used to compute the convective turnover timescale, according to Eq.~10 in \citet{Wright2011}, which in turn, together with \prot, is used to compute Ro. Ro can vary across different works, depending on the definition of the convective turnover timescale (e.g. the location where it is measured). In this review, we adopt the Ro values as determined by each work, but we focus on the observable \prot. Later in this section, we will compare the different Ro values. In what follows, the subscript of Ro indicates to which work they refer, particularly we use the initial of the first author and the year of the publication: e.g. W2011 for \citet{Wright2011}.

\begin{figure}[b]
\centering
    \includegraphics[width=\hsize]{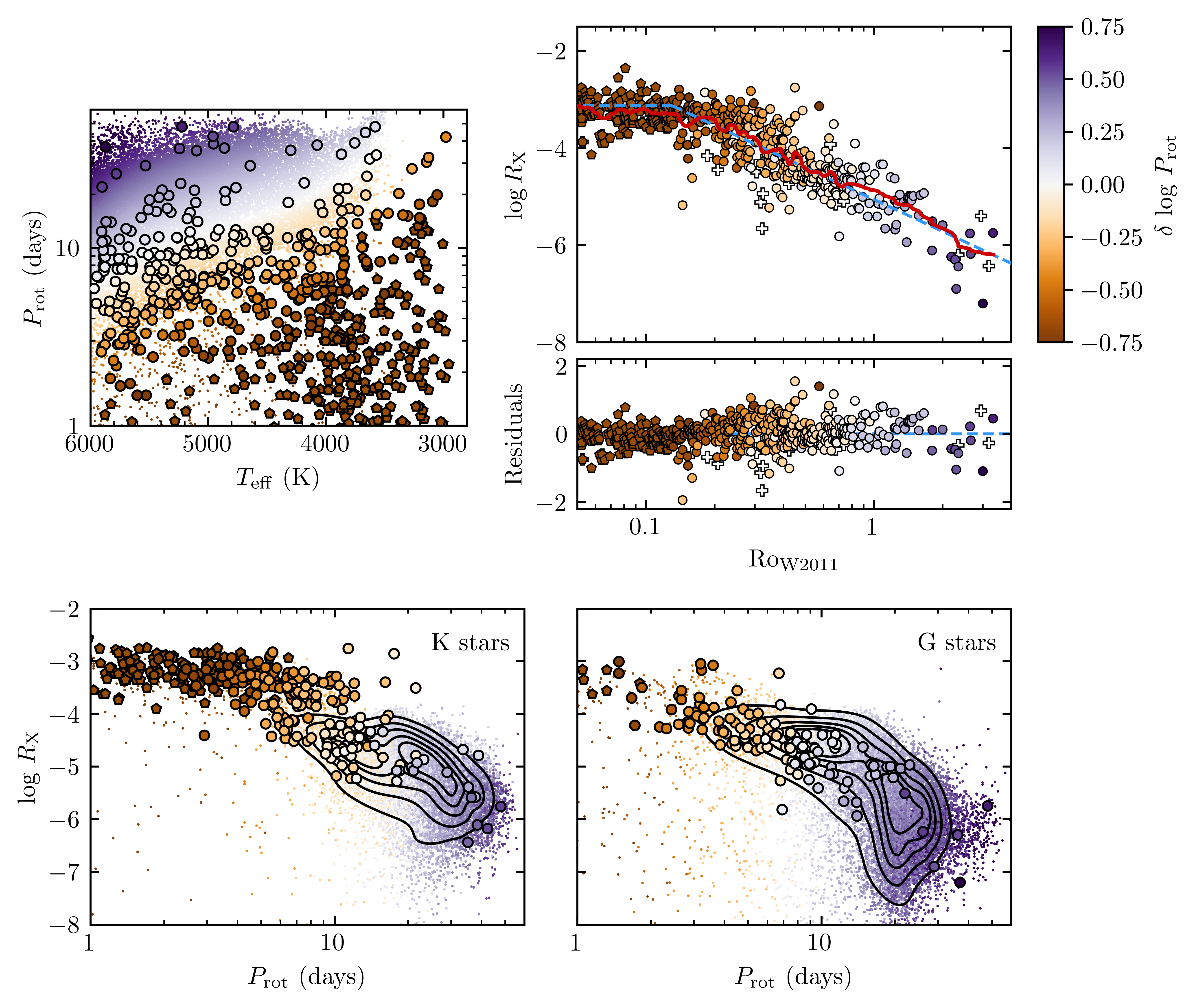}
    \caption{Comparison of the \kep\ and X-ray emission samples, with respect to the intermediate-\prot\ gap. As in Fig.~\ref{fig:Wright}, brown hexagons correspond to the stars in the X-ray saturated regime. Dots (\kep\ sample) and circles (X-ray emission unsaturated regime) are color-coded by their distance to the intermediate-\prot\ gap ($\delta \log P_\text{rot}$). The top left panel shows the \prot-\teff\ diagram. Note that the axes scale differs from the remainder figures, to better show the relevant parameter space. The top right panel shows the X-ray emission as a function of the Rossby number, where the white crosses are the F stars, committed elsewhere in this figure. The dashed blue and solid red lines show the model found by \citet{Wright2011} and the smoothed data, respectively. The residuals between the data and the model are presented in the subpanel. The bottom panels show the X-ray emission as a function of \prot\ for the K and G stars (left and right, respectively). The solid lines show the contours for the \kep\ sample.}\label{fig:Wright2}
\end{figure}

In Fig.~\ref{fig:Wright}, we split the X-ray emission sample into saturated and unsaturated regimes according to \citet{Wright2011}, with the transition $\text{Ro}_\text{W2011}$ set at 0.13 (saturated: $\text{Ro}_\text{W2011}\leq0.13$; unsaturated: $\text{Ro}_\text{W2011}>0.13$). The lower edge of the \prot\ distribution of the \kep\ sample lies above the transition between saturated and unsaturated regimes in X-ray luminosity. This comparison emphasizes the fact that, with \kep, we have access mostly to the stars in the unsaturated regime. Interestingly, it is noticeable that the relation between X-ray emission and rotation (right-hand side) shows a change in slope around the intermediate-\prot\ gap, i.e. where the shape of the \kep\ contours change. Indeed, one can notice that change in the original figures by \citet{Wright2011}. This is also shown in Fig.~\ref{fig:Wright2}, where the color code indicates the distance to the intermediate-\prot\ gap ($\delta \log P_\text{rot}$). The top right panel displays the X-ray emission against Ro$_\text{W2011}$. Stars in the saturated regime (brown hexagons) are partly omitted to better show the unsaturated regime. Around the intermediate-\prot\ gap (lighter symbols), there is a decrease in the dispersion compared to the remainder of the unsaturated regime. At \prot\ longer than the gap ($\delta \log P_\text{rot}>0$), stars seem to follow a steeper relation than before. This can be seen through the comparison between the model by \citet[][dashed blue line]{Wright2011} and the smoothed data (solid red line). In the unsaturated regime, at $\delta \log P_\text{rot}<0$, the smoothed line closely follows the model, while deviating at $\delta \log P_\text{rot}>0$. Indeed, splitting the unsaturated regime in two and fitting them separately, we find slopes of -1.92 and -2.46 for $\delta \log P_\text{rot}<0$ and $\delta \log P_\text{rot}>0$, respectively. The behavior change can also be seen in the bottom panels, where particularly for the K stars it is noticeable that stars with $\delta \log P_\text{rot}<0$ and $\delta \log P_\text{rot}>0$ follow different sequences. This transition in the unsaturated regime was naturally not seen by the authors and is noticed now thanks to the knowledge acquired through the \kep\ sample, which revealed this intermediate-\prot \ gap. It is now clear that the unsaturated regime itself presents multiple regimes. 

The transition from saturated to unsaturated regime corresponding to the ultra-fast-\prot\ bimodality can also be found in other magnetic activity proxies besides X-ray emission. For the Pleiades, with an age of $\sim125$ Myr \citep{Stauffer1998}, the saturated and unsaturated regimes are identified, for example, in the photometric magnetic activity measured from K2 light curves \citep{Brown2021} and in the spot filling factor measured from APOGEE (Apache Point Observatory for Galactic Evolution Experiment) spectra \citep{Cao2022}.

\begin{figure}[H]
    \centering
    \includegraphics[width=\hsize]{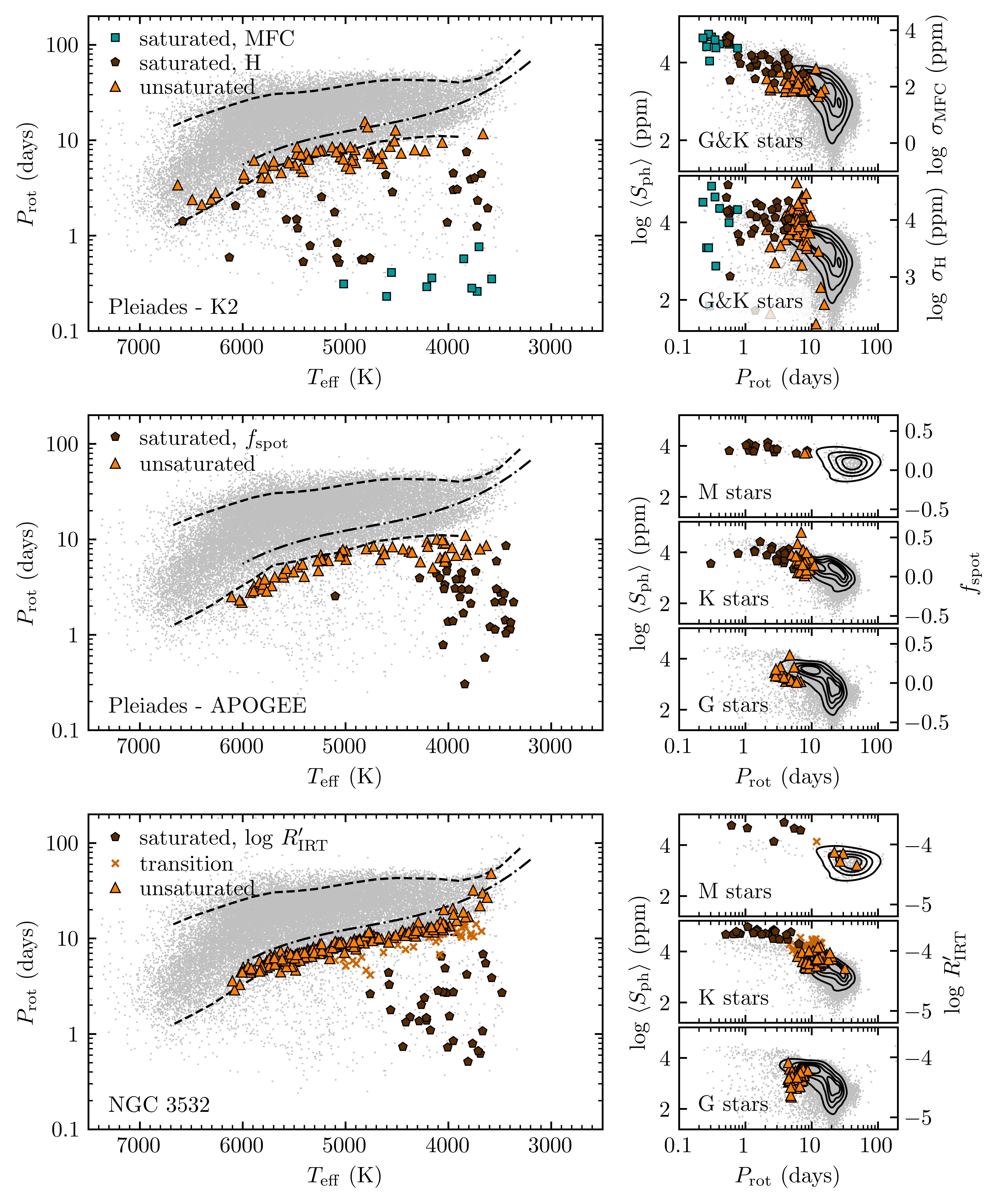}\vspace{-0.2cm}
    \caption{Same as Fig.~\ref{fig:Wright}, but for the Pleiades and NGC 3532. Their activity proxies are $\sigma_\text{H}$ (top), $f_\text{spot}$ (middle), and $\log R'_\text{IRT}$ (bottom), while $\sigma_\text{MFC}$ is the MFC amplitude (top). The top right panels show K and G dwarfs together, while the middle and bottom right panels split the spectral types. The orange and brown symbols correspond to the unsaturated and saturated regimes, respectively. The turquoise shows the MFC-saturated Pleiades members, while the orange crosses show the transitioning NGC~3532 members.}\label{fig:pleiades_ngc3532}
\end{figure}

\citet{Brown2021} discovered a new contribution to the stellar brightness variations, the mid-frequency continuum (MFC). The MFC corresponds to an excess of power between $\sim20$ and $\sim300\,\mu$Hz in comparison with the models for the acoustic background in the power spectrum, which account for photon-shot noise, activity, and two granulation components. Interestingly, the MFC scales with $\text{Ro}_\text{B2021}$, suggesting that it is related to stellar magnetism. Given its timescale, the MFC might be related to the supergranular internetwork (see \citealt{Rincon2018} for a review). However, the MFC and the photometric magnetic activity ($\sigma_\text{H}$, measured as the amplitude of the rotation harmonics, H, in the power spectrum, which is well correlated with \sph) do not vary in phase or follow a similar relation with $\text{Ro}_\text{B2021}$. \citet{Brown2021} found that the MFC also shows two regimes, with the MFC transition taking place at smaller $\text{Ro}_\text{B2021}$ (shorter \prot) than that associated with the ultra-fast-\prot\ bimodality. This is illustrated in the top panels of Fig.~\ref{fig:pleiades_ngc3532}, where the stars saturated in MFC are shown by the turquoise squares. For the photometric magnetic activity, the transition between the saturated and unsaturated regime (brown hexagons and orange triangles) is consistent with the ultra-fast-\prot\ bimodality, i.e. with the saturated and unsaturated X-ray regimes. In the top panel of Fig.~\ref{fig:pleiades_ngc3532}, we take the MFC transition from \citet[][$\log \text{Ro}_\text{B2021,MFC}=-1.65$]{Brown2021}, while for the photometric activity, we adopt $\log \text{Ro}_\text{B2021,H}=-0.7$, which is slightly shifted from the value indicated by the authors ($\log \text{Ro}_\text{B2021,H}=-0.5$). This change is motivated by the fact that the stars lying around that $\text{Ro}_\text{B2021}$ have already transitioned to the unsaturated regime. \teff, \prot, $\text{Ro}_\text{B2021}$, harmonic and MFC amplitudes ($\sigma_\text{H}$ and $\sigma_\text{MFC}$) are adopted from \citet{Brown2021}.

In comparison, \citet{Cao2022} used APOGEE spectra of the Pleiades cluster to estimate the spot filling factor ($f_\text{spot}$), based on the temperature contrast between spots and quiet surroundings. As the activity proxies estimated from K2 light curves and APOGEE spectra are both related to spots, they are expected to show the same behavior. Indeed, that is the case, except with the rotation sequence from \citet{Cao2022}, which is located at slightly shorter \prot\ than that from \citet[][see middle panel of Fig.~\ref{fig:pleiades_ngc3532}]{Brown2021}. To split the $f_\text{spot}$ sample into saturated and unsaturated regimes for Fig.~\ref{fig:pleiades_ngc3532}, we consider the transition at $\log\,\text{Ro}_{\text{C2022},f_\text{spot}}=-0.677$ as determined by \citet{Cao2022} with their power-law model. The authors identified potential binary or multiple systems, which are neglected for the graphical representation.

The saturated-unsaturated transition associated with the ultra-fast-\prot\ bimodality can also be seen in the chromospheric activity proxy measured from the emission in the Ca \textsc{ii} IRT ($\log R'_\text{IRT}$). \citet{Fritzewski2021a,Fritzewski2021b} investigated the rotation and magnetic activity of NGC 3532, whose age is estimated to be around 300 Myr old \citep{Fritzewski2019}. The rotation and magnetic-activity data of this cluster show saturated and unsaturated regimes (bottom panels in Fig.~\ref{fig:pleiades_ngc3532}), consistent with the ultra-fast-\prot\ bimodality. We adopt \teff, \prot, $\log R'_\text{IRT}$, and $\text{Ro}_\text{F2021}$ from \citet{Fritzewski2021b}. According to the authors all the stars with $\text{Ro}_\text{F2021}<0.06$ are in the saturated regime, while stars with $\text{Ro}_\text{F2021}>0.11$ have converged to the rotation sequence. Stars in between would be transitioning from one to the other regime.

\begin{figure}[b]
    \centering
    \includegraphics[width=0.8\hsize]{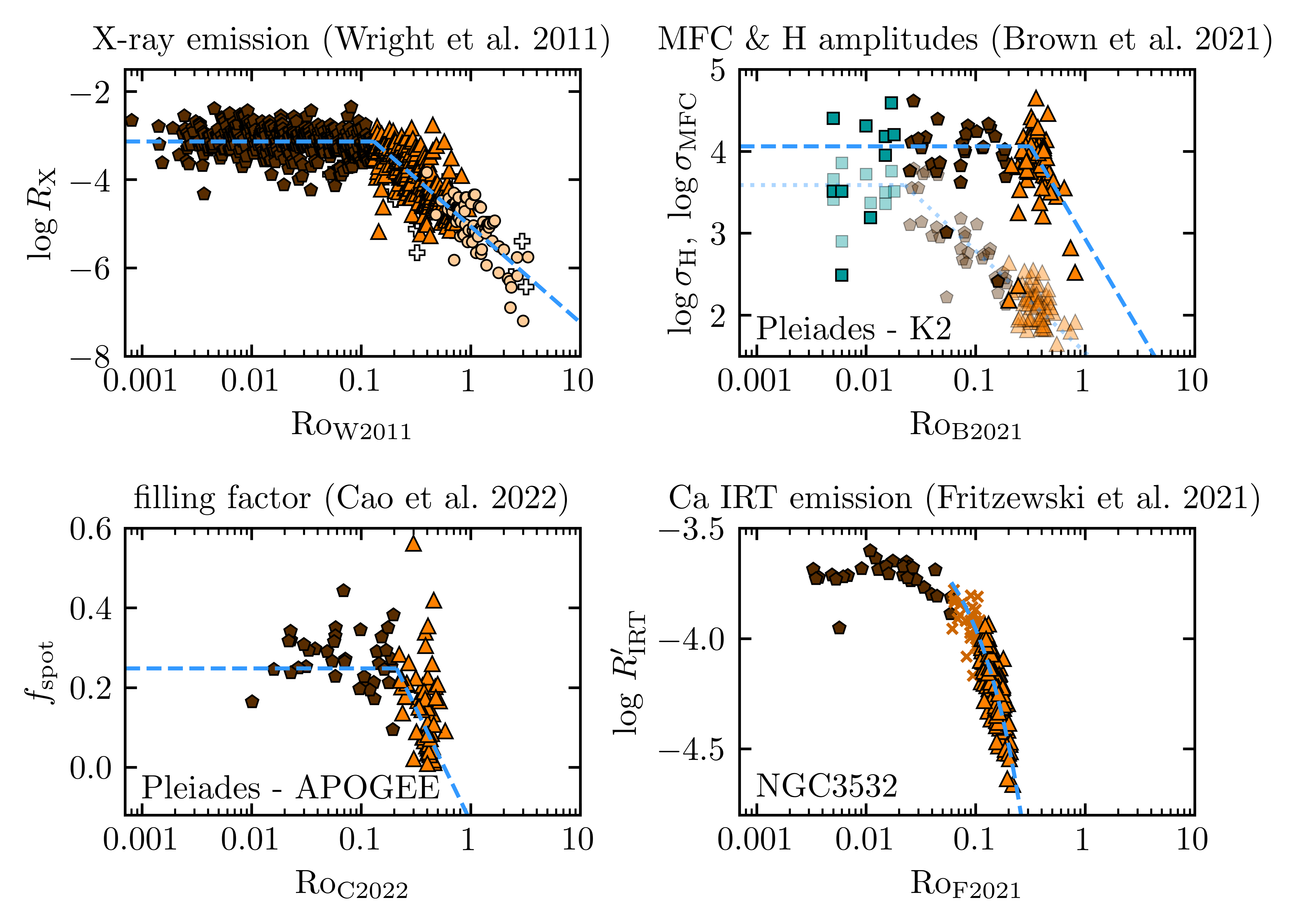}
    \caption{Comparison between the activity-Ro relations for samples shown in Figs.~\ref{fig:Wright} and \ref{fig:pleiades_ngc3532}. The symbols and colors have the same meaning as above, with the addition of the beige circles which represent stars with $\delta\log P_\text{rot}>0$, splitting the unsaturated regime in two. Ro is represented according to each work, which is indicated by the subscript. The blue dashed lines show the best fits found in the respective works. In the top right panel, the opaque symbols correspond to the rotation-harmonic component, while the small transparent symbols and dotted line correspond to the MFC. The x-scale is kept the same in all panels to better illustrate the differences in Ro.}
    \label{fig:Ro_comparison}
\end{figure}

Figure \ref{fig:Ro_comparison} compares the activity-Ro diagrams for the stellar samples in Figs.~\ref{fig:Wright} and \ref{fig:pleiades_ngc3532}. Each panel shows the Ro values as computed in the different works, where the respective best fits are overlaid. Because of the different definitions, the transition from saturated to unsaturated regime happens at different Ro values. However, as seen above, the transitions for the different activity proxies correspond to the same parameter space in terms of the observable \prot. We prefer to focus on \prot\ here, but when comparing Ro from different works, it is recommended to place them in a uniform scale (e.g., by normalizing to the solar value according to the respective definition).

\begin{figure}[h]
    \centering
    \includegraphics[width=\hsize]{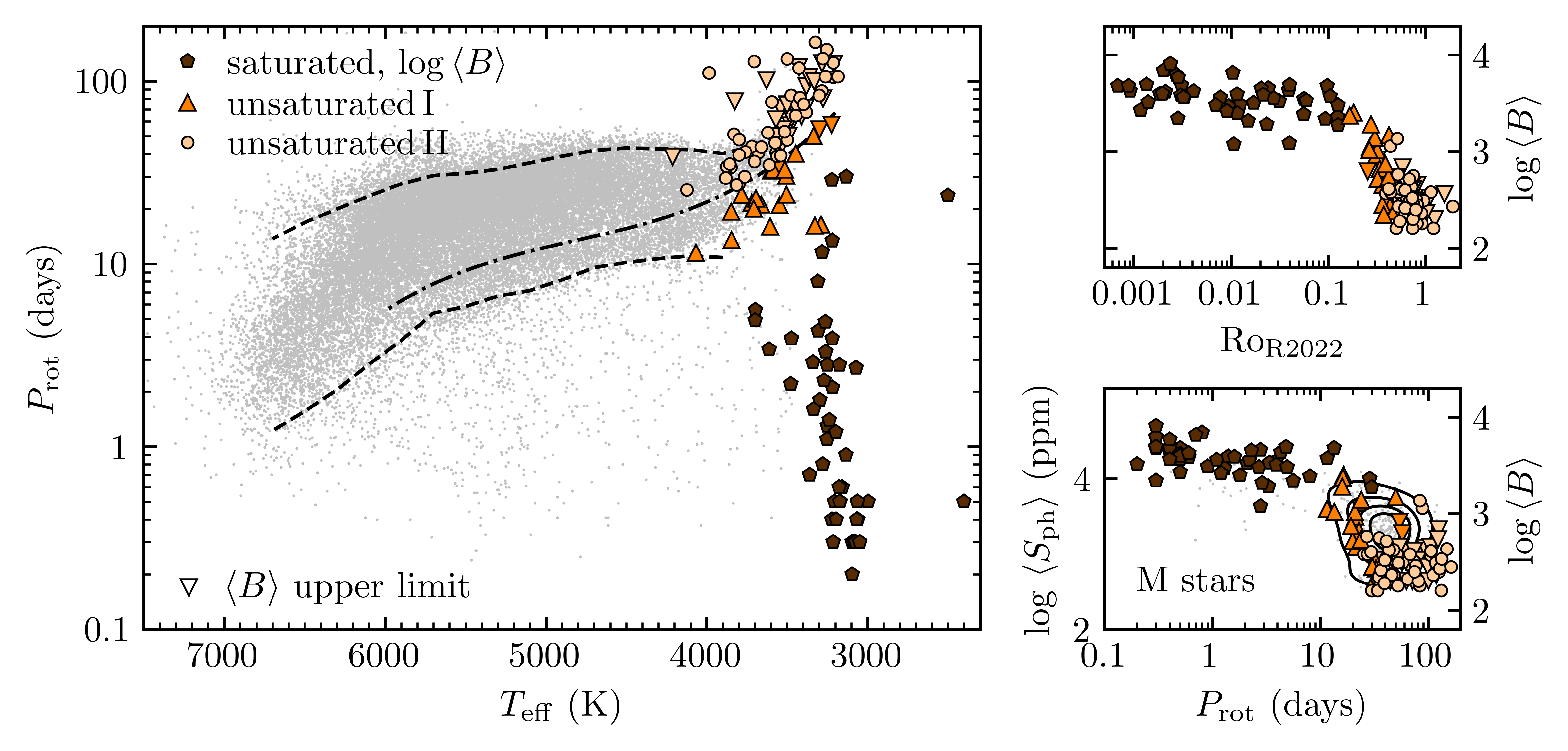}
    \caption{Same as Fig.~\ref{fig:Wright}, but for the M dwarfs in the CARMENES survey. The magnetic activity is measured through the average magnetic field strength $\langle B\rangle$. The \sph-axis scale is slightly different from the other plots. As in Fig.~\ref{fig:Ro_comparison}, the beige symbols represent stars with $\delta\log P_\text{rot}>0$. The downwards triangles identify the stars for which $\langle B \rangle$ is an upper limit.} 
    \label{fig:Reiners}
\end{figure}

\citet{Reiners2022} tracked down the transition from the saturated to unsaturated regime in the average magnetic field strength, $\langle B\rangle$, of M dwarfs (and some K) from CARMENES spectra (Fig.~\ref{fig:Reiners}). Similarly, to the magnetic activity proxies above, the transition identified by the authors at their $\text{Ro}_\text{R2022}=1.3$ takes place near the lower edge of the \prot\ distribution of \kep\ M dwarfs. For the \kep\ sample, this edge is not well defined at low \teff\ due to small sample sizes. Also, the M-dwarf sample of \citet{Reiners2022} is near the location where the gap closes for fully convective stars \citep[at $\sim3500$ K;][]{Lu2022}. Some of the $\langle B \rangle$ measurements are an upper limit. These are mostly located in the more dense region with small $\langle B\rangle$, constituting 37\% of the stars with \prot\ longer than the intermediate-\prot\ gap. We split the $\langle B \rangle$ unsaturated regime into two according to their location in relation to the intermediate-\prot\ gap of the \kep\ sample. In Fig.~\ref{fig:Reiners} we also show $\log\langle B\rangle$ as a function of $\text{Ro}_\text{R2022}$, which was computed using the convective turnover timescale and \prot\ provided by \citet{Reiners2022}.

\begin{figure}
    \centering
    \includegraphics[width=\hsize]{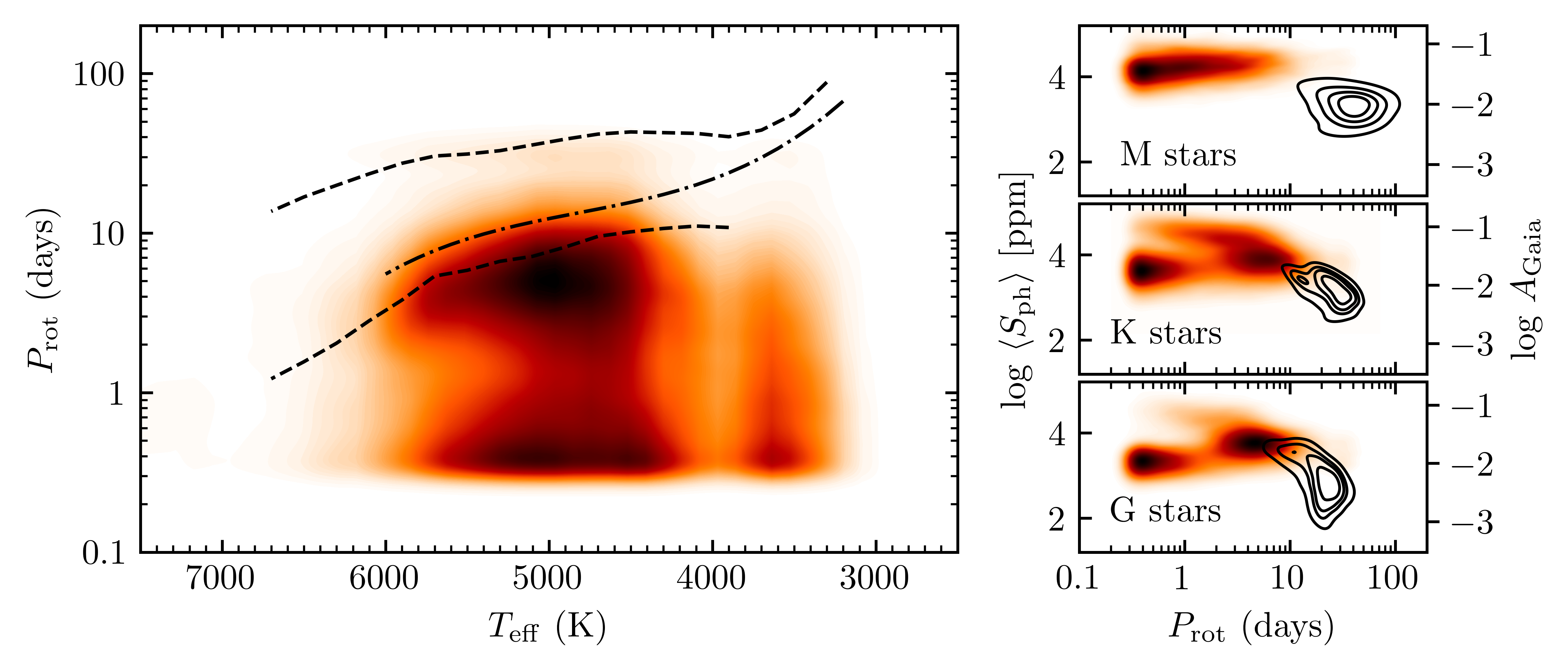}
    \caption{Same as in Fig.~\ref{fig:Wright}, but for the \textit{Gaia} DR3 rotation sample. The \textit{Gaia} sample is shown by the density map, where dark colors correspond to high-density regions. $A_\text{Gaia}$ is the amplitude of the rotation signal in the \textit{Gaia} photometric data. For the \kep\ sample, only the limiting lines and contours are shown.}
    \label{fig:gaia}
\end{figure}

\textit{Gaia} also provides photometric data that allow the measurement of rotation periods. Focusing on the targets with a relatively large number of visits and long temporal coverage, \textit{Gaia} reports \prot\ and the respective photometric activity proxy ($A_\text{Gaia}$, being the amplitude of the rotation signal) for several hundreds of thousands of stars \citep{Lanzafame2018,Distefano2023}. Indeed, the \textit{Gaia} DR3 sample, with the most reliable \prot\ estimates, includes of 474,026 stars \citep[][]{Distefano2023}. To complement the rigorous \prot\ vetting and selection by the authors, which removed evolved stars, we neglect potential binaries (52,933 stars) according to the same criteria used for the \kep\ sample (Sect.~\ref{sec:sample}). Figure~\ref{fig:gaia} compares the final \textit{Gaia} (421,093 stars) and \kep\ samples (34,898 single MS stars). \prot\ and $A_\text{Gaia}$ were taken from \citet{Distefano2023}, while \teff\ was taken from \citet{Andrea2023}. The parameter spaces probed by \text{Gaia} and the \kep\ barely overlap as noted by \citet{Lanzafame2019} and \citet{Distefano2023}. Since \text{Gaia} is not very sensitive to long \prot\ values due to its scanning pattern, all its M stars are still in the saturated regime (Fig.~\ref{fig:gaia}). \textit{Gaia} G and K stars are spread between both the saturated regime and the ``tip'' of the unsaturated regime. For those in the unsaturated regime, it is possible to recognize the expected trend with \prot\ generally decreasing with increasing \teff. \textit{Gaia} also unveiled the existence of stars with $P_\text{rot}<1$ day and very low activity (\textit{ultra-fast rotator branch}), in contrast to the stars in the saturated regime. \citet{Lanzafame2019} hypothesize that stars can either evolve from the saturated regime directly to the unsaturated regime, or to the ultra-fast rotator branch first and from this to the unsaturated regime. This region of the parameter space cannot be explored through the \kep\ sample, as its majority has already converged to the unsaturated regime.

The comparisons above suggest that the lower edge of the \kep\ distribution reflects the transition to the unsaturated regime \citep[see also][]{Matt2015}.

\section{Intermediate-P$_\text{rot}$ gap: regimes within the unsaturated regime}\label{sec:int-prot}

The intermediate-\prot\ gap and the associated \prot\ bimodality were first discovered in the \kep\ field of view by \citet{McQuillan2013a,McQuillan2014}. One of the original hypotheses was that the \prot\ bimodality resulted from two different star formation episodes and it was a feature of the \kep\ field \citep[e.g.][]{McQuillan2014,Davenport2017,Davenport2018}. However, since then it has also been recovered in the different campaigns of K2 \citep{Reinhold2020,Gordon2021}, which focused on different fields of view, and in ground-based data (from the Zwicky Transient Facility survey) covering the full northern hemisphere \citep{Lu2022}. Therefore, these findings suggest that the \prot\ bimodality and intermediate-\prot\ gap are linked to stellar evolution. 

A second hypothesis was postulated by \citet{Montet2017} and \citet{Reinhold2019}. These works found evidence for the stars below the intermediate-\prot\ gap being spot-dominated, while those above the gap being facula-dominated. This led to the proposition that this gap would be related to the transition from the spot- to facula-dominated and it would result from observational biases due to the canceling between dark spots and bright faculae \citep{Reinhold2019}. However, other observational evidence and successful modeling support a third hypothesis.  

The hypothesis proposes that this gap has its origin in the core-envelope coupling \citep{McQuillan2014,Angus2020,Spada2020,Gordon2021,Lu2022}, which leads to a stalling in the envelope's spin-down followed by a period of quick evolution once the coupling is completed \citep{Gordon2021,David2022}. Before and after this transition, stars' envelopes (and thus their surface rotation periods) would follow the Skumanich spin-down law. Starting with a decoupled core-envelope, the surface of the stars below the gap would brake due to magnetized winds. During the coupling between the fast core and the slow envelope, the surface spin-down would stall for a relatively short timescale. 
Once the coupling is completed, the spin-down would resume. In this light, fully convective stars would not face this transition. Indeed, that was what \citet{Lu2022} found in ground-based photometric data of field stars. The authors retrieve a rotational gap in the partially convective stars, but not for fully convective M dwarfs.

\begin{figure}[b]
    \centering
    \includegraphics[width=0.9\hsize]{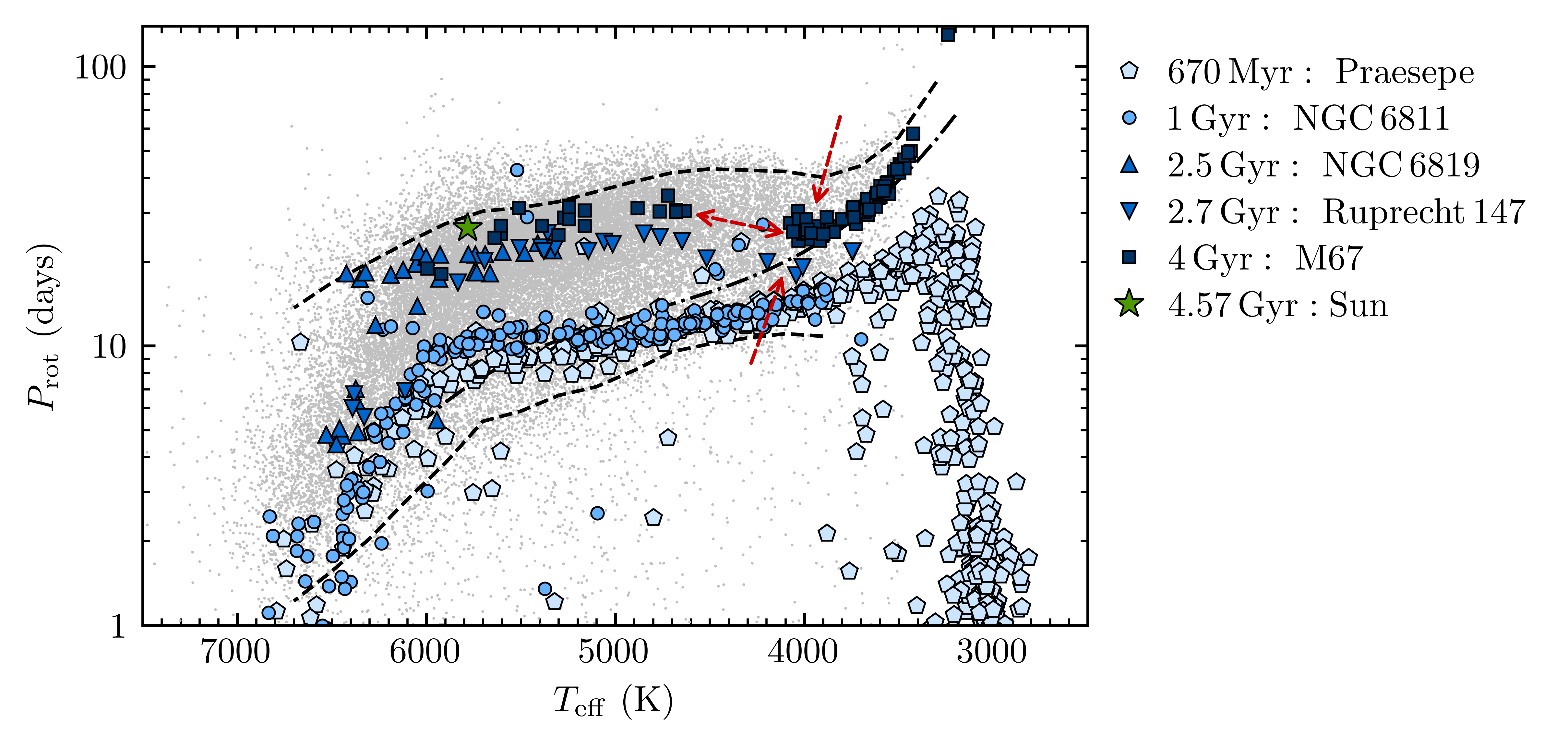}
    \caption{Same as the left panel of Fig.~\ref{fig:Wright}, but for Praesepe, NGC 6811, NGC 6819, Ruprecht 147, M67, and the Sun. The y-axis scale is slightly different from the other plots. The age references are \citep{Bahcall1995,Jeffries2013,Barnes2016,Torres2020}. The one-sided arrows indicate the approximate location of the first deviation to the monotonic behavior in the old clusters' rotation sequence. The double-sided arrow represents the observed increase of \prot\ with \teff, where data is lacking for M67.}
    \label{fig:clusters_gap}
\end{figure}

The rotational sequence of the 1~Gyr NGC 6811 open cluster is consistent with a stalling in the spin-down of K dwarfs \citep[][]{Curtis2019}, which fits the core-envelope coupling hypothesis. The stalling can be seen through the overlap between the rotational sequences of Praesepe \citep[670 Myr;][]{Douglas2019} and NGC 6811 in the regime of K-dwarfs \citep[Fig.~\ref{fig:clusters_gap}; see also][]{Curtis2019,Bouma2023}, which would not be expected according to the Skumanich spin-down law. The theoretical models by \citet{Spada2020}, which account for a mass-dependent core-envelope coupling timescale, are able to reproduce the observations. In particular, the models match the stalling of the spin-down for K dwarfs around 1 Gyr, consistent with the observations of NGC 6811. The mass-dependent coupling leads to a kink in the rotation sequences at older ages, i.e. the rotation-period sequence is no longer monotonic at these ages. As the coupling timescale increases with decreasing mass \citep{Spada2020}, this kink moves towards lower masses with age. This feature is seen in the observational data of older clusters \citep[Fig.~\ref{fig:clusters_gap};][]{Meibom2015,Barnes2016,Dungee2022,Bouma2023}.

Fig.~\ref{fig:clusters_gap} compares the rotational sequences of stellar clusters of different ages with the \kep\ sample. The rotation period and effective temperature for the clusters' members are gathered from different studies \citep{Meibom2015,Barnes2016,Godoy-Rivera2021_cluster,Dungee2022}. Praesepe members are plotted for reference. Most of Praesepe stars with $T_\text{eff}\gtrsim3500$ K have already converged to the rotation sequence, while cooler stars have not. The rotation sequence of Praesepe and younger clusters (Fig.~\ref{fig:pleiades_ngc3532}) shows a general monotonic behavior, with \prot\ decreasing with \teff. The remainder of the clusters show evidence of a stalling in the spin-down associated with the intermediate-\prot\ gap, as discussed above: NGC 6811 partially overlaps with Praesepe; and the older clusters show a kink in the rotation sequence at relatively low temperatures. For the 2.7 Gyr Ruprecht 147, \prot\ increases with \teff\ between $\sim 4000$ and $\sim 4800$ K, while elsewhere it decreases with \teff. For the 4 Gyr M67,  the rotation sequence starts by decreasing with \teff\ until $\sim 3900$ K, where the behavior inverts. The \prot\ values at $\sim 3900$ K are shorter than those at $\sim 4600$ K, suggesting that \prot\ increases within this interval, despite the lack of data.

Finally, we remind that the imprint of the transition associated with the intermediate-\prot\ gap can be seen in the relation between X-ray emission and rotation period (Figs.~\ref{fig:Wright} and \ref{fig:Wright2}). Similarly, it can also be identified in the relation between chromospheric emission and rotation (Fig.~\ref{fig:Egeland}, discussed below).

\section{Vaughan-Preston gap: discontinuity in the magnetic activity }\label{sec:vp}

Focusing on stars with longer periods than the intermediate-\prot\ gap, another gap has been observed, possibly related to a transition at later stages of stellar evolution. Contrary to the other sections, this transition is not seen in the rotation period. Instead, this transition is seen in the chromospheric activity, where there is a scarceness of stars with intermediary Ca \textsc{ii} H \& K emission \citep[e.g.][]{VaughanPreston1980,Vaughan1980,Henry1996,GomesDaSilva2021}. This is known as the Vaughan-Preston (VP) gap. The origin of the VP gap has been contested since its finding: is it of astrophysical origin or an observational bias? Indeed, when investigating larger stellar samples, the gap is not observed anymore \citep{BoroSaikia2018,EBrown2022}. Nevertheless, \citet{EBrown2022} concluded that their results are consistent with chromospheric-emission bimodality and associated transitions.

\begin{figure}[h!]
    \centering
    \includegraphics[width=\hsize]{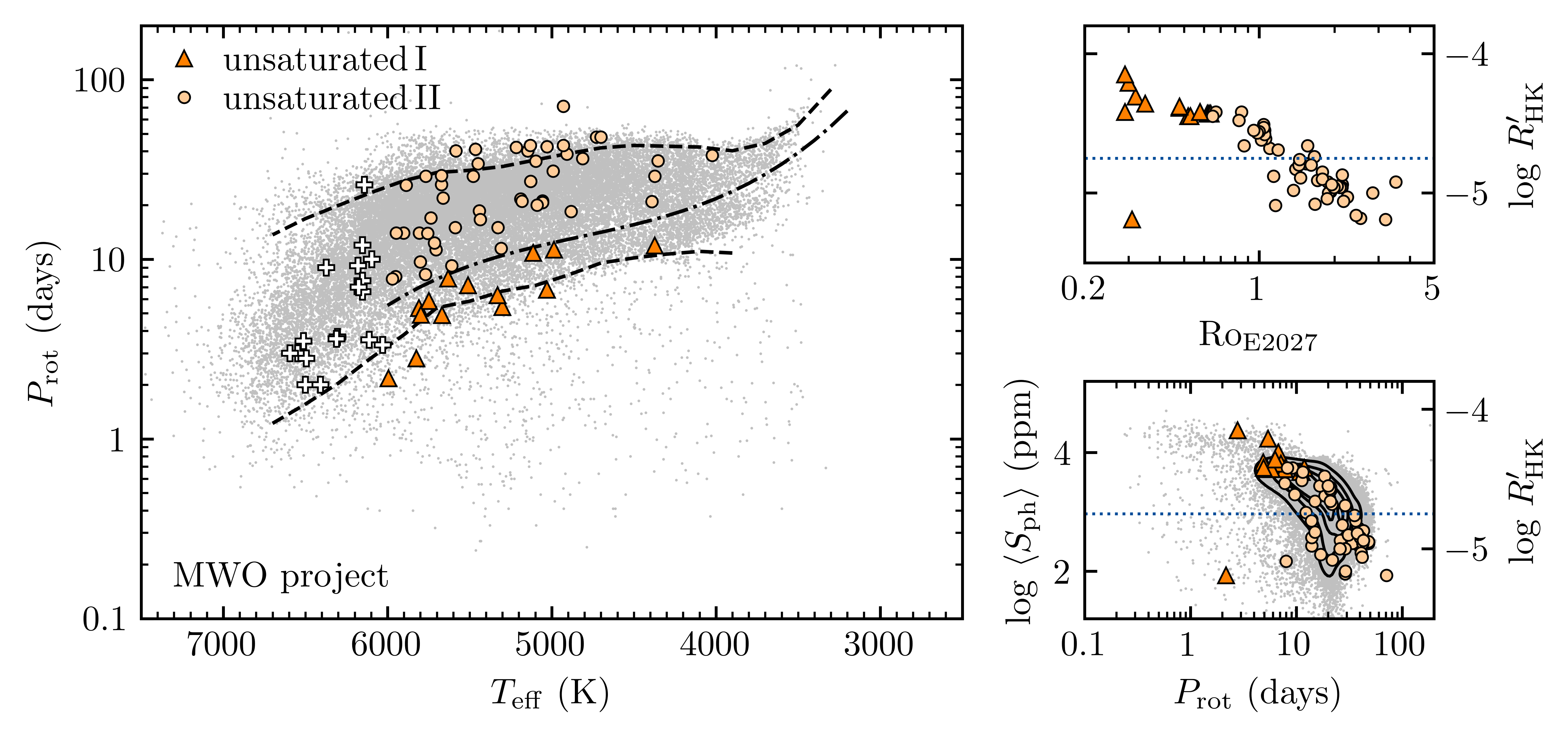}
    \caption{Same as Fig.~\ref{fig:Reiners}, but for the MWO stars. The dotted blue line marks the observed VP gap in the chromospheric emission, $\log\,R'_\text{HK}$. In the bottom right panel, only the G and K \kep\ dwarfs are shown in gray, with the respective contours. As in Fig.~\ref{fig:Wright2}, the white crosses denote F stars.}
    \label{fig:Egeland}
\end{figure}

Figure~\ref{fig:Egeland} compares the Mount Wilson Observatory \citep[MWO; e.g.][]{Wilson1968,Wilson1978,Baliunas1995} sample with the \kep\ sample: \prot, $\text{Ro}_\text{E2017}$, $\log\, R'_\text{HK}$, and (B-V) were adopted from \citet[][see also \citealt{Lehtinen2021}]{EgelandThesis}. For the MWO sample (monitored over two decades), we transform the color index (B-V) into \teff, taking into account the metallicity when available. We retrieve metallicity estimates from \citet{Valenti2005}, \citet{EgelandThesis}, and APOGEE \citep{Abdurro'uf2022} for 60 stars, with a mean value of -0.07 dex. For the remainder, we opt to assume solar metallicity. For the (B-V)-\teff\ conversion, we use the publicly available routine in \texttt{PyAstronomy}\footnote{\href{https://github.com/sczesla/PyAstronomy}{https://github.com/sczesla/PyAstronomy}} \citep{pya} based on \citet{Ramirez2005}\footnote{\href{https://pyastronomy.readthedocs.io/en/latest/pyaslDoc/aslDoc/aslExt_1Doc/ramirez2005.html}{https://pyastronomy.readthedocs.io/en/latest/pyaslDoc/aslDoc/aslExt\_1Doc/ramirez2005.html}}. One of the targets is ignored because it is outside of the valid parameter space (B-V$>$1.51). The stars below and above the intermediate-\prot\ gap (Sect.~\ref{sec:int-prot}) are shown in orange and beige, respectively: the dash-dotted line obtained for \kep\ was used to split the stars. The VP gap located at $\log\,R'_\text{HK}=-4.75$ is shown by the blue dotted line. As shown by \citet{Santos2021ApJS,Santos2023}, there is no evidence of the VP gap in the \kep\ data, which is several times larger than the ground-based chromospheric emission samples. On the one hand, this could mean that indeed there is an observational bias in the ground-based surveys, and when increasing the sample size, the gap is no longer found. On the other hand, although \kep\ observations are relatively long-term, their 4-yr length is still limiting especially when dealing with a variable property such as magnetic activity. Taking solar data,  \citet{Santos2023} showed how the inferences from 4-yr observations on magnetic activity and its variation can change over time, depending on the phase of the cycle. The \sph\ variation strongly depends on the average activity level of the star and on its rotation rate. Consequently, the limiting 4-year monitoring of stars can lead to a smearing of the data points, potentially hiding the possible VP gap. In this case, the large sample size can contribute to a greater concealment.

\section{Midlife transition: Weakening of the magnetic braking}

In recent years, evidence for a possible weakening of the magnetic braking was found \citep[e.g.][]{vanSaders2016,Metcalfe2016,Hall2021,Masuda2022b}. The surface rotation of the \kep\ MS asteroseismic targets was found to be too fast in comparison to what was expected given their asteroseismic age \citep{Angus2015,vanSaders2016}. To explain the observations, \citet{vanSaders2016} proposed that the magnetic braking would become less efficient at a critical Ro, the so-called weakening of the magnetic braking (WMB). In this case, \prot\ evolves according to the Skumanich relation until reaching the critical Ro where its evolution is no longer driven by magnetic braking.

The origin of this transition is presently uncertain. One theory suggests that stellar differential rotation weakens around the critical Ro \citep{Metcalfe2016}, disrupting the stellar dynamo process \citep[see the review by][]{Brun2017}. In numerical simulations, this can be associated with a shift from solar-like differential rotation, which has a fast equator and slow poles, to solid body or even anti-solar rotation \citep{Brun2022, Noraz2022}. It has been suggested that this transition could result in a change in magnetic topology from large-scale to small-scale fields \citep{Metcalfe2017, Metcalfe2022}. However, as the large-scale magnetic field typically governs the efficiency of the magnetic braking \citep{Finley2018a}, and spectropolarimetric observations have shown it does not abruptly disappear at the critical Ro \citep{See2019b}, it appears more likely that the overall stellar magnetic field strength weakens \citep{Metcalfe2023}. This decrease may also reduce the stellar wind mass-loss rate \citep{Shoda2023}, further weakening the wind braking. In other theories, the evolution of the latitudinal differential rotation is sufficient to stall the rotation-evolution of stars around the critical Ro, due to the active latitudes that couple surface rotation to the stellar wind \citep{Tokuno2023, Finley2023}.

The magnetic cycles of Sun-like stars are also observed to become longer \citep{Soon1993,Brandenburg1998,Bohm-Vitense2007}, with some appearing to lose all cyclic variability akin to the Maunder minimum observed for the Sun \citep{Baum2022}. Interestingly, the Sun is near this critical Ro, which raised the question of whether the Sun could be in transition \citep{Metcalfe2016} and whether the Maunder Minimum could be a symptom of this. Observations point towards the Sun's magnetic braking being two to three times smaller than required by the Skumanich relation \citep{Finley2018b,Kasper2021}. However, the magnetic braking timescale for the Sun is around 10-100 Myrs, so this disagreement could be explained by long-term variation in solar activity that exceeds the available $\sim$10,000 years of cosmogenic radionuclide records \citep{Finley2019b}.

\citet{David2022} and \citet{Metcalfe2023} identified an over-density of stars close to the upper edge of the \kep\ \prot\ distribution. This might support the WMB as old stars would remain with a largely unchanged surface rotation, which in turn would result in a pileup of stars with different activity levels near this edge. This upper edge is also recovered in \citet{vanSaders2019} who simulated the \kep\ field population using angular momentum evolution models accounting for the WMB. However, we must note that the \kep\ surface rotation sample does not probe very well this region of the parameter space as it coincides approximately with the current detection limit, which depends on the magnitude \citep[for reference, at \kep\ magnitude 14, it lies around 100 ppm;][]{Mathur2023}. In both cases, if the stars follow the Skumanich spin-down law or if they face a WMB, stars beyond this point in the main sequence are weakly active and have small amplitude brightness variations due to active regions. Thus, the \kep\ \prot\ upper edge can be either the result of one of these effects or the combination of both.

\begin{figure}[h]
    \centering
    \includegraphics[width=\hsize]{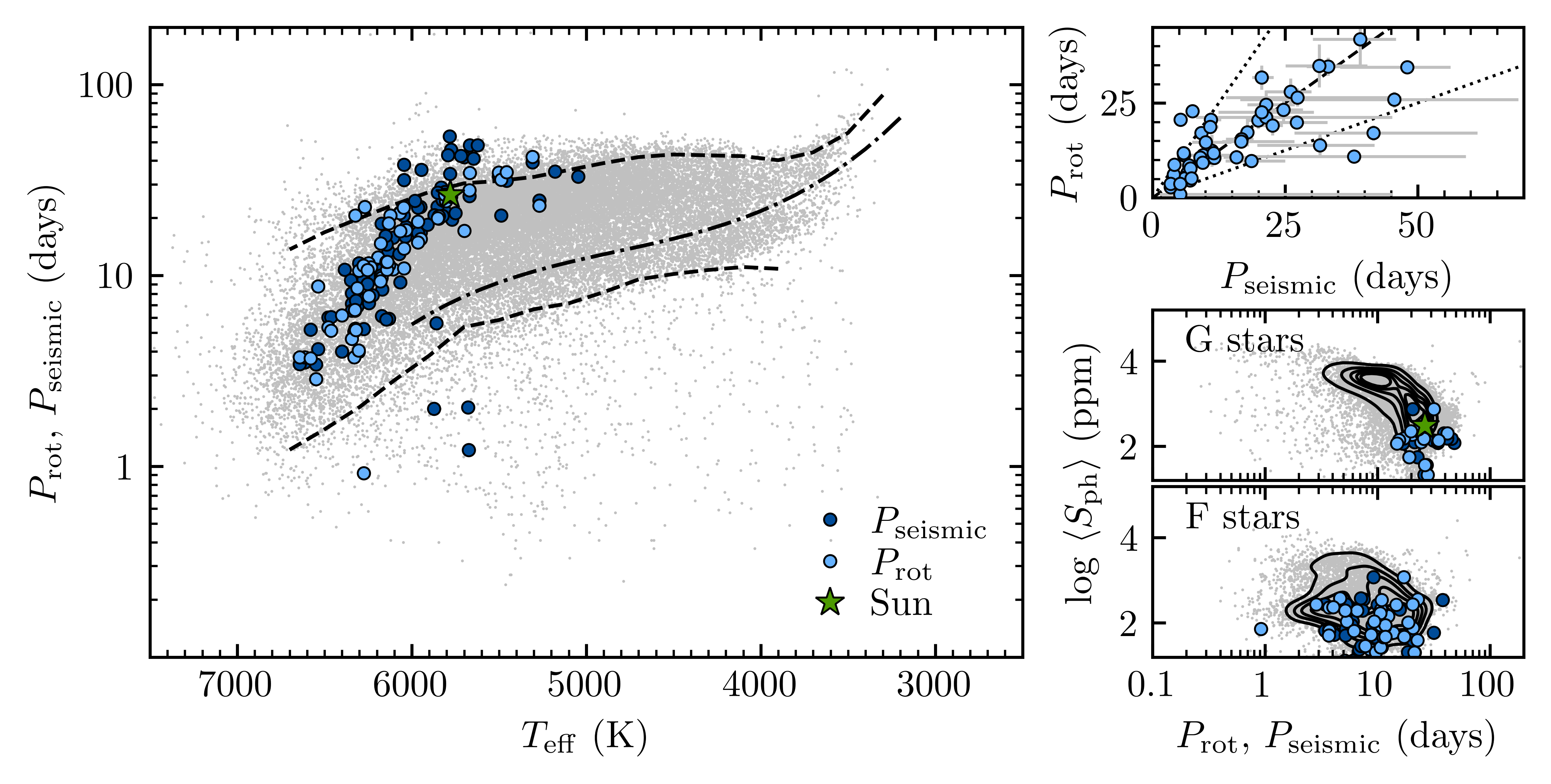}
    \caption{Same as Fig.~\ref{fig:Wright}, but for the \kep\ asteroseismic sample and the Sun. Dark blue indicates the seismic periods $P_\text{seismic}$ as determined by \citet{Hall2021}, while the light blue indicates the surface periods \prot, which were taken by cross-matching with \citet{Santos2019a,Santos2021ApJS}, from where we also took \sph. The top right panel compares the surface (\prot) and seismic ($P_\text{seismic}$) periods for the stars with both estimates (56 stars). The dashed and dotted lines mark the 1:1, 1:2, and 2:1 lines. }%
    \label{fig:Hall}
\end{figure}

Asteroseismology could come to the rescue as it is easier to detect acoustic oscillations in weakly active stars \citep[e.g.][]{Chaplin2000,Jimenez2002,Santos2018,Mathur2019}. This is also illustrated in Fig.~\ref{fig:Hall}, as the seismic sample tends to be close to the upper edge of \prot\ distribution with small \sph\ values. Sadly, the \kep\ main-sequence asteroseismic sample with a high enough signal-to-noise ratio to measure rotational splittings is relatively small (94 stars of different mass and metallicity). Nevertheless, \citet{Hall2021} was able to retrieve seismic rotation periods, $P_\text{seismic}$, for this sample confirming that old stars are rotating faster than expected. Therefore, this discrepancy cannot be solely explained by an observational bias in the \prot\ sample. 

Figure~\ref{fig:Hall} compares the \kep\ seismic-rotation and surface-rotation samples. The 94 stars from \citet{Hall2021} are shown in dark blue, where their rotation period corresponds to that measured through asteroseismology. Cross-matching this sample with \citet{Santos2019a, Santos2021ApJS}, we find 56 stars (light blue), out of the 94, with measured surface rotation and \sph. In the right, only the 56 stars are shown (twice, in light and dark blue), as \sph\ is not available for the remainder. We opt to show both \prot\ and $P_\text{seismic}$ to demonstrate that both sets of measurements occupy the same parameter space. However, we note that there are some discrepancies between the seismic surface rotation periods \citep[top right panel of Fig.~\ref{fig:Hall};][see also \citealt{Benomar2015} and \citealt{Breton2023}]{Hall2021}. Some differences are expected, arising from differential rotation, as the observed acoustic modes are most sensitive to the subsurface layers \citep[e.g.][]{Basu2012,Benomar2015} and might be sensitive to different latitudes in comparison to the active-region latitudes. Additionally, there are uncertainties inherent to both techniques. While seismic detections are easier for weakly active stars, which tend to be slow rotators, these are associated with small effects on the acoustic modes, hampering the estimation of $P_\text{seismic}$. Furthermore, mode linewidths increase with effective temperature \citep[e.g.][]{Appourchaux2012,Appourchaux2014,Corsaro2012,Lund2017}, which poses a challenge because, depending on their values, they can lead to an overlap between azimuthal components, preventing constraints on $P_\text{seismic}$. There is also a correlation between rotational splitting and inclination angle \citep[e.g.][]{Ballot2006}. As our sensitivity to the active latitudes depends on the inclination angle, inclination has an impact on the amplitude of the light curve modulation due to active regions. Nevertheless, the impact on surface rotation estimate should not be significant. Surface rotation measurements can also be biased toward the half of the true \prot\ \citep[e.g.][]{McQuillan2013a,McQuillan2014}, but this corresponds to a small percentage of the targets and efforts were made to avoid such misestimations \citep{Santos2019a,Santos2021ApJS,Breton2021}.
Finally, photometric pollution by nearby stars can also influence the \prot\ estimate.

While, as described above, there is evidence supporting the WMB, we emphasize that this regime is at the \kep\ detection limit of the surface rotation and that the seismic sample is relatively small. Factors such as the spectral type (many are F stars, with thin convective layers), differential rotation, active latitudes, and chemical composition \citep[see also][]{Claytor2020} can impact rotation and magnetic-activity evolution. Based on ground-based spectroscopic data of solar twins, \citet{Lorenzo-Oliveira2019} concluded that their results favor a gradual rotation evolution, rather than a broken spin-down law. Future observations are needed to shed light on the evolution of these properties beyond the middle of the star's lifetime.

\section{Conclusions and perspectives}

Undoubtedly pivotal for exoplanetary and stellar physics, with the end of its second life --the K2 mission--, \kep\ concluded its operations in the fall of 2018. Despite this, to this day, \kep\ data continue to be the focal point of numerous studies, consistently contributing to significant discoveries in the field. Considering stellar rotation and magnetic activity, CoRoT \citep{Baglin2006a} had already provided an important sneak-peak into the potential of space-based photometry \citep[e.g.][]{Garcia2010,Affer2012}. But \kep\ allowed us to investigate for the first time the rotation and magnetic activity of tens of thousands of solar-like stars. In particular, despite its observational limitations, \kep\ extended the number of old weakly active stars, like the Sun, with measured rotation and magnetic activity.

In this review, concerning \kep, we focused mostly on the information from the rotation modulation and briefly on asteroseismic inferences. However, it is important to acknowledge that flares can also provide information on the magnetic activity of stars and the environment around them \citep[e.g.][]{Davenport2016,Yang2019,Notsu2019}. We place the \kep\ sample in the context of established or potential transitions in the evolution of rotation and magnetic activity of MS solar-like stars. \kep\ revealed two unexpected transitions: the intermediate-\prot\ gap and the WMB. The former is thought to be a transition in the stellar evolution related to the core-envelope coupling, while the physical mechanisms driving the latter are still under debate. One of the main reasons for this debate arises from observational limitations, because of the small number of stars in this regime with measured rotation periods, as these stars are weakly active and below the photometric threshold detection.

Stars are born with strong magnetic activity and rapid rotation rates. As they spin down due to magnetized winds, they eventually converge onto a narrow rotation sequence, transitioning from a saturated regime to an unsaturated regime in magnetic activity. \kep\ however did not probe the saturated regime well, as most of its stars are relatively old. The transition between saturated and unsaturated regimes coincides roughly with the lower edge of the \kep\ \prot\ distribution ($P_\text{rot}\sim 8$ days at $T_\text{eff}=5000$ K).

The unsaturated regime itself splits into different regimes separated by the intermediate-\prot\ gap ($P_\text{rot}\sim 12$ days at $T_\text{eff}=5000$ K). Currently, the most plausible hypothesis for this observed transition is the core-envelope coupling. This coupling would momentarily stall the surface spin-down due to the angular momentum transfer between the fast-rotating core and the slow-rotating envelope. This transition can also be noticed in the X-ray emission sample in \citet{Wright2011}, which was not established before.

In ground-based data, evidence of a possible gap or transition at intermediate magnetic activity has also been found, corresponding to the Vaughan-Preston (VP) gap ($P_\text{rot}\sim 30$ days at $T_\text{eff}=5000$ K). Although there is no evidence of the VP gap in \kep\ data, it cannot be discarded. On one hand, \kep\ could support the case for observational bias and the nonexistence of such a gap. On the other hand, the 4-year timespan of \kep\ observations may be insufficient to fully cover magnetic cycles, especially of solar analogs, leading to a cumulative dispersion that may be enough to conceal the VP gap. This shows the need for longer-term observations to reach a more complete knowledge of the magnetic activity.

The \kep\ rotation distribution is characterized by a well-defined upper edge ($P_\text{rot}\sim 40$ days at $T_\text{eff}=5000$ K), with the Sun lying near this edge. In recent years, signs of a change in stellar evolution around this edge have been found in seismic observations of solar-like stars, leading to the postulation of the weakening of magnetic braking (WMB). However, the WMB is still under debate, due to observational limitations arising from the hard-to-detect small-amplitude signals. In addition, instrumental artifacts hamper the recovery of rotation periods for slow rotators \citep[e.g.][]{Breton2021,Santos2021ApJS}. If these artifacts can be better mitigated, it will be possible to extend the number of detections in this regime.

This review also shows that the observed behavior of the different activity proxies is similar. In spite of being sensitive to different layers of the stellar atmosphere and different magnetic features, the activity proxies still relate to each other. As a consequence, the well-established transitions in the stellar evolution are found in all. This also reinforces the validity of photometry to measure and investigate magnetic activity.

\textit{Gaia} has prompted an extraordinary expansion of the number of stars with known rotation and magnetic activity proxies, allowing these properties to be measured for several hundreds of thousands of stars. However, the parameter spaces of \kep\ and \textit{Gaia} samples do not overlap significantly. Indeed they are complementary: with \textit{Gaia} it is possible to investigate young ultra-fast rotators, which were mostly absent from \kep\ data. \textit{Gaia} rotation sample yielded the discovery of ultra-fast rotators with unexpected weak magnetic activity. Furthermore, as can be seen above, it is also possible that some of the \textit{Gaia} stars are crossing the intermediate-\prot\ gap. Thus, in spite of the small sample size in this region of the parameter space, \textit{Gaia} might also provide insights into this transition.

Currently, NASA's TESS \citep[Transiting Exoplanet Survey Satellite;][]{Ricker2014} is mostly sensitive to the fast strongly active rotators. Nevertheless, future TESS extended missions can expand the reach of this mission, but the photometric precision still will not allow the detection of small amplitude signals.

Future ESA's PLATO \citep[PLAnetary Transits and Oscillations of stars;][]{Rauer2014} mission, planned to be launched in 2026, will observe stars with high precision and continuously for at least two years. PLATO is expected to greatly increase the number of MS solar-like stars and subgiants with seismic detections. The relatively long-term observations will also allow us to measure rotation and magnetic activity both from rotational modulation due to active regions and from asteroseismology. However, as discussed above, one can argue that observations longer than two years are needed to provide a better characterization of magnetic activity. Still, thanks to its high precision, PLATO will be crucial in expanding the observational limit towards more weakly active slower rotators, moving the upper edge of the observed \prot\ distribution towards longer \prot, which is currently a major challenge in the field. This will provide more information on the rotation and magnetic-activity evolution beyond the age of the Sun, particularly in the regime of the proposed WMB. PLATO is also expected to increase the overlap between seismic and surface rotation samples, both by increasing seismic detections with high SNR and by pushing the detection limit of surface rotation. This overlap will provide complementary information about the stars, particularly accurate ages and different measures of magnetic activity and rotation. This will greatly improve our understanding of stellar evolution and also differential rotation. Moreover, PLATO will observe bright stars, which is a clear advantage in relation to \kep, meaning that complementary ground-based spectroscopic observations will also be possible. This will allow having independent constraints on rotation and magnetic activity, as well as better characterization of the atmospheric parameters of the stars, namely effective temperature and metallicity. Particularly, better metallicity measurements will help to improve our knowledge of the impact of chemical composition on the magnetic-activity and rotation evolution.

Having observations in different wavelengths will also bring complementary information on the magnetic activity in the different layers of the stellar atmosphere. Furthermore, while dark spots dominate in the passband of the photometric missions mentioned above at the rotation timescale \citep{Shapiro2016,Li_Basri2024}, in other wavelengths the bright facula or plage become the most prevalent. As these bright features live longer than spots, they have the potential to produce more stable signals. Therefore, moving away from the optical can increase the rotation yields for stars like the Sun and older \citep{Li_Basri2024}. Again, this will be particularly important to investigate the WMB and understand the evolution beyond the solar age. Still, long-term observations are required to properly characterize magnetic activity.

PLATO will also be important to investigate the existence of the VP gap, however, the smearing effect described above can be worse in the case of the 2-year observations (in comparison to the \kep\ 4-year observations). PLATO will also provide more observations for stars around the intermediate-\prot\ gap and perhaps provide better age constraints for these than what we had before. This can help us to better depict and understand the transition associated with this gap.

One of the most relevant applications of rotation and magnetic activity is age-dating stars. Asteroseismology provides the most precise way to estimate stellar ages, but it is available for a small number of stars. For example, magnetic activity is known to suppress the already small amplitudes of acoustic modes, making seismic detections not possible for relatively high activity levels. In contrast, the detection and characterization of rotation modulation thanks to active regions is easier for such active stars. In principle, this would allow us to provide stellar ages for a large number of stars through gyrochronology, magnetochronology, and/or gyromagnetochronology. However, one needs to better understand the evolution of rotation and magnetic activity. Therefore, unless we understand (or at the very least calibrate) all of these transitions discussed above, the power of age-dating using rotation and magnetic-activity proxies is heavily endangered \citep[e.g.][]{Silva-Beyer2023}. Furthermore, a better understanding of stellar magnetism is also important for exoplanetary physics. Magnetic activity and rotation can disguise or even mimic planetary signals, hampering the detection and characterization of planets \citep[e.g.][]{Queloz2001,Oshagh2013,Meunier2020}. Also, these stellar properties have a significant impact on the habitability of the planets, leading for example to the loss of their atmospheres and shaping the orbital architecture of the systems \citep[e.g.][]{Kaltenegger2017,Strugarek2018,Owen2019}. This also reinforces the need for reliable stellar ages, including those determined from rotation and/or magnetic activity, as they will shed light on the evolution of the planetary systems and on the evolution of exoplanet atmospheres.

\section*{Author Contributions}

ARGS: Conceptualization, Formal analysis, Writing – original draft; DGR: Formal analysis, Writing – review \& editing; AJF: Writing – original draft; SM: Conceptualization, Writing – review \& editing; RAG: Conceptualization, Writing – review \& editing; SNB: Writing – review \& editing; A-MB: Conceptualization, Writing – review \& editing.

\section*{Funding}
This work was supported by Funda\c{c}\~ao para a Ci\^encia e a Tecnologia (FCT) through national funds and by Fundo Europeu de Desenvolvimento Regional (FEDER) through COMPETE2020 - Programa Operacional Competitividade e Internacionaliza\c{c}\~ao by these grants: UIDB/04434/2020, UIDP/04434/2020, \& 2022.03993.PTDC. 
ARGS acknowledges the support from the FCT through the work contract No. 2020.02480.CEECIND/CP1631/CT0001. DGR and SM acknowledge support from the Spanish Ministry of Science and Innovation (MICINN) grant no.~PID2019-107187GB-I00. AJF acknowledges support from the European Research Council (ERC) under the European Union’s Horizon 2020 research and innovation programme (grant agreement No 810218 WHOLESUN). SM acknowledges support from the Spanish Ministry of Science and Innovation (MICINN) with the Ram\'on y Cajal fellowship no.~RYC-2015-17697, PID2019-107061GB-C66 for PLATO, and through AEI under the Severo Ochoa Centres of Excellence Programme 2020--2023 (CEX2019-000920-S). RAG acknowledges the support from PLATO and GOLF CNES grants. SNB acknowledges support from PLATO ASI-INAF agreement n. 2015-019-R.1-2018. A-MB acknowledges the support from STFC consolidated grant ST/T000252/1. 


\section*{Data Availability Statement}

The data used in this review is publically available in the respective original studies as reported.



\bibliographystyle{Frontiers-Harvard} 
\bibliography{skumanich}

\begin{thebibliography}{227}
\providecommand{\natexlab}[1]{#1}
\expandafter\ifx\csname urlstyle\endcsname\relax
  \providecommand{\doi}[1]{doi:\discretionary{}{}{}#1}\else
  \providecommand{\doi}{doi:\discretionary{}{}{}\begingroup
  \urlstyle{rm}\Url}\fi
\providecommand{\selectlanguage}[1]{\relax}
\providecommand{\bibAnnoteFile}[1]{%
  \IfFileExists{#1}{\begin{quotation}\noindent\textsc{Key:} #1\\
  \textsc{Annotation:}\ \input{#1}\end{quotation}}{}}
\providecommand{\bibAnnote}[2]{%
  \begin{quotation}\noindent\textsc{Key:} #1\\
  \textsc{Annotation:}\ #2\end{quotation}}

\bibitem[{{Abdurro'uf} et~al.(2022){Abdurro'uf}, Accetta, Aerts, Silva~Aguirre,
  Ahumada, Ajgaonkar et~al.}]{Abdurro'uf2022}
{Abdurro'uf}, Accetta, K., Aerts, C., Silva~Aguirre, V., Ahumada, R.,
  Ajgaonkar, N., et~al. (2022).
\newblock The {Seventeenth} {Data} {Release} of the {Sloan} {Digital} {Sky}
  {Surveys}: {Complete} {Release} of {MaNGA}, {MaStar}, and {APOGEE}-2 {Data}.
\newblock \emph{ApJS} 259, 35.
\newblock \doi{10.3847/1538-4365/ac4414}
\bibAnnoteFile{Abdurro'uf2022}

\bibitem[{Affer et~al.(2012)Affer, Micela, Favata, and Flaccomio}]{Affer2012}
Affer, L., Micela, G., Favata, F., and Flaccomio, E. (2012).
\newblock The rotation of field stars from {CoRoT} data.
\newblock \emph{MNRAS} 424, 11--22.
\newblock \doi{10.1111/j.1365-2966.2012.20802.x}
\bibAnnoteFile{Affer2012}

\bibitem[{Amard and Matt(2020)}]{Amard2020}
Amard, L. and Matt, S.~P. (2020).
\newblock The {Impact} of {Metallicity} on the {Evolution} of the {Rotation}
  and {Magnetic} {Activity} of {Sun}-like {Stars}.
\newblock \emph{ApJ} 889, 108.
\newblock \doi{10.3847/1538-4357/ab6173}
\bibAnnoteFile{Amard2020}

\bibitem[{Amard et~al.(2020)Amard, Roquette, and Matt}]{Amard2020b}
Amard, L., Roquette, J., and Matt, S.~P. (2020).
\newblock Evidence for metallicity-dependent spin evolution in the {Kepler}
  field.
\newblock \emph{MNRAS} 499, 3481--3493.
\newblock \doi{10.1093/mnras/staa3038}
\bibAnnoteFile{Amard2020b}

\bibitem[{Andrae et~al.(2023)Andrae, Fouesneau, Sordo, Bailer-Jones,
  Dharmawardena, Rybizki et~al.}]{Andrea2023}
Andrae, R., Fouesneau, M., Sordo, R., Bailer-Jones, C. A.~L., Dharmawardena,
  T.~E., Rybizki, J., et~al. (2023).
\newblock Gaia {Data} {Release} 3. {Analysis} of the {Gaia} {BP}/{RP} spectra
  using the {General} {Stellar} {Parameterizer} from {Photometry}.
\newblock \emph{A\&A} 674, A27.
\newblock \doi{10.1051/0004-6361/202243462}
\bibAnnoteFile{Andrea2023}

\bibitem[{Angus et~al.(2015)Angus, Aigrain, Foreman-Mackey, and
  McQuillan}]{Angus2015}
Angus, R., Aigrain, S., Foreman-Mackey, D., and McQuillan, A. (2015).
\newblock Calibrating gyrochronology using {Kepler} asteroseismic targets.
\newblock \emph{MNRAS} 450, 1787--1798.
\newblock \doi{10.1093/mnras/stv423}
\bibAnnoteFile{Angus2015}

\bibitem[{Angus et~al.(2020)Angus, Beane, Price-Whelan, Newton, Curtis, Berger
  et~al.}]{Angus2020}
Angus, R., Beane, A., Price-Whelan, A.~M., Newton, E., Curtis, J.~L., Berger,
  T., et~al. (2020).
\newblock Exploring the {Evolution} of {Stellar} {Rotation} {Using} {Galactic}
  {Kinematics}.
\newblock \emph{AJ} 160, 90.
\newblock \doi{10.3847/1538-3881/ab91b2}
\bibAnnoteFile{Angus2020}

\bibitem[{Appourchaux et~al.(2014)Appourchaux, Antia, Benomar, Campante,
  Davies, Handberg et~al.}]{Appourchaux2014}
Appourchaux, T., Antia, H.~M., Benomar, O., Campante, T.~L., Davies, G.~R.,
  Handberg, R., et~al. (2014).
\newblock Oscillation mode linewidths and heights of 23 main-sequence stars
  observed by {Kepler}.
\newblock \emph{A\&A} 566, A20.
\newblock \doi{10.1051/0004-6361/201323317}
\bibAnnoteFile{Appourchaux2014}

\bibitem[{Appourchaux et~al.(2012)Appourchaux, Benomar, Gruberbauer, Chaplin,
  García, Handberg et~al.}]{Appourchaux2012}
Appourchaux, T., Benomar, O., Gruberbauer, M., Chaplin, W.~J., García, R.~A.,
  Handberg, R., et~al. (2012).
\newblock Oscillation mode linewidths of main-sequence and subgiant stars
  observed by {Kepler}.
\newblock \emph{A\&A} 537, A134.
\newblock \doi{10.1051/0004-6361/201118496}
\bibAnnoteFile{Appourchaux2012}

\bibitem[{Baglin et~al.(2006)Baglin, Auvergne, Barge, Deleuil, Catala, Michel
  et~al.}]{Baglin2006a}
Baglin, A., Auvergne, M., Barge, P., Deleuil, M., Catala, C., Michel, E.,
  et~al. (2006).
\newblock Scientific {Objectives} for a {Minisat}: {CoRoT} (Proceedings of "The
  CoRoT Mission Pre-Launch Status - Stellar Seismology and Planet Finding" (ESA
  SP-1306). Editors: M. Fridlund, A. Baglin, J. Lochard and L. Conroy. ISBN
  92-9092-465-9.), vol. 1306, 33
\bibAnnoteFile{Baglin2006a}

\bibitem[{Bahcall et~al.(1995)Bahcall, Pinsonneault, and
  Wasserburg}]{Bahcall1995}
Bahcall, J.~N., Pinsonneault, M.~H., and Wasserburg, G.~J. (1995).
\newblock Solar models with helium and heavy-element diffusion.
\newblock \emph{RMP} 67, 781--808.
\newblock \doi{10.1103/RevModPhys.67.781}
\bibAnnoteFile{Bahcall1995}

\bibitem[{Baliunas et~al.(1995)Baliunas, Donahue, Soon, Horne, Frazer,
  Woodard-Eklund et~al.}]{Baliunas1995}
Baliunas, S.~L., Donahue, R.~A., Soon, W.~H., Horne, J.~H., Frazer, J.,
  Woodard-Eklund, L., et~al. (1995).
\newblock Chromospheric variations in main-sequence stars.
\newblock \emph{ApJ} 438, 269--287.
\newblock \doi{10.1086/175072}
\bibAnnoteFile{Baliunas1995}

\bibitem[{Ballot et~al.(2006)Ballot, García, and Lambert}]{Ballot2006}
Ballot, J., García, R.~A., and Lambert, P. (2006).
\newblock Rotation speed and stellar axis inclination from p modes: how {CoRoT}
  would see other suns.
\newblock \emph{MNRAS} 369, 1281--1286.
\newblock \doi{10.1111/j.1365-2966.2006.10375.x}
\bibAnnoteFile{Ballot2006}

\bibitem[{Balona(2015)}]{Balona2015}
Balona, L.~A. (2015).
\newblock Flare stars across the {H}-{R} diagram.
\newblock \emph{MNRAS} 447, 2714--2725.
\newblock \doi{10.1093/mnras/stu2651}
\bibAnnoteFile{Balona2015}

\bibitem[{Balona(2019)}]{Balona2019}
Balona, L.~A. (2019).
\newblock Evidence for spots on hot stars suggests major revision of stellar
  physics.
\newblock \emph{MNRAS} 490, 2112--2116.
\newblock \doi{10.1093/mnras/stz2808}
\bibAnnoteFile{Balona2019}

\bibitem[{Barnes(2003{\natexlab{a}})}]{Barnes2003L}
Barnes, S.~A. (2003{\natexlab{a}}).
\newblock A {Connection} between the {Morphology} of the {X}-{Ray} {Emission}
  and {Rotation} for {Solar}-{Type} {Stars} in {Open} {Clusters}.
\newblock \emph{ApJL} 586, L145--L147.
\newblock \doi{10.1086/374681}
\bibAnnoteFile{Barnes2003L}

\bibitem[{Barnes(2003{\natexlab{b}})}]{Barnes2003}
Barnes, S.~A. (2003{\natexlab{b}}).
\newblock On the {Rotational} {Evolution} of {Solar}- and {Late}-{Type}
  {Stars}, {Its} {Magnetic} {Origins}, and the {Possibility} of {Stellar}
  {Gyrochronology}.
\newblock \emph{ApJ} 586, 464--479.
\newblock \doi{10.1086/367639}
\bibAnnoteFile{Barnes2003}

\bibitem[{Barnes(2007)}]{Barnes2007}
Barnes, S.~A. (2007).
\newblock Ages for {Illustrative} {Field} {Stars} {Using} {Gyrochronology}:
  {Viability}, {Limitations}, and {Errors}.
\newblock \emph{ApJ} 669, 1167--1189.
\newblock \doi{10.1086/519295}
\bibAnnoteFile{Barnes2007}

\bibitem[{Barnes et~al.(2016)Barnes, Weingrill, Fritzewski, Strassmeier, and
  Platais}]{Barnes2016}
Barnes, S.~A., Weingrill, J., Fritzewski, D., Strassmeier, K.~G., and Platais,
  I. (2016).
\newblock Rotation {Periods} for {Cool} {Stars} in the 4 {Gyr} old {Open}
  {Cluster} {M67}, {The} {Solar}-{Stellar} {Connection}, and the
  {Applicability} of {Gyrochronology} to at least {Solar} {Age}.
\newblock \emph{ApJ} 823, 16.
\newblock \doi{10.3847/0004-637X/823/1/16}
\bibAnnoteFile{Barnes2016}

\bibitem[{Basri et~al.(2010)Basri, Walkowicz, Batalha, Gilliland, Jenkins,
  Borucki et~al.}]{Basri2010}
Basri, G., Walkowicz, L.~M., Batalha, N., Gilliland, R.~L., Jenkins, J.,
  Borucki, W.~J., et~al. (2010).
\newblock Photometric {Variability} in {Kepler} {Target} {Stars}: {The} {Sun}
  {Among} {Stars}—a {First} {Look}.
\newblock \emph{ApJL} 713, L155--L159.
\newblock \doi{10.1088/2041-8205/713/2/L155}
\bibAnnoteFile{Basri2010}

\bibitem[{Basu et~al.(2012)Basu, Broomhall, Chaplin, and Elsworth}]{Basu2012}
Basu, S., Broomhall, A.-M., Chaplin, W.~J., and Elsworth, Y. (2012).
\newblock Thinning of the {Sun}'s {Magnetic} {Layer}: {The} {Peculiar} {Solar}
  {Minimum} {Could} {Have} {Been} {Predicted}.
\newblock \emph{ApJ} 758, 43.
\newblock \doi{10.1088/0004-637X/758/1/43}
\bibAnnoteFile{Basu2012}

\bibitem[{Baum et~al.(2022)Baum, Wright, Luhn, and Isaacson}]{Baum2022}
Baum, A.~C., Wright, J.~T., Luhn, J.~K., and Isaacson, H. (2022).
\newblock Five decades of chromospheric activity in 59 sun-like stars and new
  maunder minimum candidate hd 166620.
\newblock \emph{The Astronomical Journal} 163, 183
\bibAnnoteFile{Baum2022}

\bibitem[{Bazot et~al.(2019)Bazot, Benomar, Christensen-Dalsgaard, Gizon,
  Hanasoge, Nielsen et~al.}]{Bazot2019}
Bazot, M., Benomar, O., Christensen-Dalsgaard, J., Gizon, L., Hanasoge, S.,
  Nielsen, M., et~al. (2019).
\newblock Latitudinal differential rotation in the solar analogues 16 {Cygni}
  {A} and {B}.
\newblock \emph{A\&A} 623, A125.
\newblock \doi{10.1051/0004-6361/201834594}
\bibAnnoteFile{Bazot2019}

\bibitem[{Benomar et~al.(2018)Benomar, Bazot, Nielsen, Gizon, Sekii, Takata
  et~al.}]{Benomar2018}
Benomar, O., Bazot, M., Nielsen, M.~B., Gizon, L., Sekii, T., Takata, M.,
  et~al. (2018).
\newblock Asteroseismic detection of latitudinal differential rotation in 13
  {Sun}-like stars.
\newblock \emph{Science} 361, 1231--1234.
\newblock \doi{10.1126/science.aao6571}
\bibAnnoteFile{Benomar2018}

\bibitem[{Benomar et~al.(2015)Benomar, Takata, Shibahashi, Ceillier, and
  García}]{Benomar2015}
Benomar, O., Takata, M., Shibahashi, H., Ceillier, T., and García, R.~A.
  (2015).
\newblock Nearly uniform internal rotation of solar-like main-sequence stars
  revealed by space-based asteroseismology and spectroscopic measurements.
\newblock \emph{MNRAS} 452, 2654--2674.
\newblock \doi{10.1093/mnras/stv1493}
\bibAnnoteFile{Benomar2015}

\bibitem[{Bonanno and Corsaro(2022)}]{Bonanno&Corsaro2022}
Bonanno, A. and Corsaro, E. (2022).
\newblock On the {Origin} of the {Dichotomy} of {Stellar} {Activity} {Cycles}.
\newblock \emph{ApJ} 939, L26.
\newblock \doi{10.3847/2041-8213/ac9c05}
\bibAnnoteFile{Bonanno&Corsaro2022}

\bibitem[{Boro~Saikia et~al.(2018)Boro~Saikia, Marvin, Jeffers, Reiners,
  Cameron, Marsden et~al.}]{BoroSaikia2018}
Boro~Saikia, S., Marvin, C.~J., Jeffers, S.~V., Reiners, A., Cameron, R.,
  Marsden, S.~C., et~al. (2018).
\newblock Chromospheric activity catalogue of 4454 cool stars. {Questioning}
  the active branch of stellar activity cycles.
\newblock \emph{A\&A} 616, A108.
\newblock \doi{10.1051/0004-6361/201629518}
\bibAnnoteFile{BoroSaikia2018}

\bibitem[{Borucki et~al.(2010)Borucki, Koch, Basri, Batalha, Brown, Caldwell
  et~al.}]{Borucki2010}
Borucki, W.~J., Koch, D., Basri, G., Batalha, N., Brown, T., Caldwell, D.,
  et~al. (2010).
\newblock Kepler {Planet}-{Detection} {Mission}: {Introduction} and {First}
  {Results}.
\newblock \emph{Science} 327, 977.
\newblock \doi{10.1126/science.1185402}
\bibAnnoteFile{Borucki2010}

\bibitem[{Bouma et~al.(2023)Bouma, Palumbo, and Hillenbrand}]{Bouma2023}
Bouma, L.~G., Palumbo, E.~K., and Hillenbrand, L.~A. (2023).
\newblock The {Empirical} {Limits} of {Gyrochronology}.
\newblock \emph{ApJ} 947, L3.
\newblock \doi{10.3847/2041-8213/acc589}
\bibAnnoteFile{Bouma2023}

\bibitem[{Brandenburg et~al.(1998)Brandenburg, Saar, and
  Turpin}]{Brandenburg1998}
Brandenburg, A., Saar, S.~H., and Turpin, C.~R. (1998).
\newblock Time {Evolution} of the {Magnetic} {Activity} {Cycle} {Period}.
\newblock \emph{ApJL} 498, L51--L54.
\newblock \doi{10.1086/311297}
\bibAnnoteFile{Brandenburg1998}

\bibitem[{Breton et~al.(2023)Breton, Dhouib, García, Brun, Mathis,
  Pérez~Hernández et~al.}]{Breton2023}
Breton, S.~N., Dhouib, H., García, R.~A., Brun, A.~S., Mathis, S.,
  Pérez~Hernández, F., et~al. (2023).
\newblock In search of gravity mode signatures in main sequence solar-type
  stars observed by {Kepler}.
\newblock \emph{A\&A} 679, A104.
\newblock \doi{10.1051/0004-6361/202346601}
\bibAnnoteFile{Breton2023}

\bibitem[{Breton et~al.(2021)Breton, Santos, Bugnet, Mathur, García, and
  Pallé}]{Breton2021}
Breton, S.~N., Santos, A. R.~G., Bugnet, L., Mathur, S., García, R.~A., and
  Pallé, P.~L. (2021).
\newblock {ROOSTER}: a machine-learning analysis tool for {Kepler} stellar
  rotation periods.
\newblock \emph{A\&A} 647, A125.
\newblock \doi{10.1051/0004-6361/202039947}
\bibAnnoteFile{Breton2021}

\bibitem[{Broomhall et~al.(2012)Broomhall, Chaplin, Elsworth, and
  Simoniello}]{Broomhall2012}
Broomhall, A.-M., Chaplin, W.~J., Elsworth, Y., and Simoniello, R. (2012).
\newblock Quasi-biennial variations in helioseismic frequencies: can the source
  of the variation be localized?
\newblock \emph{MNRAS} 420, 1405--1414.
\newblock \doi{10.1111/j.1365-2966.2011.20123.x}
\bibAnnoteFile{Broomhall2012}

\bibitem[{Broomhall et~al.(2014)Broomhall, Chatterjee, Howe, Norton, and
  Thompson}]{Broomhall2014}
Broomhall, A.-M., Chatterjee, P., Howe, R., Norton, A.~A., and Thompson, M.~J.
  (2014).
\newblock The {Sun}'s {Interior} {Structure} and {Dynamics}, and the {Solar}
  {Cycle}.
\newblock \emph{Space Sci. Rev.} 186, 191--225.
\newblock \doi{10.1007/s11214-014-0101-3}
\bibAnnoteFile{Broomhall2014}

\bibitem[{Brown et~al.(2022)Brown, Jeffers, Marsden, Morin, Boro~Saikia, Petit
  et~al.}]{EBrown2022}
Brown, E.~L., Jeffers, S.~V., Marsden, S.~C., Morin, J., Boro~Saikia, S.,
  Petit, P., et~al. (2022).
\newblock Linking chromospheric activity and magnetic field properties for
  late-type dwarf stars.
\newblock \emph{MNRAS} 514, 4300--4319.
\newblock \doi{10.1093/mnras/stac1291}
\bibAnnoteFile{EBrown2022}

\bibitem[{Brown(2014)}]{Brown2014}
Brown, T.~M. (2014).
\newblock The {Metastable} {Dynamo} {Model} of {Stellar} {Rotational}
  {Evolution}.
\newblock \emph{ApJ} 789, 101.
\newblock \doi{10.1088/0004-637X/789/2/101}
\bibAnnoteFile{Brown2014}

\bibitem[{Brown et~al.(2021)Brown, García, Mathur, Metcalfe, and
  Santos}]{Brown2021}
Brown, T.~M., García, R.~A., Mathur, S., Metcalfe, T.~S., and Santos, A. R.~G.
  (2021).
\newblock Brightness {Fluctuation} {Spectra} of {Sun}-like {Stars}. {I}. {The}
  {Mid}-frequency {Continuum}.
\newblock \emph{ApJ} 916, 66.
\newblock \doi{10.3847/1538-4357/ac0635}
\bibAnnoteFile{Brown2021}

\bibitem[{Brun and Browning(2017)}]{Brun2017}
Brun, A.~S. and Browning, M.~K. (2017).
\newblock Magnetism, dynamo action and the solar-stellar connection.
\newblock \emph{Living Rev. Solar Phys.} 14, 4.
\newblock \doi{10.1007/s41116-017-0007-8}
\bibAnnoteFile{Brun2017}

\bibitem[{Brun et~al.(2022)Brun, Strugarek, Noraz, Perri, Varela, Augustson
  et~al.}]{Brun2022}
Brun, A.~S., Strugarek, A., Noraz, Q., Perri, B., Varela, J., Augustson, K.,
  et~al. (2022).
\newblock Powering stellar magnetism: Energy transfers in cyclic dynamos of
  sun-like stars.
\newblock \emph{The Astrophysical Journal} 926, 21
\bibAnnoteFile{Brun2022}

\bibitem[{Böhm-Vitense(2007)}]{Bohm-Vitense2007}
Böhm-Vitense, E. (2007).
\newblock Chromospheric {Activity} in {G} and {K} {Main}-{Sequence} {Stars},
  and {What} {It} {Tells} {Us} about {Stellar} {Dynamos}.
\newblock \emph{ApJ} 657, 486--493.
\newblock \doi{10.1086/510482}
\bibAnnoteFile{Bohm-Vitense2007}

\bibitem[{Cao and Pinsonneault(2022)}]{Cao2022}
Cao, L. and Pinsonneault, M.~H. (2022).
\newblock Star-spots and magnetism: testing the activity paradigm in the
  {Pleiades} and {M67}.
\newblock \emph{MNRAS} 517, 2165--2189.
\newblock \doi{10.1093/mnras/stac2706}
\bibAnnoteFile{Cao2022}

\bibitem[{Ceillier et~al.(2017)Ceillier, Tayar, Mathur, Salabert, García,
  Stello et~al.}]{Ceillier2017}
Ceillier, T., Tayar, J., Mathur, S., Salabert, D., García, R.~A., Stello, D.,
  et~al. (2017).
\newblock Surface rotation of {Kepler} red giant stars.
\newblock \emph{A\&A} 605, A111.
\newblock \doi{10.1051/0004-6361/201629884}
\bibAnnoteFile{Ceillier2017}

\bibitem[{Chaplin et~al.(2011)Chaplin, Bedding, Bonanno, Broomhall, García,
  Hekker et~al.}]{Chaplin2011b}
Chaplin, W.~J., Bedding, T.~R., Bonanno, A., Broomhall, A.-M., García, R.~A.,
  Hekker, S., et~al. (2011).
\newblock Evidence for the {Impact} of {Stellar} {Activity} on the
  {Detectability} of {Solar}-like {Oscillations} {Observed} by {Kepler}.
\newblock \emph{ApJL} 732, L5.
\newblock \doi{10.1088/2041-8205/732/1/L5}
\bibAnnoteFile{Chaplin2011b}

\bibitem[{Chaplin et~al.(2000)Chaplin, Elsworth, Isaak, Miller, and
  New}]{Chaplin2000}
Chaplin, W.~J., Elsworth, Y., Isaak, G.~R., Miller, B.~A., and New, R. (2000).
\newblock Variations in the excitation and damping of low-l solar p modes over
  the solar activity cycle*.
\newblock \emph{MNRAS} 313, 32--42.
\newblock \doi{10.1046/j.1365-8711.2000.03176.x}
\bibAnnoteFile{Chaplin2000}

\bibitem[{Chaplin et~al.(2004)Chaplin, Elsworth, Isaak, Miller, and
  New}]{Chaplin2004}
Chaplin, W.~J., Elsworth, Y., Isaak, G.~R., Miller, B.~A., and New, R. (2004).
\newblock The solar cycle as seen by low-l p-mode frequencies: comparison with
  global and decomposed activity proxies.
\newblock \emph{MNRAS} 352, 1102--1108.
\newblock \doi{10.1111/j.1365-2966.2004.07998.x}
\bibAnnoteFile{Chaplin2004}

\bibitem[{Claytor et~al.(2023)Claytor, van Saders, Cao, Pinsonneault, Teske,
  and Beaton}]{Claytor2023}
[Dataset] Claytor, Z.~R., van Saders, J.~L., Cao, L., Pinsonneault, M.~H.,
  Teske, J., and Beaton, R.~L. (2023).
\newblock {TESS} {Stellar} {Rotation} up to 80 days in the {Southern}
  {Continuous} {Viewing} {Zone}.
\newblock \doi{10.48550/arXiv.2307.05664}
\bibAnnoteFile{Claytor2023}

\bibitem[{Claytor et~al.(2020)Claytor, van Saders, Santos, García, Mathur,
  Tayar et~al.}]{Claytor2020}
Claytor, Z.~R., van Saders, J.~L., Santos, A. R.~G., García, R.~A., Mathur,
  S., Tayar, J., et~al. (2020).
\newblock Chemical {Evolution} in the {Milky} {Way}: {Rotation}-based {Ages}
  for {APOGEE}-{Kepler} {Cool} {Dwarf} {Stars}.
\newblock \emph{ApJ} 888, 43.
\newblock \doi{10.3847/1538-4357/ab5c24}
\bibAnnoteFile{Claytor2020}

\bibitem[{Corsaro et~al.(2012)Corsaro, Stello, Huber, Bedding, Bonanno,
  Brogaard et~al.}]{Corsaro2012}
Corsaro, E., Stello, D., Huber, D., Bedding, T.~R., Bonanno, A., Brogaard, K.,
  et~al. (2012).
\newblock Asteroseismology of the {Open} {Clusters} {NGC} 6791, {NGC} 6811, and
  {NGC} 6819 from 19 {Months} of {Kepler} {Photometry}.
\newblock \emph{ApJ} 757, 190.
\newblock \doi{10.1088/0004-637X/757/2/190}
\bibAnnoteFile{Corsaro2012}

\bibitem[{Curtis et~al.(2019)Curtis, Agüeros, Douglas, and
  Meibom}]{Curtis2019}
Curtis, J.~L., Agüeros, M.~A., Douglas, S.~T., and Meibom, S. (2019).
\newblock A {Temporary} {Epoch} of {Stalled} {Spin}-down for {Low}-mass
  {Stars}: {Insights} from {NGC} 6811 with {Gaia} and {Kepler}.
\newblock \emph{ApJ} 879, 49.
\newblock \doi{10.3847/1538-4357/ab2393}
\bibAnnoteFile{Curtis2019}

\bibitem[{{Czesla} et~al.(2019){Czesla}, {Schr{\"o}ter}, {Schneider}, {Huber},
  {Pfeifer}, {Andreasen} et~al.}]{pya}
[Dataset] {Czesla}, S., {Schr{\"o}ter}, S., {Schneider}, C.~P., {Huber}, K.~F.,
  {Pfeifer}, F., {Andreasen}, D.~T., et~al. (2019).
\newblock {PyA: Python astronomy-related packages}
\bibAnnoteFile{pya}

\bibitem[{Davenport(2016)}]{Davenport2016}
Davenport, J. R.~A. (2016).
\newblock The {Kepler} {Catalog} of {Stellar} {Flares}.
\newblock \emph{ApJ} 829, 23.
\newblock \doi{10.3847/0004-637X/829/1/23}
\bibAnnoteFile{Davenport2016}

\bibitem[{Davenport(2017)}]{Davenport2017}
Davenport, J. R.~A. (2017).
\newblock Rotating {Stars} from {Kepler} {Observed} with {Gaia} {DR1}.
\newblock \emph{ApJ} 835, 16.
\newblock \doi{10.3847/1538-4357/835/1/16}
\bibAnnoteFile{Davenport2017}

\bibitem[{Davenport and Covey(2018)}]{Davenport2018}
Davenport, J. R.~A. and Covey, K.~R. (2018).
\newblock Rotating {Stars} from {Kepler} {Observed} with {Gaia} {DR2}.
\newblock \emph{ApJ} 868, 151.
\newblock \doi{10.3847/1538-4357/aae842}
\bibAnnoteFile{Davenport2018}

\bibitem[{David et~al.(2022)David, Angus, Curtis, van Saders, Colman, Contardo
  et~al.}]{David2022}
David, T.~J., Angus, R., Curtis, J.~L., van Saders, J.~L., Colman, I.~L.,
  Contardo, G., et~al. (2022).
\newblock Further {Evidence} of {Modified} {Spin}-down in {Sun}-like {Stars}:
  {Pileups} in the {Temperature}-{Period} {Distribution}.
\newblock \emph{ApJ} 933, 114.
\newblock \doi{10.3847/1538-4357/ac6dd3}
\bibAnnoteFile{David2022}

\bibitem[{Davies et~al.(2015)Davies, Chaplin, Farr, García, Lund, Mathis
  et~al.}]{Davies2015}
Davies, G.~R., Chaplin, W.~J., Farr, W.~M., García, R.~A., Lund, M.~N.,
  Mathis, S., et~al. (2015).
\newblock Asteroseismic inference on rotation, gyrochronology and planetary
  system dynamics of 16 {Cygni}.
\newblock \emph{MNRAS} 446, 2959--2966.
\newblock \doi{10.1093/mnras/stu2331}
\bibAnnoteFile{Davies2015}

\bibitem[{Distefano et~al.(2023)Distefano, Lanzafame, Brugaletta, Holl, Lanza,
  Messina et~al.}]{Distefano2023}
Distefano, E., Lanzafame, A.~C., Brugaletta, E., Holl, B., Lanza, A.~F.,
  Messina, S., et~al. (2023).
\newblock Gaia {Data} {Release} 3. {Rotational} modulation and patterns of
  colour variation in solar-like variables.
\newblock \emph{A\&A} 674, A20.
\newblock \doi{10.1051/0004-6361/202244178}
\bibAnnoteFile{Distefano2023}

\bibitem[{Donati and Brown(1997)}]{Donati1997}
Donati, J.~F. and Brown, S.~F. (1997).
\newblock Zeeman-{Doppler} imaging of active stars. {V}. {Sensitivity} of
  maximum entropy magnetic maps to field orientation.
\newblock \emph{A\&A} 326, 1135--1142
\bibAnnoteFile{Donati1997}

\bibitem[{Douglas et~al.(2019)Douglas, Curtis, Agüeros, Cargile, Brewer,
  Meibom et~al.}]{Douglas2019}
Douglas, S.~T., Curtis, J.~L., Agüeros, M.~A., Cargile, P.~A., Brewer, J.~M.,
  Meibom, S., et~al. (2019).
\newblock K2 {Rotation} {Periods} for {Low}-mass {Hyads} and a {Quantitative}
  {Comparison} of the {Distribution} of {Slow} {Rotators} in the {Hyades} and
  {Praesepe}.
\newblock \emph{ApJ} 879, 100.
\newblock \doi{10.3847/1538-4357/ab2468}
\bibAnnoteFile{Douglas2019}

\bibitem[{Dungee et~al.(2022)Dungee, van Saders, Gaidos, Chun, García, Magnier
  et~al.}]{Dungee2022}
Dungee, R., van Saders, J., Gaidos, E., Chun, M., García, R.~A., Magnier,
  E.~A., et~al. (2022).
\newblock A 4 {Gyr} {M}-dwarf {Gyrochrone} from {CFHT}/{MegaPrime} {Monitoring}
  of the {Open} {Cluster} {M67}.
\newblock \emph{ApJ} 938, 118.
\newblock \doi{10.3847/1538-4357/ac90be}
\bibAnnoteFile{Dungee2022}

\bibitem[{Egeland(2017)}]{EgelandThesis}
Egeland, R. (2017).
\newblock Long-{Term} {Variability} of the {Sun} in the {Context} of
  {Solar}-{Analog} {Stars}.
\newblock \emph{Ph.D. Thesis}
  \doi{http://adsabs.harvard.edu/abs/2017PhDT.........3E}
\bibAnnoteFile{EgelandThesis}

\bibitem[{Elsworth et~al.(1990)Elsworth, Howe, Isaak, McLeod, and
  New}]{Elsworth1990}
Elsworth, Y., Howe, R., Isaak, G.~R., McLeod, C.~P., and New, R. (1990).
\newblock Variation of low-order acoustic solar oscillations over the solar
  cycle.
\newblock \emph{Nat.} 345, 322--324.
\newblock \doi{10.1038/345322a0}
\bibAnnoteFile{Elsworth1990}

\bibitem[{Findeisen et~al.(2011)Findeisen, Hillenbrand, and
  Soderblom}]{Findeisen2011}
Findeisen, K., Hillenbrand, L., and Soderblom, D. (2011).
\newblock Stellar {Activity} in the {Broadband} {Ultraviolet}.
\newblock \emph{AJ} 142, 23.
\newblock \doi{10.1088/0004-6256/142/1/23}
\bibAnnoteFile{Findeisen2011}

\bibitem[{Finley and Brun(2023)}]{Finley2023}
Finley, A.~J. and Brun, A.~S. (2023).
\newblock Accounting for differential rotation in calculations of the sun’s
  angular momentum-loss rate.
\newblock \emph{Astronomy \& Astrophysics} 674, A42
\bibAnnoteFile{Finley2023}

\bibitem[{Finley et~al.(2019{\natexlab{a}})Finley, Deshmukh, Matt, Owens, and
  Wu}]{Finley2019b}
Finley, A.~J., Deshmukh, S., Matt, S.~P., Owens, M., and Wu, C.-J.
  (2019{\natexlab{a}}).
\newblock Solar angular momentum loss over the past several millennia.
\newblock \emph{The Astrophysical Journal} 883, 67
\bibAnnoteFile{Finley2019b}

\bibitem[{Finley et~al.(2019{\natexlab{b}})Finley, Hewitt, Matt, Owens, Pinto,
  and R{\'e}ville}]{Finley2019c}
Finley, A.~J., Hewitt, A.~L., Matt, S.~P., Owens, M., Pinto, R.~F., and
  R{\'e}ville, V. (2019{\natexlab{b}}).
\newblock Direct detection of solar angular momentum loss with the wind
  spacecraft.
\newblock \emph{The Astrophysical Journal Letters} 885, L30
\bibAnnoteFile{Finley2019c}

\bibitem[{Finley and Matt(2018)}]{Finley2018a}
Finley, A.~J. and Matt, S.~P. (2018).
\newblock The effect of combined magnetic geometries on thermally driven winds.
  ii. dipolar, quadrupolar, and octupolar topologies.
\newblock \emph{The Astrophysical Journal} 854, 78
\bibAnnoteFile{Finley2018a}

\bibitem[{Finley et~al.(2018)Finley, Matt, and See}]{Finley2018b}
Finley, A.~J., Matt, S.~P., and See, V. (2018).
\newblock The {Effect} of {Magnetic} {Variability} on {Stellar} {Angular}
  {Momentum} {Loss}. {I}. {The} {Solar} {Wind} {Torque} during {Sunspot}
  {Cycles} 23 and 24.
\newblock \emph{ApJ} 864, 125.
\newblock \doi{10.3847/1538-4357/aad7b6}
\bibAnnoteFile{Finley2018b}

\bibitem[{Fritzewski et~al.(2019)Fritzewski, Barnes, James, Geller, Meibom, and
  Strassmeier}]{Fritzewski2019}
Fritzewski, D.~J., Barnes, S.~A., James, D.~J., Geller, A.~M., Meibom, S., and
  Strassmeier, K.~G. (2019).
\newblock Spectroscopic membership for the populous 300 {Myr}-old open cluster
  {NGC} 3532.
\newblock \emph{A\&A} 622, A110.
\newblock \doi{10.1051/0004-6361/201833587}
\bibAnnoteFile{Fritzewski2019}

\bibitem[{Fritzewski et~al.(2021{\natexlab{a}})Fritzewski, Barnes, James,
  Järvinen, and Strassmeier}]{Fritzewski2021b}
Fritzewski, D.~J., Barnes, S.~A., James, D.~J., Järvinen, S.~P., and
  Strassmeier, K.~G. (2021{\natexlab{a}}).
\newblock A detailed understanding of the rotation-activity relationship using
  the 300 {Myr} old open cluster {NGC} 3532.
\newblock \emph{A\&A} 656, A103.
\newblock \doi{10.1051/0004-6361/202140896}
\bibAnnoteFile{Fritzewski2021b}

\bibitem[{Fritzewski et~al.(2021{\natexlab{b}})Fritzewski, Barnes, James, and
  Strassmeier}]{Fritzewski2021a}
Fritzewski, D.~J., Barnes, S.~A., James, D.~J., and Strassmeier, K.~G.
  (2021{\natexlab{b}}).
\newblock Rotation periods for cool stars in the open cluster {NGC} 3532. {The}
  transition from fast to slow rotation.
\newblock \emph{A\&A} 652, A60.
\newblock \doi{10.1051/0004-6361/202140894}
\bibAnnoteFile{Fritzewski2021a}

\bibitem[{{Gaia Collaboration} et~al.(2022){Gaia Collaboration}, Arenou,
  Babusiaux, Barstow, Faigler, Jorissen et~al.}]{Binaries_GaiaDR3}
{Gaia Collaboration}, Arenou, F., Babusiaux, C., Barstow, M.~A., Faigler, S.,
  Jorissen, A., et~al. (2022).
\newblock \emph{Gaia {Data} {Release} 3: {Stellar} multiplicity, a teaser for
  the hidden treasure}.
\newblock Tech. rep.
\bibAnnoteFile{Binaries_GaiaDR3}

\bibitem[{{Gaia Collaboration} et~al.(2023){Gaia Collaboration}, Vallenari,
  Brown, Prusti, de~Bruijne, Arenou et~al.}]{Gaia_DR3}
{Gaia Collaboration}, Vallenari, A., Brown, A. G.~A., Prusti, T., de~Bruijne,
  J. H.~J., Arenou, F., et~al. (2023).
\newblock Gaia {Data} {Release} 3. {Summary} of the content and survey
  properties.
\newblock \emph{A\&A} 674, A1.
\newblock \doi{10.1051/0004-6361/202243940}
\bibAnnoteFile{Gaia_DR3}

\bibitem[{Gaidos et~al.(2000)Gaidos, Henry, and Henry}]{Gaidos2000}
Gaidos, E.~J., Henry, G.~W., and Henry, S.~M. (2000).
\newblock Spectroscopy and {Photometry} of {Nearby} {Young} {Solar} {Analogs}.
\newblock \emph{AJ} 120, 1006--1013.
\newblock \doi{10.1086/301488}
\bibAnnoteFile{Gaidos2000}

\bibitem[{Gallet and Bouvier(2013)}]{Gallet2013}
Gallet, F. and Bouvier, J. (2013).
\newblock Improved angular momentum evolution model for solar-like stars.
\newblock \emph{A\&A} 556, A36.
\newblock \doi{10.1051/0004-6361/201321302}
\bibAnnoteFile{Gallet2013}

\bibitem[{García et~al.(2014{\natexlab{a}})García, Ceillier, Salabert,
  Mathur, van Saders, Pinsonneault et~al.}]{Garcia2014}
García, R.~A., Ceillier, T., Salabert, D., Mathur, S., van Saders, J.~L.,
  Pinsonneault, M., et~al. (2014{\natexlab{a}}).
\newblock Rotation and magnetism of {Kepler} pulsating solar-like stars.
  {Towards} asteroseismically calibrated age-rotation relations.
\newblock \emph{A\&A} 572, A34.
\newblock \doi{10.1051/0004-6361/201423888}
\bibAnnoteFile{Garcia2014}

\bibitem[{García et~al.(2023)García, Gourvès, Santos, Strugarek,
  Godoy-Rivera, Mathur et~al.}]{Garcia2023}
García, R.~A., Gourvès, C., Santos, A. R.~G., Strugarek, A., Godoy-Rivera,
  D., Mathur, S., et~al. (2023).
\newblock Stellar spectral-type (mass) dependence of the dearth of close-in
  planets around fast-rotating stars. {Architecture} of {Kepler} confirmed
  single-exoplanet systems compared to star-planet evolution models.
\newblock \emph{A\&A} 679, L12.
\newblock \doi{10.1051/0004-6361/202346933}
\bibAnnoteFile{Garcia2023}

\bibitem[{García et~al.(2011)García, Hekker, Stello, Gutiérrez-Soto,
  Handberg, Huber et~al.}]{Garcia2011}
García, R.~A., Hekker, S., Stello, D., Gutiérrez-Soto, J., Handberg, R.,
  Huber, D., et~al. (2011).
\newblock Preparation of {Kepler} light curves for asteroseismic analyses.
\newblock \emph{MNRAS} 414, L6--L10.
\newblock \doi{10.1111/j.1745-3933.2011.01042.x}
\bibAnnoteFile{Garcia2011}

\bibitem[{García et~al.(2014{\natexlab{b}})García, Mathur, Pires, Régulo,
  Bellamy, Pallé et~al.}]{Garcia2014a}
García, R.~A., Mathur, S., Pires, S., Régulo, C., Bellamy, B., Pallé, P.~L.,
  et~al. (2014{\natexlab{b}}).
\newblock Impact on asteroseismic analyses of regular gaps in {Kepler} data.
\newblock \emph{A\&A} 568, A10.
\newblock \doi{10.1051/0004-6361/201323326}
\bibAnnoteFile{Garcia2014a}

\bibitem[{García et~al.(2010)García, Mathur, Salabert, Ballot, Regulo,
  Metcalfe et~al.}]{Garcia2010}
García, R.~A., Mathur, S., Salabert, D., Ballot, J., Regulo, C., Metcalfe,
  T.~S., et~al. (2010).
\newblock {CoRoT} {Reveals} a {Magnetic} {Activity} {Cycle} in a {Sun}-{Like}
  {Star}.
\newblock \emph{Science} 329, 1032.
\newblock \doi{10.1126/science.1191064}
\bibAnnoteFile{Garcia2010}

\bibitem[{Garraffo et~al.(2016)Garraffo, Drake, and Cohen}]{Garraffo2016}
Garraffo, C., Drake, J.~J., and Cohen, O. (2016).
\newblock The missing magnetic morphology term in stellar rotation evolution.
\newblock \emph{Astronomy \& Astrophysics} 595, A110
\bibAnnoteFile{Garraffo2016}

\bibitem[{Garraffo et~al.(2018)Garraffo, Drake, Dotter, Choi, Burke, Moschou
  et~al.}]{Garraffo2018}
Garraffo, C., Drake, J.~J., Dotter, A., Choi, J., Burke, D.~J., Moschou, S.~P.,
  et~al. (2018).
\newblock The {Revolution} {Revolution}: {Magnetic} {Morphology} {Driven}
  {Spin}-down.
\newblock \emph{ApJ} 862, 90.
\newblock \doi{10.3847/1538-4357/aace5d}
\bibAnnoteFile{Garraffo2018}

\bibitem[{Gehan et~al.(2022)Gehan, Gaulme, and Yu}]{Gehan2022}
Gehan, C., Gaulme, P., and Yu, J. (2022).
\newblock Surface magnetism of rapidly rotating red giants: {Single} versus
  close binary stars.
\newblock \emph{A\&A} 668, A116.
\newblock \doi{10.1051/0004-6361/202245083}
\bibAnnoteFile{Gehan2022}

\bibitem[{Gehan et~al.(2024)Gehan, Godoy-Rivera, and Gaulme}]{Gehan2024}
[Dataset] Gehan, C., Godoy-Rivera, D., and Gaulme, P. (2024).
\newblock Magnetic activity of red giants: a near-{UV} and
  {H}\${\textbackslash}alpha\$ view, and the enhancing role of tidal
  interactions.
\newblock \doi{10.48550/arXiv.2401.13549}
\bibAnnoteFile{Gehan2024}

\bibitem[{Gizon et~al.(2013)Gizon, Ballot, Michel, Stahn, Vauclair, Bruntt
  et~al.}]{Gizon2013}
Gizon, L., Ballot, J., Michel, E., Stahn, T., Vauclair, G., Bruntt, H., et~al.
  (2013).
\newblock Seismic constraints on rotation of {Sun}-like star and mass of
  exoplanet.
\newblock \emph{Proc. Natl. Acad. Sci.} 110, 13267--13271.
\newblock \doi{10.1073/pnas.1303291110}
\bibAnnoteFile{Gizon2013}

\bibitem[{Gizon and Solanki(2003)}]{Gizon2003}
Gizon, L. and Solanki, S.~K. (2003).
\newblock Determining the {Inclination} of the {Rotation} {Axis} of a
  {Sun}-like {Star}.
\newblock \emph{ApJ} 589, 1009--1019.
\newblock \doi{10.1086/374715}
\bibAnnoteFile{Gizon2003}

\bibitem[{Godoy-Rivera et~al.(2021{\natexlab{a}})Godoy-Rivera, Pinsonneault,
  and Rebull}]{Godoy-Rivera2021_cluster}
Godoy-Rivera, D., Pinsonneault, M.~H., and Rebull, L.~M. (2021{\natexlab{a}}).
\newblock Stellar {Rotation} in the {Gaia} {Era}: {Revised} {Open} {Clusters}'
  {Sequences}.
\newblock \emph{ApJS} 257, 46.
\newblock \doi{10.3847/1538-4365/ac2058}
\bibAnnoteFile{Godoy-Rivera2021_cluster}

\bibitem[{Godoy-Rivera et~al.(2021{\natexlab{b}})Godoy-Rivera, Tayar,
  Pinsonneault, Rodríguez~Martínez, Stassun, van Saders
  et~al.}]{Godoy-Rivera2021}
Godoy-Rivera, D., Tayar, J., Pinsonneault, M.~H., Rodríguez~Martínez, R.,
  Stassun, K.~G., van Saders, J.~L., et~al. (2021{\natexlab{b}}).
\newblock Testing the {Limits} of {Precise} {Subgiant} {Characterization} with
  {APOGEE} and {Gaia}: {Opening} a {Window} to {Unprecedented} {Astrophysical}
  {Studies}.
\newblock \emph{ApJ} 915, 19.
\newblock \doi{10.3847/1538-4357/abf8ba}
\bibAnnoteFile{Godoy-Rivera2021}

\bibitem[{Gomes~da Silva et~al.(2021)Gomes~da Silva, Santos, Adibekyan, Sousa,
  Campante, Figueira et~al.}]{GomesDaSilva2021}
Gomes~da Silva, J., Santos, N.~C., Adibekyan, V., Sousa, S.~G., Campante,
  T.~L., Figueira, P., et~al. (2021).
\newblock Stellar chromospheric activity of 1674 {FGK} stars from the
  {AMBRE}-{HARPS} sample. {I}. {A} catalogue of homogeneous chromospheric
  activity.
\newblock \emph{A\&A} 646, A77.
\newblock \doi{10.1051/0004-6361/202039765}
\bibAnnoteFile{GomesDaSilva2021}

\bibitem[{Gordon et~al.(2021)Gordon, Davenport, Angus, Foreman-Mackey, Agol,
  Covey et~al.}]{Gordon2021}
Gordon, T.~A., Davenport, J. R.~A., Angus, R., Foreman-Mackey, D., Agol, E.,
  Covey, K.~R., et~al. (2021).
\newblock Stellar {Rotation} in the {K2} {Sample}: {Evidence} for {Modified}
  {Spin}-down.
\newblock \emph{ApJ} 913, 70.
\newblock \doi{10.3847/1538-4357/abf63e}
\bibAnnoteFile{Gordon2021}

\bibitem[{Gosnell et~al.(2022)Gosnell, Gully-Santiago, Leiner, and
  Tofflemire}]{Gosnell2022}
Gosnell, N.~M., Gully-Santiago, M.~A., Leiner, E.~M., and Tofflemire, B.~M.
  (2022).
\newblock Observationally {Constraining} the {Starspot} {Properties} of
  {Magnetically} {Active} {M67} {Sub}-subgiant {S1063}.
\newblock \emph{ApJ} 925, 5.
\newblock \doi{10.3847/1538-4357/ac3668}
\bibAnnoteFile{Gosnell2022}

\bibitem[{Gully-Santiago et~al.(2017)Gully-Santiago, Herczeg, Czekala, Somers,
  Grankin, Covey et~al.}]{Gully-Santiago2017}
Gully-Santiago, M.~A., Herczeg, G.~J., Czekala, I., Somers, G., Grankin, K.,
  Covey, K.~R., et~al. (2017).
\newblock Placing the {Spotted} {T} {Tauri} {Star} {LkCa} 4 on an {HR}
  {Diagram}.
\newblock \emph{ApJ} 836, 200.
\newblock \doi{10.3847/1538-4357/836/2/200}
\bibAnnoteFile{Gully-Santiago2017}

\bibitem[{Günther et~al.(2020)Günther, Zhan, Seager, Rimmer, Ranjan, Stassun
  et~al.}]{Gunther2020}
Günther, M.~N., Zhan, Z., Seager, S., Rimmer, P.~B., Ranjan, S., Stassun,
  K.~G., et~al. (2020).
\newblock Stellar {Flares} from the {First} {TESS} {Data} {Release}:
  {Exploring} a {New} {Sample} of {M} {Dwarfs}.
\newblock \emph{AJ} 159, 60.
\newblock \doi{10.3847/1538-3881/ab5d3a10.48550/arXiv.1901.00443}
\bibAnnoteFile{Gunther2020}

\bibitem[{Hale et~al.(1919)Hale, Ellerman, Nicholson, and Joy}]{Hale1919}
Hale, G.~E., Ellerman, F., Nicholson, S.~B., and Joy, A.~H. (1919).
\newblock The {Magnetic} {Polarity} of {Sun}-{Spots}.
\newblock \emph{ApJ} 49, 153.
\newblock \doi{10.1086/142452}
\bibAnnoteFile{Hale1919}

\bibitem[{Hall et~al.(2021)Hall, Davies, van Saders, Nielsen, Lund, Chaplin
  et~al.}]{Hall2021}
Hall, O.~J., Davies, G.~R., van Saders, J., Nielsen, M.~B., Lund, M.~N.,
  Chaplin, W.~J., et~al. (2021).
\newblock Weakened magnetic braking supported by asteroseismic rotation rates
  of {Kepler} dwarfs.
\newblock \emph{Nature Astronomy} 5, 707--714.
\newblock \doi{10.1038/s41550-021-01335-x}
\bibAnnoteFile{Hall2021}

\bibitem[{Hathaway(2015)}]{Hathaway2015}
Hathaway, D.~H. (2015).
\newblock The {Solar} {Cycle}.
\newblock \emph{Living Rev. Sol. Phys.} 12, 4.
\newblock \doi{10.1007/lrsp-2015-4}
\bibAnnoteFile{Hathaway2015}

\bibitem[{Henriksen et~al.(2023)Henriksen, Antoci, Saio, Cantiello, Kjeldsen,
  Kurtz et~al.}]{Henriksen2023}
Henriksen, A.~I., Antoci, V., Saio, H., Cantiello, M., Kjeldsen, H., Kurtz,
  D.~W., et~al. (2023).
\newblock Rotational modulation in {A} and {F} stars: magnetic stellar spots or
  convective core rotation?
\newblock \emph{MNRAS} 520, 216--232.
\newblock \doi{10.1093/mnras/stad153}
\bibAnnoteFile{Henriksen2023}

\bibitem[{Henry et~al.(1996)Henry, Soderblom, Donahue, and
  Baliunas}]{Henry1996}
Henry, T.~J., Soderblom, D.~R., Donahue, R.~A., and Baliunas, S.~L. (1996).
\newblock A {Survey} of {Ca} {II} {H} and {K} {Chromospheric} {Emission} in
  {Southern} {Solar}-{Type} {Stars}.
\newblock \emph{AJ} 111, 439.
\newblock \doi{10.1086/117796}
\bibAnnoteFile{Henry1996}

\bibitem[{Howell et~al.(2014)Howell, Sobeck, Haas, Still, Barclay, Mullally
  et~al.}]{Howell2014}
Howell, S.~B., Sobeck, C., Haas, M., Still, M., Barclay, T., Mullally, F.,
  et~al. (2014).
\newblock The {K2} {Mission}: {Characterization} and {Early} {Results}.
\newblock \emph{PASP} 126, 398.
\newblock \doi{10.1086/676406}
\bibAnnoteFile{Howell2014}

\bibitem[{Ilin et~al.(2019)Ilin, Schmidt, Davenport, and
  Strassmeier}]{Ilin2019}
Ilin, E., Schmidt, S.~J., Davenport, J. R.~A., and Strassmeier, K.~G. (2019).
\newblock Flares in open clusters with {K2} . {I}. {M} 45 ({Pleiades}), {M} 44
  ({Praesepe}), and {M} 67.
\newblock \emph{A\&A} 622, A133.
\newblock \doi{10.1051/0004-6361/201834400}
\bibAnnoteFile{Ilin2019}

\bibitem[{Ilin et~al.(2021)Ilin, Schmidt, Poppenhäger, Davenport, Kristiansen,
  and Omohundro}]{Ilin2021}
Ilin, E., Schmidt, S.~J., Poppenhäger, K., Davenport, J. R.~A., Kristiansen,
  M.~H., and Omohundro, M. (2021).
\newblock Flares in open clusters with {K2}. {II}. {Pleiades}, {Hyades},
  {Praesepe}, {Ruprecht} 147, and {M} 67.
\newblock \emph{A\&A} 645, A42.
\newblock \doi{10.1051/0004-6361/202039198}
\bibAnnoteFile{Ilin2021}

\bibitem[{Jain et~al.(2009)Jain, Tripathy, and Hill}]{Jain2009}
Jain, K., Tripathy, S.~C., and Hill, F. (2009).
\newblock Solar {Activity} {Phases} and {Intermediate}-{Degree} {Mode}
  {Frequencies}.
\newblock \emph{ApJ} 695, 1567--1576.
\newblock \doi{10.1088/0004-637X/695/2/1567}
\bibAnnoteFile{Jain2009}

\bibitem[{Jeffries et~al.(2013)Jeffries, Sandquist, Mathieu, Geller, Orosz,
  Milliman et~al.}]{Jeffries2013}
Jeffries, M.~W., Jr., Sandquist, E.~L., Mathieu, R.~D., Geller, A.~M., Orosz,
  J.~A., Milliman, K.~E., et~al. (2013).
\newblock {WOCS} 40007: {A} {Detached} {Eclipsing} {Binary} near the {Turnoff}
  of the {Open} {Cluster} {NGC} 6819.
\newblock \emph{AJ} 146, 58.
\newblock \doi{10.1088/0004-6256/146/3/58}
\bibAnnoteFile{Jeffries2013}

\bibitem[{Jenkins et~al.(2010)Jenkins, Caldwell, Chandrasekaran, Twicken,
  Bryson, Quintana et~al.}]{Jenkins2010}
Jenkins, J.~M., Caldwell, D.~A., Chandrasekaran, H., Twicken, J.~D., Bryson,
  S.~T., Quintana, E.~V., et~al. (2010).
\newblock Overview of the {Kepler} {Science} {Processing} {Pipeline}.
\newblock \emph{ApJL} 713, L87--L91.
\newblock \doi{10.1088/2041-8205/713/2/L87}
\bibAnnoteFile{Jenkins2010}

\bibitem[{Jiménez et~al.(2002)Jiménez, Roca~Cortés, and
  Jiménez-Reyes}]{Jimenez2002}
Jiménez, A., Roca~Cortés, T., and Jiménez-Reyes, S.~J. (2002).
\newblock Variation of the low-degree solar acoustic mode parameters over the
  solar cycle.
\newblock \emph{Sol. Phys.} 209, 247--263.
\newblock \doi{10.1023/A:1021226503589}
\bibAnnoteFile{Jimenez2002}

\bibitem[{Jiménez-Reyes et~al.(1998)Jiménez-Reyes, Régulo, Pallé, and
  Roca~Cortes}]{Jimenez-Reyes1998}
Jiménez-Reyes, S.~J., Régulo, C., Pallé, P.~L., and Roca~Cortes, T. (1998).
\newblock Solar activity cycle frequency shifts of low-degree p-modes.
\newblock \emph{A\&A} 329, 1119--1124
\bibAnnoteFile{Jimenez-Reyes1998}

\bibitem[{Kaltenegger(2017)}]{Kaltenegger2017}
Kaltenegger, L. (2017).
\newblock How to {Characterize} {Habitable} {Worlds} and {Signs} of {Life}.
\newblock \emph{ARA\&A} 55, 433.
\newblock \doi{10.1146/annurev-astro-082214-122238}
\bibAnnoteFile{Kaltenegger2017}

\bibitem[{Karoff et~al.(2016)Karoff, Knudsen, De~Cat, Bonanno, Fogtmann-Schulz,
  Fu et~al.}]{Karoff2016}
Karoff, C., Knudsen, M.~F., De~Cat, P., Bonanno, A., Fogtmann-Schulz, A., Fu,
  J., et~al. (2016).
\newblock Observational evidence for enhanced magnetic activity of superflare
  stars.
\newblock \emph{Nature Commun.} 7, 11058.
\newblock \doi{10.1038/ncomms11058}
\bibAnnoteFile{Karoff2016}

\bibitem[{Karoff et~al.(2018)Karoff, Metcalfe, Santos, Montet, Isaacson, Witzke
  et~al.}]{Karoff2018}
Karoff, C., Metcalfe, T.~S., Santos, A. R.~G., Montet, B.~T., Isaacson, H.,
  Witzke, V., et~al. (2018).
\newblock The {Influence} of {Metallicity} on {Stellar} {Differential}
  {Rotation} and {Magnetic} {Activity}.
\newblock \emph{ApJ} 852, 46.
\newblock \doi{10.3847/1538-4357/aaa026}
\bibAnnoteFile{Karoff2018}

\bibitem[{Kasper et~al.(2021)Kasper, Klein, Lichko, Huang, Chen, Badman
  et~al.}]{Kasper2021}
Kasper, J., Klein, K., Lichko, E., Huang, J., Chen, C., Badman, S., et~al.
  (2021).
\newblock Parker solar probe enters the magnetically dominated solar corona.
\newblock \emph{Physical review letters} 127, 255101
\bibAnnoteFile{Kasper2021}

\bibitem[{Katz et~al.(2023)Katz, Sartoretti, Guerrier, Panuzzo, Seabroke,
  Thévenin et~al.}]{Katz2023}
Katz, D., Sartoretti, P., Guerrier, A., Panuzzo, P., Seabroke, G.~M.,
  Thévenin, F., et~al. (2023).
\newblock Gaia {Data} {Release} 3. {Properties} and validation of the radial
  velocities.
\newblock \emph{A\&A} 674, A5.
\newblock \doi{10.1051/0004-6361/202244220}
\bibAnnoteFile{Katz2023}

\bibitem[{Kawaler(1988)}]{Kawaler1988}
Kawaler, S.~D. (1988).
\newblock Angular {Momentum} {Loss} in {Low}-{Mass} {Stars}.
\newblock \emph{ApJ} 333, 236.
\newblock \doi{10.1086/166740}
\bibAnnoteFile{Kawaler1988}

\bibitem[{Kiefer et~al.(2017)Kiefer, Schad, Davies, and Roth}]{Kiefer2017}
Kiefer, R., Schad, A., Davies, G., and Roth, M. (2017).
\newblock Stellar magnetic activity and variability of oscillation parameters:
  {An} investigation of 24 solar-like stars observed by {Kepler}.
\newblock \emph{A\&A} 598, A77.
\newblock \doi{10.1051/0004-6361/201628469}
\bibAnnoteFile{Kiefer2017}

\bibitem[{Kirk et~al.(2016)Kirk, Conroy, Prša, Abdul-Masih, Kochoska,
  Matijevič et~al.}]{Kirk2016}
Kirk, B., Conroy, K., Prša, A., Abdul-Masih, M., Kochoska, A., Matijevič, G.,
  et~al. (2016).
\newblock Kepler {Eclipsing} {Binary} {Stars}. {VII}. {The} {Catalog} of
  {Eclipsing} {Binaries} {Found} in the {Entire} {Kepler} {Data} {Set}.
\newblock \emph{AJ} 151, 68.
\newblock \doi{10.3847/0004-6256/151/3/68}
\bibAnnoteFile{Kirk2016}

\bibitem[{Kochukhov(2021)}]{Kochukhov2021}
Kochukhov, O. (2021).
\newblock Magnetic fields of {M} dwarfs.
\newblock \emph{A\&ARv} 29, 1.
\newblock \doi{10.1007/s00159-020-00130-3}
\bibAnnoteFile{Kochukhov2021}

\bibitem[{Kraft(1967)}]{Kraft1967}
Kraft, R.~P. (1967).
\newblock Studies of {Stellar} {Rotation}. {V}. {The} {Dependence} of
  {Rotation} on {Age} among {Solar}-{Type} {Stars}.
\newblock \emph{ApJ} 150, 551.
\newblock \doi{10.1086/149359}
\bibAnnoteFile{Kraft1967}

\bibitem[{Lanzafame et~al.(2019)Lanzafame, Distefano, Barnes, and
  Spada}]{Lanzafame2019}
Lanzafame, A.~C., Distefano, E., Barnes, S.~A., and Spada, F. (2019).
\newblock Evidence of {New} {Magnetic} {Transitions} in {Late}-type {Dwarfs}
  from {Gaia} {DR2}.
\newblock \emph{ApJ} 877, 157.
\newblock \doi{10.3847/1538-4357/ab1aa2}
\bibAnnoteFile{Lanzafame2019}

\bibitem[{Lanzafame et~al.(2018)Lanzafame, Distefano, Messina, Pagano, Lanza,
  Eyer et~al.}]{Lanzafame2018}
Lanzafame, A.~C., Distefano, E., Messina, S., Pagano, I., Lanza, A.~F., Eyer,
  L., et~al. (2018).
\newblock Gaia {Data} {Release} 2 - {Rotational} modulation in late-type
  dwarfs.
\newblock \emph{A\&A} 616, A16.
\newblock \doi{10.1051/0004-6361/201833334}
\bibAnnoteFile{Lanzafame2018}

\bibitem[{Lehtinen et~al.(2021)Lehtinen, Käpylä, Olspert, and
  Spada}]{Lehtinen2021}
Lehtinen, J.~J., Käpylä, M.~J., Olspert, N., and Spada, F. (2021).
\newblock A {Knee} {Point} in the {Rotation}-{Activity} {Scaling} of
  {Late}-type {Stars} with a {Connection} to {Dynamo} {Transitions}.
\newblock \emph{ApJ} 910, 110.
\newblock \doi{10.3847/1538-4357/abe621}
\bibAnnoteFile{Lehtinen2021}

\bibitem[{Leighton(1959)}]{Leighton1959}
Leighton, R.~B. (1959).
\newblock Observations of {Solar} {Magnetic} {Fields} in {Plage} {Regions}.
\newblock \emph{ApJ} 130, 366.
\newblock \doi{10.1086/146727}
\bibAnnoteFile{Leighton1959}

\bibitem[{Li and Basri(2024)}]{Li_Basri2024}
[Dataset] Li, C. and Basri, G. (2024).
\newblock Do {Faculae} {Affect} {Autocorrelation} {Rotation} {Periods} in
  {Sun}-like {Stars}?
\newblock \doi{10.48550/arXiv.2401.13003}
\bibAnnoteFile{Li_Basri2024}

\bibitem[{Lorenzo-Oliveira et~al.(2018)Lorenzo-Oliveira, Freitas, Meléndez,
  Bedell, Ramírez, Bean et~al.}]{Lorenzo-Oliveira2018}
Lorenzo-Oliveira, D., Freitas, F.~C., Meléndez, J., Bedell, M., Ramírez, I.,
  Bean, J.~L., et~al. (2018).
\newblock The {Solar} {Twin} {Planet} {Search}. {The} age-chromospheric
  activity relation.
\newblock \emph{A\&A} 619, A73.
\newblock \doi{10.1051/0004-6361/201629294}
\bibAnnoteFile{Lorenzo-Oliveira2018}

\bibitem[{Lorenzo-Oliveira et~al.(2019)Lorenzo-Oliveira, Meléndez,
  Yana~Galarza, Ponte, dos Santos, Spina et~al.}]{Lorenzo-Oliveira2019}
Lorenzo-Oliveira, D., Meléndez, J., Yana~Galarza, J., Ponte, G., dos Santos,
  L.~A., Spina, L., et~al. (2019).
\newblock Constraining the evolution of stellar rotation using solar twins.
\newblock \emph{MNRAS} 485, L68--L72.
\newblock \doi{10.1093/mnrasl/slz034}
\bibAnnoteFile{Lorenzo-Oliveira2019}

\bibitem[{Lu et~al.(2023)Lu, Angus, Foreman-Mackey, and Hattori}]{Lu2023}
[Dataset] Lu, Y., Angus, R., Foreman-Mackey, D., and Hattori, S. (2023).
\newblock In this {Day} and {Age}: {An} {Empirical} {Gyrochronology} {Relation}
  for {Partially} and {Fully} {Convective} {Single} {Field} {Stars}.
\newblock \doi{10.48550/arXiv.2310.14990}
\bibAnnoteFile{Lu2023}

\bibitem[{Lu et~al.(2022)Lu, Curtis, Angus, David, and Hattori}]{Lu2022}
Lu, Y.~L., Curtis, J.~L., Angus, R., David, T.~J., and Hattori, S. (2022).
\newblock Bridging the {Gap}-{The} {Disappearance} of the {Intermediate}
  {Period} {Gap} for {Fully} {Convective} {Stars}, {Uncovered} by {New} {ZTF}
  {Rotation} {Periods}.
\newblock \emph{AJ} 164, 251.
\newblock \doi{10.3847/1538-3881/ac9bee}
\bibAnnoteFile{Lu2022}

\bibitem[{Lund et~al.(2017)Lund, Silva~Aguirre, Davies, Chaplin,
  Christensen-Dalsgaard, Houdek et~al.}]{Lund2017}
Lund, M.~N., Silva~Aguirre, V., Davies, G.~R., Chaplin, W.~J.,
  Christensen-Dalsgaard, J., Houdek, G., et~al. (2017).
\newblock Standing on the {Shoulders} of {Dwarfs}: the {Kepler} {Asteroseismic}
  {LEGACY} {Sample}. {I}. {Oscillation} {Mode} {Parameters}.
\newblock \emph{ApJ} 835, 172.
\newblock \doi{10.3847/1538-4357/835/2/172}
\bibAnnoteFile{Lund2017}

\bibitem[{Mamajek and Hillenbrand(2008)}]{Mamajek2008}
Mamajek, E.~E. and Hillenbrand, L.~A. (2008).
\newblock Improved {Age} {Estimation} for {Solar}-{Type} {Dwarfs} {Using}
  {Activity}-{Rotation} {Diagnostics}.
\newblock \emph{ApJ} 687, 1264--1293.
\newblock \doi{10.1086/591785}
\bibAnnoteFile{Mamajek2008}

\bibitem[{Masuda et~al.(2022)Masuda, Petigura, and Hall}]{Masuda2022b}
Masuda, K., Petigura, E.~A., and Hall, O.~J. (2022).
\newblock Inferring the rotation period distribution of stars from their
  projected rotation velocities and radii: {Application} to late-{F}/early-{G}
  {Kepler} stars.
\newblock \emph{MNRAS} 510, 5623--5638.
\newblock \doi{10.1093/mnras/stab3650}
\bibAnnoteFile{Masuda2022b}

\bibitem[{Mathur et~al.(2023)Mathur, Claytor, Santos, García, Amard, Bugnet
  et~al.}]{Mathur2023}
Mathur, S., Claytor, Z.~R., Santos, A. R.~G., García, R.~A., Amard, L.,
  Bugnet, L., et~al. (2023).
\newblock Magnetic {Activity} {Evolution} of {Solar}-like {Stars}. {I}. {S}
  ph-{Age} {Relation} {Derived} from {Kepler} {Observations}.
\newblock \emph{ApJ} 952, 131.
\newblock \doi{10.3847/1538-4357/acd118}
\bibAnnoteFile{Mathur2023}

\bibitem[{Mathur et~al.(2014{\natexlab{a}})Mathur, García, Ballot, Ceillier,
  Salabert, Metcalfe et~al.}]{Mathur2014}
Mathur, S., García, R.~A., Ballot, J., Ceillier, T., Salabert, D., Metcalfe,
  T.~S., et~al. (2014{\natexlab{a}}).
\newblock Magnetic activity of {F} stars observed by {Kepler}.
\newblock \emph{A\&A} 562, A124.
\newblock \doi{10.1051/0004-6361/201322707}
\bibAnnoteFile{Mathur2014}

\bibitem[{Mathur et~al.(2019)Mathur, García, Bugnet, Santos, Santiago, and
  Beck}]{Mathur2019}
Mathur, S., García, R.~A., Bugnet, L., Santos, A. R.~G., Santiago, N., and
  Beck, P.~G. (2019).
\newblock Revisiting the impact of stellar magnetic activity on the detection
  of solar-like oscillations by {Kepler}.
\newblock \emph{FrASS} 6, 46.
\newblock \doi{10.3389/fspas.2019.00046}
\bibAnnoteFile{Mathur2019}

\bibitem[{Mathur et~al.(2014{\natexlab{b}})Mathur, Salabert, García, and
  Ceillier}]{Mathur2014b}
Mathur, S., Salabert, D., García, R.~A., and Ceillier, T.
  (2014{\natexlab{b}}).
\newblock Photometric magnetic-activity metrics tested with the {Sun}:
  application to {Kepler} {M} dwarfs.
\newblock \emph{JSWSC} 4, A15.
\newblock \doi{10.1051/swsc/2014011}
\bibAnnoteFile{Mathur2014b}

\bibitem[{Mathys(2017)}]{Mathys2017}
Mathys, G. (2017).
\newblock Ap stars with resolved magnetically split lines: {Magnetic} field
  determinations from {Stokes} {I} and {V} spectra.
\newblock \emph{A\&A} 601, A14.
\newblock \doi{10.1051/0004-6361/201628429}
\bibAnnoteFile{Mathys2017}

\bibitem[{Matt et~al.(2015)Matt, Brun, Baraffe, Bouvier, and
  Chabrier}]{Matt2015}
Matt, S.~P., Brun, A.~S., Baraffe, I., Bouvier, J., and Chabrier, G. (2015).
\newblock The {Mass}-dependence of {Angular} {Momentum} {Evolution} in
  {Sun}-like {Stars}.
\newblock \emph{ApJ} 799, L23.
\newblock \doi{10.1088/2041-8205/799/2/L23}
\bibAnnoteFile{Matt2015}

\bibitem[{McQuillan et~al.(2013)McQuillan, Aigrain, and Mazeh}]{McQuillan2013a}
McQuillan, A., Aigrain, S., and Mazeh, T. (2013).
\newblock Measuring the rotation period distribution of field {M} dwarfs with
  {Kepler}.
\newblock \emph{MNRAS} 432, 1203--1216.
\newblock \doi{10.1093/mnras/stt536}
\bibAnnoteFile{McQuillan2013a}

\bibitem[{McQuillan et~al.(2014)McQuillan, Mazeh, and Aigrain}]{McQuillan2014}
McQuillan, A., Mazeh, T., and Aigrain, S. (2014).
\newblock Rotation {Periods} of 34,030 {Kepler} {Main}-sequence {Stars}: {The}
  {Full} {Autocorrelation} {Sample}.
\newblock \emph{ApJS} 211, 24.
\newblock \doi{10.1088/0067-0049/211/2/24}
\bibAnnoteFile{McQuillan2014}

\bibitem[{Meibom et~al.(2015)Meibom, Barnes, Platais, Gilliland, Latham, and
  Mathieu}]{Meibom2015}
Meibom, S., Barnes, S.~A., Platais, I., Gilliland, R.~L., Latham, D.~W., and
  Mathieu, R.~D. (2015).
\newblock A spin-down clock for cool stars from observations of a
  2.5-billion-year-old cluster.
\newblock \emph{Nature} 517, 589--591.
\newblock \doi{10.1038/nature14118}
\bibAnnoteFile{Meibom2015}

\bibitem[{Metcalfe et~al.(2016)Metcalfe, Egeland, and van
  Saders}]{Metcalfe2016}
Metcalfe, T.~S., Egeland, R., and van Saders, J. (2016).
\newblock Stellar {Evidence} {That} the {Solar} {Dynamo} {May} {Be} in
  {Transition}.
\newblock \emph{ApJL} 826, L2.
\newblock \doi{10.3847/2041-8205/826/1/L2}
\bibAnnoteFile{Metcalfe2016}

\bibitem[{Metcalfe et~al.(2022)Metcalfe, Finley, Kochukhov, See, Ayres, Stassun
  et~al.}]{Metcalfe2022}
Metcalfe, T.~S., Finley, A.~J., Kochukhov, O., See, V., Ayres, T.~R., Stassun,
  K.~G., et~al. (2022).
\newblock The {Origin} of {Weakened} {Magnetic} {Braking} in {Old} {Solar}
  {Analogs}.
\newblock \emph{ApJL} 933, L17.
\newblock \doi{10.3847/2041-8213/ac794d}
\bibAnnoteFile{Metcalfe2022}

\bibitem[{Metcalfe et~al.(2023)Metcalfe, Strassmeier, Ilyin, van Saders, Ayres,
  Finley et~al.}]{Metcalfe2023}
Metcalfe, T.~S., Strassmeier, K.~G., Ilyin, I.~V., van Saders, J.~L., Ayres,
  T.~R., Finley, A.~J., et~al. (2023).
\newblock Constraints on {Magnetic} {Braking} from the {G8} {Dwarf} {Stars} 61
  {UMa} and \${\textbackslash}tau\$ {Cet}.
\newblock \emph{ApJL} 948, L6.
\newblock \doi{10.3847/2041-8213/acce38}
\bibAnnoteFile{Metcalfe2023}

\bibitem[{Metcalfe and van Saders(2017)}]{Metcalfe2017}
Metcalfe, T.~S. and van Saders, J. (2017).
\newblock Magnetic {Evolution} and the {Disappearance} of {Sun}-{Like}
  {Activity} {Cycles}.
\newblock \emph{Sol. Phys.} 292, 126.
\newblock \doi{10.1007/s11207-017-1157-5}
\bibAnnoteFile{Metcalfe2017}

\bibitem[{Metcalfe et~al.(2021)Metcalfe, van Saders, Basu, Buzasi, Drake,
  Egeland et~al.}]{Metcalfe2021}
Metcalfe, T.~S., van Saders, J.~L., Basu, S., Buzasi, D., Drake, J.~J.,
  Egeland, R., et~al. (2021).
\newblock Magnetic and {Rotational} {Evolution} of \${\textbackslash}rho\$
  {CrB} from {Asteroseismology} with {TESS}.
\newblock \emph{ApJ} 921, 122.
\newblock \doi{10.3847/1538-4357/ac1f19}
\bibAnnoteFile{Metcalfe2021}

\bibitem[{Meunier et~al.(2020)Meunier, Lagrange, and Borgniet}]{Meunier2020}
Meunier, N., Lagrange, A.~M., and Borgniet, S. (2020).
\newblock Activity time series of old stars from late {F} to early {K}. {V}.
  {Effect} on exoplanet detectability with high-precision astrometry.
\newblock \emph{A\&A} 644, A77.
\newblock \doi{10.1051/0004-6361/202038710}
\bibAnnoteFile{Meunier2020}

\bibitem[{Montet et~al.(2017)Montet, Tovar, and Foreman-Mackey}]{Montet2017}
Montet, B.~T., Tovar, G., and Foreman-Mackey, D. (2017).
\newblock Long-term {Photometric} {Variability} in {Kepler} {Full}-frame
  {Images}: {Magnetic} {Cycles} of {Sun}-like {Stars}.
\newblock \emph{ApJ} 851, 116.
\newblock \doi{10.3847/1538-4357/aa9e00}
\bibAnnoteFile{Montet2017}

\bibitem[{Newton et~al.(2017)Newton, Irwin, Charbonneau, Berlind, Calkins, and
  Mink}]{Newton2017}
Newton, E.~R., Irwin, J., Charbonneau, D., Berlind, P., Calkins, M.~L., and
  Mink, J. (2017).
\newblock The {H$\alpha$} {Emission} of {Nearby} {M} {Dwarfs} and its
  {Relation} to {Stellar} {Rotation}.
\newblock \emph{ApJ} 834, 85.
\newblock \doi{10.3847/1538-4357/834/1/85}
\bibAnnoteFile{Newton2017}

\bibitem[{Nielsen et~al.(2013)Nielsen, Gizon, Schunker, and
  Karoff}]{Nielsen2013}
Nielsen, M.~B., Gizon, L., Schunker, H., and Karoff, C. (2013).
\newblock Rotation periods of 12 000 main-sequence {Kepler} stars: {Dependence}
  on stellar spectral type and comparison with v sin i observations.
\newblock \emph{A\&A} 557, L10.
\newblock \doi{10.1051/0004-6361/201321912}
\bibAnnoteFile{Nielsen2013}

\bibitem[{Nielsen et~al.(2017)Nielsen, Schunker, Gizon, Schou, and
  Ball}]{Nielsen2017}
Nielsen, M.~B., Schunker, H., Gizon, L., Schou, J., and Ball, W.~H. (2017).
\newblock Limits on radial differential rotation in {Sun}-like stars from
  parametric fits to oscillation power spectra.
\newblock \emph{A\&A} 603, A6.
\newblock \doi{10.1051/0004-6361/201730896}
\bibAnnoteFile{Nielsen2017}

\bibitem[{Noraz et~al.(2022)Noraz, Brun, Strugarek, and Depambour}]{Noraz2022}
Noraz, Q., Brun, A.~S., Strugarek, A., and Depambour, G. (2022).
\newblock Impact of anti-solar differential rotation in mean-field solar-type
  dynamos-exploring possible magnetic cycles in slowly rotating stars.
\newblock \emph{Astronomy \& Astrophysics} 658, A144
\bibAnnoteFile{Noraz2022}

\bibitem[{Notsu et~al.(2019)Notsu, Maehara, Honda, Hawley, Davenport, Namekata
  et~al.}]{Notsu2019}
Notsu, Y., Maehara, H., Honda, S., Hawley, S.~L., Davenport, J. R.~A.,
  Namekata, K., et~al. (2019).
\newblock Do {Kepler} {Superflare} {Stars} {Really} {Include} {Slowly}
  {Rotating} {Sun}-like {Stars}?—{Results} {Using} {APO} 3.5 m {Telescope}
  {Spectroscopic} {Observations} and {Gaia}-{DR2} {Data}.
\newblock \emph{The Astrophysical Journal} 876, 58.
\newblock \doi{10.3847/1538-4357/ab14e6}
\bibAnnoteFile{Notsu2019}

\bibitem[{Noyes et~al.(1984)Noyes, Hartmann, Baliunas, Duncan, and
  Vaughan}]{Noyes1984b}
Noyes, R.~W., Hartmann, L.~W., Baliunas, S.~L., Duncan, D.~K., and Vaughan,
  A.~H. (1984).
\newblock Rotation, convection, and magnetic activity in lower main-sequence
  stars.
\newblock \emph{ApJ} 279, 763--777.
\newblock \doi{10.1086/161945}
\bibAnnoteFile{Noyes1984b}

\bibitem[{Oláh et~al.(2009)Oláh, Kolláth, Granzer, Strassmeier, Lanza,
  Järvinen et~al.}]{Olah2009}
Oláh, K., Kolláth, Z., Granzer, T., Strassmeier, K.~G., Lanza, A.~F.,
  Järvinen, S., et~al. (2009).
\newblock Multiple and changing cycles of active stars. {II}. {Results}.
\newblock \emph{A\&A} 501, 703--713.
\newblock \doi{10.1051/0004-6361/200811304}
\bibAnnoteFile{Olah2009}

\bibitem[{Oshagh et~al.(2013)Oshagh, Santos, Boisse, Boué, Montalto, Dumusque
  et~al.}]{Oshagh2013}
Oshagh, M., Santos, N.~C., Boisse, I., Boué, G., Montalto, M., Dumusque, X.,
  et~al. (2013).
\newblock Effect of stellar spots on high-precision transit light-curve.
\newblock \emph{A\&A} 556, A19.
\newblock \doi{10.1051/0004-6361/201321309}
\bibAnnoteFile{Oshagh2013}

\bibitem[{Owen(2019)}]{Owen2019}
Owen, J.~E. (2019).
\newblock Atmospheric {Escape} and the {Evolution} of {Close}-{In}
  {Exoplanets}.
\newblock \emph{AREPS} 47, 67--90.
\newblock \doi{10.1146/annurev-earth-053018-060246}
\bibAnnoteFile{Owen2019}

\bibitem[{Pace(2013)}]{Pace2013}
Pace, G. (2013).
\newblock Chromospheric activity as age indicator. {An} {L}-shaped
  chromospheric-activity versus age diagram.
\newblock \emph{A\&A} 551, L8.
\newblock \doi{10.1051/0004-6361/201220364}
\bibAnnoteFile{Pace2013}

\bibitem[{Pallavicini et~al.(1981)Pallavicini, Golub, Rosner, Vaiana, Ayres,
  and Linsky}]{Pallavicini1981}
Pallavicini, R., Golub, L., Rosner, R., Vaiana, G.~S., Ayres, T., and Linsky,
  J.~L. (1981).
\newblock Relations among stellar {X}-ray emission observed from {Einstein},
  stellar rotation and bolometric luminosity.
\newblock \emph{ApJ} 248, 279--290.
\newblock \doi{10.1086/159152}
\bibAnnoteFile{Pallavicini1981}

\bibitem[{Pillitteri et~al.(2006)Pillitteri, Micela, Damiani, and
  Sciortino}]{Pillitteri2006}
Pillitteri, I., Micela, G., Damiani, F., and Sciortino, S. (2006).
\newblock Deep {X}-ray survey of the young open cluster {NGC} 2516 with
  {XMM}-{Newton}.
\newblock \emph{A\&A} 450, 993--1004.
\newblock \doi{10.1051/0004-6361:20054003}
\bibAnnoteFile{Pillitteri2006}

\bibitem[{Pinsonneault et~al.(1989)Pinsonneault, Kawaler, Sofia, and
  Demarque}]{Pinsonneault1989}
Pinsonneault, M.~H., Kawaler, S.~D., Sofia, S., and Demarque, P. (1989).
\newblock Evolutionary {Models} of the {Rotating} {Sun}.
\newblock \emph{ApJ} 338, 424.
\newblock \doi{10.1086/167210}
\bibAnnoteFile{Pinsonneault1989}

\bibitem[{Pizzolato et~al.(2003)Pizzolato, Maggio, Micela, Sciortino, and
  Ventura}]{Pizzolato2003}
Pizzolato, N., Maggio, A., Micela, G., Sciortino, S., and Ventura, P. (2003).
\newblock The stellar activity-rotation relationship revisited: {Dependence} of
  saturated and non-saturated {X}-ray emission regimes on stellar mass for
  late-type dwarfs.
\newblock \emph{A\&A} 397, 147--157.
\newblock \doi{10.1051/0004-6361:20021560}
\bibAnnoteFile{Pizzolato2003}

\bibitem[{Ponte et~al.(2023)Ponte, Lorenzo-Oliveira, Melendez, Yana~Galarza,
  and Valio}]{Ponte2023}
Ponte, G., Lorenzo-Oliveira, D., Melendez, J., Yana~Galarza, J., and Valio, A.
  (2023).
\newblock Photometric variations from stellar activity as an age indicator for
  solar-twins.
\newblock \emph{MNRAS} 522, 2675--2682.
\newblock \doi{10.1093/mnras/stad1085}
\bibAnnoteFile{Ponte2023}

\bibitem[{Queloz et~al.(2001)Queloz, Henry, Sivan, Baliunas, Beuzit, Donahue
  et~al.}]{Queloz2001}
Queloz, D., Henry, G.~W., Sivan, J.~P., Baliunas, S.~L., Beuzit, J.~L.,
  Donahue, R.~A., et~al. (2001).
\newblock No planet for {HD} 166435.
\newblock \emph{A\&A} 379, 279--287.
\newblock \doi{10.1051/0004-6361:20011308}
\bibAnnoteFile{Queloz2001}

\bibitem[{Ramírez and Meléndez(2005)}]{Ramirez2005}
Ramírez, I. and Meléndez, J. (2005).
\newblock The {Effective} {Temperature} {Scale} of {FGK} {Stars}. {II}.
  {Teff}:{Color}:[{Fe}/{H}] {Calibrations}.
\newblock \emph{ApJ} 626, 465--485.
\newblock \doi{10.1086/430102}
\bibAnnoteFile{Ramirez2005}

\bibitem[{Rauer et~al.(2014)Rauer, Catala, Aerts, Appourchaux, Benz, Brandeker
  et~al.}]{Rauer2014}
Rauer, H., Catala, C., Aerts, C., Appourchaux, T., Benz, W., Brandeker, A.,
  et~al. (2014).
\newblock The {PLATO} 2.0 mission.
\newblock \emph{Exp. Astron.} 38, 249--330.
\newblock \doi{10.1007/s10686-014-9383-4}
\bibAnnoteFile{Rauer2014}

\bibitem[{Reiners et~al.(2022)Reiners, Shulyak, Käpylä, Ribas, Nagel,
  Zechmeister et~al.}]{Reiners2022}
Reiners, A., Shulyak, D., Käpylä, P.~J., Ribas, I., Nagel, E., Zechmeister,
  M., et~al. (2022).
\newblock Magnetism, rotation, and nonthermal emission in cool stars -
  {Average} magnetic field measurements in 292 {M} dwarfs.
\newblock \emph{A\&A} 662, A41.
\newblock \doi{10.1051/0004-6361/202243251}
\bibAnnoteFile{Reiners2022}

\bibitem[{Reinhold et~al.(2019)Reinhold, Bell, Kuszlewicz, Hekker, and
  Shapiro}]{Reinhold2019}
Reinhold, T., Bell, K.~J., Kuszlewicz, J., Hekker, S., and Shapiro, A.~I.
  (2019).
\newblock Transition from spot to faculae domination. {An} alternate
  explanation for the dearth of intermediate {Kepler} rotation periods.
\newblock \emph{A\&A} 621, A21.
\newblock \doi{10.1051/0004-6361/201833754}
\bibAnnoteFile{Reinhold2019}

\bibitem[{Reinhold and Hekker(2020)}]{Reinhold2020}
Reinhold, T. and Hekker, S. (2020).
\newblock Stellar rotation periods from {K2} {Campaigns} 0-18. {Evidence} for
  rotation period bimodality and simultaneous variability decrease.
\newblock \emph{A\&A} 635, A43.
\newblock \doi{10.1051/0004-6361/201936887}
\bibAnnoteFile{Reinhold2020}

\bibitem[{Reinhold et~al.(2013)Reinhold, Reiners, and Basri}]{Reinhold2013a}
Reinhold, T., Reiners, A., and Basri, G. (2013).
\newblock Rotation and differential rotation of active {Kepler} stars.
\newblock \emph{A\&A} 560, A4.
\newblock \doi{10.1051/0004-6361/201321970}
\bibAnnoteFile{Reinhold2013a}

\bibitem[{Reinhold et~al.(2023)Reinhold, Shapiro, Solanki, and
  Basri}]{Reinhold2023}
Reinhold, T., Shapiro, A.~I., Solanki, S.~K., and Basri, G. (2023).
\newblock New rotation period measurements of 67 163 {Kepler} stars.
\newblock \emph{A\&A} 678, A24.
\newblock \doi{10.1051/0004-6361/202346789}
\bibAnnoteFile{Reinhold2023}

\bibitem[{Ricker et~al.(2014)Ricker, Winn, Vanderspek, Latham, Bakos, Bean
  et~al.}]{Ricker2014}
Ricker, G.~R., Winn, J.~N., Vanderspek, R., Latham, D.~W., Bakos, G.~A., Bean,
  J.~L., et~al. (2014).
\newblock Transiting {Exoplanet} {Survey} {Satellite} ({TESS}) 9143, 914320.
\newblock \doi{10.1117/12.2063489}.
\newblock Conference Name: Space Telescopes and Instrumentation 2014: Optical,
  Infrared, and Millimeter Wave Place: eprint: arXiv:1406.0151
\bibAnnoteFile{Ricker2014}

\bibitem[{Rincon and Rieutord(2018)}]{Rincon2018}
Rincon, F. and Rieutord, M. (2018).
\newblock The {Sun}’s supergranulation.
\newblock \emph{Living Rev. Sol. Phys.} 15, 6.
\newblock \doi{10.1007/s41116-018-0013-5}
\bibAnnoteFile{Rincon2018}

\bibitem[{Réville et~al.(2015)Réville, Brun, Matt, Strugarek, and
  Pinto}]{Reville2015}
Réville, V., Brun, A.~S., Matt, S.~P., Strugarek, A., and Pinto, R.~F. (2015).
\newblock The {Effect} of {Magnetic} {Topology} on {Thermally} {Driven} {Wind}:
  {Toward} a {General} {Formulation} of the {Braking} {Law}.
\newblock \emph{ApJ} 798, 116.
\newblock \doi{10.1088/0004-637X/798/2/116}
\bibAnnoteFile{Reville2015}

\bibitem[{Saikia et~al.(2018)Saikia, Lueftinger, Jeffers, Folsom, See, Petit
  et~al.}]{Saikia2018L}
Saikia, S.~B., Lueftinger, T., Jeffers, S., Folsom, C., See, V., Petit, P.,
  et~al. (2018).
\newblock Direct evidence of a full dipole flip during the magnetic cycle of a
  sun-like star.
\newblock \emph{A\&A} 620, L11
\bibAnnoteFile{Saikia2018L}

\bibitem[{Salabert et~al.(2016)Salabert, García, Beck, Egeland, Pallé, Mathur
  et~al.}]{Salabert2016a}
Salabert, D., García, R.~A., Beck, P.~G., Egeland, R., Pallé, P.~L., Mathur,
  S., et~al. (2016).
\newblock Photospheric and chromospheric magnetic activity of seismic solar
  analogs. {Observational} inputs on the solar-stellar connection from {Kepler}
  and {Hermes}.
\newblock \emph{A\&A} 596, A31.
\newblock \doi{10.1051/0004-6361/201628583}
\bibAnnoteFile{Salabert2016a}

\bibitem[{Salabert et~al.(2017)Salabert, García, Jiménez, Bertello, Corsaro,
  and Pallé}]{Salabert2017}
Salabert, D., García, R.~A., Jiménez, A., Bertello, L., Corsaro, E., and
  Pallé, P.~L. (2017).
\newblock Photospheric activity of the {Sun} with {VIRGO} and {GOLF}.
  {Comparison} with standard activity proxies.
\newblock \emph{A\&A} 608, A87.
\newblock \doi{10.1051/0004-6361/201731560}
\bibAnnoteFile{Salabert2017}

\bibitem[{Salabert et~al.(2015)Salabert, García, and
  Turck-Chièze}]{Salabert2015}
Salabert, D., García, R.~A., and Turck-Chièze, S. (2015).
\newblock Seismic sensitivity to sub-surface solar activity from 18 yr of
  {GOLF}/{SoHO} observations.
\newblock \emph{A\&A} 578, A137.
\newblock \doi{10.1051/0004-6361/201425236}
\bibAnnoteFile{Salabert2015}

\bibitem[{Santos et~al.(2021)Santos, Breton, Mathur, and
  García}]{Santos2021ApJS}
Santos, A. R.~G., Breton, S.~N., Mathur, S., and García, R.~A. (2021).
\newblock Surface {Rotation} and {Photometric} {Activity} for {Kepler}
  {Targets}. {II}. {G} and {F} {Main}-sequence {Stars} and {Cool} {Subgiant}
  {Stars}.
\newblock \emph{ApJS} 255, 17.
\newblock \doi{10.3847/1538-4365/ac033f}
\bibAnnoteFile{Santos2021ApJS}

\bibitem[{Santos et~al.(2018)Santos, Campante, Chaplin, Cunha, Lund, Kiefer
  et~al.}]{Santos2018}
Santos, A. R.~G., Campante, T.~L., Chaplin, W.~J., Cunha, M.~S., Lund, M.~N.,
  Kiefer, R., et~al. (2018).
\newblock Signatures of {Magnetic} {Activity} in the {Seismic} {Data} of
  {Solar}-type {Stars} {Observed} by {Kepler}.
\newblock \emph{ApJS} 237, 17.
\newblock \doi{10.3847/1538-4365/aac9b6}
\bibAnnoteFile{Santos2018}

\bibitem[{Santos et~al.(2019)Santos, García, Mathur, Bugnet, Saders, Metcalfe
  et~al.}]{Santos2019a}
Santos, A. R.~G., García, R.~A., Mathur, S., Bugnet, L., Saders, J. L.~v.,
  Metcalfe, T.~S., et~al. (2019).
\newblock Surface {Rotation} and {Photometric} {Activity} for {Kepler}
  {Targets}. {I}. {M} and {K} {Main}-sequence {Stars}.
\newblock \emph{ApJS} 244, 21.
\newblock \doi{10.3847/1538-4365/ab3b56}
\bibAnnoteFile{Santos2019a}

\bibitem[{Santos et~al.(2023)Santos, Mathur, García, Broomhall, Egeland,
  Jiménez et~al.}]{Santos2023}
Santos, A. R.~G., Mathur, S., García, R.~A., Broomhall, A.~M., Egeland, R.,
  Jiménez, A., et~al. (2023).
\newblock Temporal variation of the photometric magnetic activity for the {Sun}
  and {Kepler} solar-like stars.
\newblock \emph{A\&A} 672, A56.
\newblock \doi{10.1051/0004-6361/202245430}
\bibAnnoteFile{Santos2023}

\bibitem[{Saunders et~al.(2023)Saunders, van Saders, Lyttle, Metcalfe, Li,
  Davies et~al.}]{Saunders2023}
[Dataset] Saunders, N., van Saders, J.~L., Lyttle, A.~J., Metcalfe, T.~S., Li,
  T., Davies, G.~R., et~al. (2023).
\newblock Stellar {Cruise} {Control}: {Weakened} {Magnetic} {Braking} {Leads}
  to {Sustained} {Rapid} {Rotation} of {Old} {Stars}.
\newblock \doi{10.48550/arXiv.2309.05666}
\bibAnnoteFile{Saunders2023}

\bibitem[{Schmitt et~al.(1995)Schmitt, Fleming, and Giampapa}]{Schmitt1995}
Schmitt, J. H. M.~M., Fleming, T.~A., and Giampapa, M.~S. (1995).
\newblock The {X}-{Ray} {View} of the {Low}-{Mass} {Stars} in the {Solar}
  {Neighborhood}.
\newblock \emph{ApJ} 450, 392.
\newblock \doi{10.1086/176149}
\bibAnnoteFile{Schmitt1995}

\bibitem[{Schunker et~al.(2016{\natexlab{a}})Schunker, Schou, and
  Ball}]{Schunker2016a}
Schunker, H., Schou, J., and Ball, W.~H. (2016{\natexlab{a}}).
\newblock Asteroseismic inversions for radial differential rotation of
  {Sun}-like stars: {Sensitivity} to uncertainties.
\newblock \emph{A\&A} 586, A24.
\newblock \doi{10.1051/0004-6361/201525937}
\bibAnnoteFile{Schunker2016a}

\bibitem[{Schunker et~al.(2016{\natexlab{b}})Schunker, Schou, Ball, Nielsen,
  and Gizon}]{Schunker2016}
Schunker, H., Schou, J., Ball, W.~H., Nielsen, M.~B., and Gizon, L.
  (2016{\natexlab{b}}).
\newblock Asteroseismic inversions for radial differential rotation of
  {Sun}-like stars: ensemble fits.
\newblock \emph{A\&A} 586, A79.
\newblock \doi{10.1051/0004-6361/201527485}
\bibAnnoteFile{Schunker2016}

\bibitem[{See et~al.(2019{\natexlab{a}})See, Matt, Finley, Folsom, Boro~Saikia,
  Donati et~al.}]{See2019a}
See, V., Matt, S.~P., Finley, A.~J., Folsom, C.~P., Boro~Saikia, S., Donati,
  J.-F., et~al. (2019{\natexlab{a}}).
\newblock Do {Non}-dipolar {Magnetic} {Fields} {Contribute} to {Spin}-down
  {Torques}?
\newblock \emph{ApJ} 886, 120.
\newblock \doi{10.3847/1538-4357/ab46b2}
\bibAnnoteFile{See2019a}

\bibitem[{See et~al.(2019{\natexlab{b}})See, Matt, Folsom, Boro~Saikia, Donati,
  Fares et~al.}]{See2019b}
See, V., Matt, S.~P., Folsom, C.~P., Boro~Saikia, S., Donati, J.-F., Fares, R.,
  et~al. (2019{\natexlab{b}}).
\newblock Estimating {Magnetic} {Filling} {Factors} from {Zeeman}-{Doppler}
  {Magnetograms}.
\newblock \emph{ApJ} 876, 118.
\newblock \doi{10.3847/1538-4357/ab1096}
\bibAnnoteFile{See2019b}

\bibitem[{See et~al.(2023)See, Roquette, Amard, and Matt}]{See2023}
See, V., Roquette, J., Amard, L., and Matt, S. (2023).
\newblock Further evidence of the link between activity and metallicity using
  the flaring properties of stars in the {Kepler} field.
\newblock \emph{MNRAS} 524, 5781--5786.
\newblock \doi{10.1093/mnras/stad2020}.
\newblock ADS Bibcode: 2023MNRAS.524.5781S
\bibAnnoteFile{See2023}

\bibitem[{See et~al.(2021)See, Roquette, Amard, and Matt}]{See2021}
See, V., Roquette, J., Amard, L., and Matt, S.~P. (2021).
\newblock Photometric {Variability} as a {Proxy} for {Magnetic} {Activity} and
  {Its} {Dependence} on {Metallicity}.
\newblock \emph{ApJ} 912, 127.
\newblock \doi{10.3847/1538-4357/abed47}
\bibAnnoteFile{See2021}

\bibitem[{Semel(1989)}]{Semel1989}
Semel, M. (1989).
\newblock Zeeman-{Doppler} imaging of active stars. {I} - {Basic} principles.
\newblock \emph{A\&A} 225, 456--466
\bibAnnoteFile{Semel1989}

\bibitem[{Shapiro et~al.(2016)Shapiro, Solanki, Krivova, Yeo, and
  Schmutz}]{Shapiro2016}
Shapiro, A.~I., Solanki, S.~K., Krivova, N.~A., Yeo, K.~L., and Schmutz, W.~K.
  (2016).
\newblock Are solar brightness variations faculae- or spot-dominated?
\newblock \emph{ArXiv e-prints} 1602, arXiv:1602.04447
\bibAnnoteFile{Shapiro2016}

\bibitem[{Shoda et~al.(2023)Shoda, Cranmer, and Toriumi}]{Shoda2023}
Shoda, M., Cranmer, S.~R., and Toriumi, S. (2023).
\newblock Formulating mass-loss rates for sun-like stars: A hybrid model
  approach.
\newblock \emph{The Astrophysical Journal} 957, 71
\bibAnnoteFile{Shoda2023}

\bibitem[{Shoda et~al.(2020)Shoda, Suzuki, Matt, Cranmer, Vidotto, Strugarek
  et~al.}]{Shoda2020}
Shoda, M., Suzuki, T.~K., Matt, S.~P., Cranmer, S.~R., Vidotto, A.~A.,
  Strugarek, A., et~al. (2020).
\newblock Alfv{\'e}n-wave-driven magnetic rotator winds from low-mass stars. i.
  rotation dependences of magnetic braking and mass-loss rate.
\newblock \emph{The Astrophysical Journal} 896, 123
\bibAnnoteFile{Shoda2020}

\bibitem[{Silva-Beyer et~al.(2023)Silva-Beyer, Godoy-Rivera, and
  Chanamé}]{Silva-Beyer2023}
Silva-Beyer, J., Godoy-Rivera, D., and Chanamé, J. (2023).
\newblock The breakdown of current gyrochronology as evidenced by old coeval
  stars.
\newblock \emph{MNRAS} 523, 5947--5961.
\newblock \doi{10.1093/mnras/stad1803}
\bibAnnoteFile{Silva-Beyer2023}

\bibitem[{Simonian et~al.(2019)Simonian, Pinsonneault, and
  Terndrup}]{Simonian2019}
Simonian, G. V.~A., Pinsonneault, M.~H., and Terndrup, D.~M. (2019).
\newblock Rapid {Rotation} in the {Kepler} {Field}: {Not} a {Single} {Star}
  {Phenomenon}.
\newblock \emph{ApJ} 871, 174.
\newblock \doi{10.3847/1538-4357/aaf97c}
\bibAnnoteFile{Simonian2019}

\bibitem[{Skumanich(1972)}]{Skumanich1972}
Skumanich, A. (1972).
\newblock Time {Scales} for {CA} {II} {Emission} {Decay}, {Rotational}
  {Braking}, and {Lithium} {Depletion}.
\newblock \emph{ApJ} 171, 565.
\newblock \doi{10.1086/151310}
\bibAnnoteFile{Skumanich1972}

\bibitem[{Soderblom et~al.(1991)Soderblom, Duncan, and Johnson}]{Soderblom1991}
Soderblom, D.~R., Duncan, D.~K., and Johnson, D. R.~H. (1991).
\newblock The {Chromospheric} {Emission}--{Age} {Relation} for {Stars} of the
  {Lower} {Main} {Sequence} and {Its} {Implications} for the {Star} {Formation}
  {Rate}.
\newblock \emph{ApJ} 375, 722.
\newblock \doi{10.1086/170238}
\bibAnnoteFile{Soderblom1991}

\bibitem[{Soderblom et~al.(1993)Soderblom, Stauffer, Hudon, and
  Jones}]{Soderblom1993}
Soderblom, D.~R., Stauffer, J.~R., Hudon, J.~D., and Jones, B.~F. (1993).
\newblock Rotation and {Chromospheric} {Emission} among {F}, {G}, and {K}
  {Dwarfs} of the {Pleiades}.
\newblock \emph{ApJS} 85, 315.
\newblock \doi{10.1086/191767}
\bibAnnoteFile{Soderblom1993}

\bibitem[{Solanki(2003)}]{Solanki2003}
Solanki, S.~K. (2003).
\newblock Sunspots: {An} overview.
\newblock \emph{A\&ARv} 11, 153--286.
\newblock \doi{10.1007/s00159-003-0018-4}
\bibAnnoteFile{Solanki2003}

\bibitem[{Soon et~al.(1993)Soon, Baliunas, and Zhang}]{Soon1993}
Soon, W.~H., Baliunas, S.~L., and Zhang, Q. (1993).
\newblock An {Interpretation} of {Cycle} {Periods} of {Stellar} {Chromospheric}
  {Activity}.
\newblock \emph{ApJ} 414, L33.
\newblock \doi{10.1086/186989}
\bibAnnoteFile{Soon1993}

\bibitem[{Spada and Lanzafame(2020)}]{Spada2020}
Spada, F. and Lanzafame, A.~C. (2020).
\newblock Competing effect of wind braking and interior coupling in the
  rotational evolution of solar-like stars.
\newblock \emph{A\&A} 636, A76.
\newblock \doi{10.1051/0004-6361/201936384}
\bibAnnoteFile{Spada2020}

\bibitem[{Stauffer and Hartmann(1987)}]{Stauffer1987}
Stauffer, J.~R. and Hartmann, L.~W. (1987).
\newblock The {Distribution} of {Rotational} {Velocities} for {Low}-{Mass}
  {Stars} in the {Pleiades}.
\newblock \emph{ApJ} 318, 337.
\newblock \doi{10.1086/165371}
\bibAnnoteFile{Stauffer1987}

\bibitem[{Stauffer et~al.(1998)Stauffer, Schultz, and
  Kirkpatrick}]{Stauffer1998}
Stauffer, J.~R., Schultz, G., and Kirkpatrick, J.~D. (1998).
\newblock Keck {Spectra} of {Pleiades} {Brown} {Dwarf} {Candidates} and a
  {Precise} {Determination} of the {Lithium} {Depletion} {Edge} in the
  {Pleiades}.
\newblock \emph{ApJ} 499, L199--L203.
\newblock \doi{10.1086/311379}
\bibAnnoteFile{Stauffer1998}

\bibitem[{Strugarek(2018)}]{Strugarek2018}
Strugarek, A. (2018).
\newblock Models of {Star}-{Planet} {Magnetic} {Interaction}.
\newblock In \emph{Handbook of {Exoplanets}}, eds. H.~J. Deeg and J.~A.
  Belmonte (Cham: Springer International Publishing). 1833--1855.
\newblock \doi{10.1007/978-3-319-55333-7_25}
\bibAnnoteFile{Strugarek2018}

\bibitem[{Suárez~Mascareño et~al.(2017)Suárez~Mascareño, Rebolo,
  González~Hernández, and Esposito}]{Suarez-Mascareno2017}
Suárez~Mascareño, A., Rebolo, R., González~Hernández, J.~I., and Esposito,
  M. (2017).
\newblock Characterization of the radial velocity signal induced by rotation in
  late-type dwarfs.
\newblock \emph{MNRAS} 468, 4772--4781.
\newblock \doi{10.1093/mnras/stx771}
\bibAnnoteFile{Suarez-Mascareno2017}

\bibitem[{Thomas et~al.(2019)Thomas, Chaplin, Davies, Howe, Santos, Elsworth
  et~al.}]{Thomas2019}
Thomas, A. E.~L., Chaplin, W.~J., Davies, G.~R., Howe, R., Santos, A. R.~G.,
  Elsworth, Y., et~al. (2019).
\newblock Asteroseismic constraints on active latitudes of solar-type stars:
  {HD} 173701 has active bands at higher latitudes than the {Sun}.
\newblock \emph{MNRAS} 485, 3857--3868.
\newblock \doi{10.1093/mnras/stz672}
\bibAnnoteFile{Thomas2019}

\bibitem[{Tokuno et~al.(2023)Tokuno, Suzuki, and Shoda}]{Tokuno2023}
Tokuno, T., Suzuki, T.~K., and Shoda, M. (2023).
\newblock Transition of latitudinal differential rotation as a possible cause
  of weakened magnetic braking of solar-type stars.
\newblock \emph{MNRAS} 520, 418--436.
\newblock \doi{10.1093/mnras/stad103}
\bibAnnoteFile{Tokuno2023}

\bibitem[{Torres et~al.(2020)Torres, Vanderburg, Curtis, Kraus, Rizzuto, and
  Ireland}]{Torres2020}
Torres, G., Vanderburg, A., Curtis, J.~L., Kraus, A.~L., Rizzuto, A.~C., and
  Ireland, M.~J. (2020).
\newblock Eclipsing {Binaries} in the {Open} {Cluster} {Ruprecht} 147. {III}.
  {The} {Triple} {System} {EPIC} 219552514 at the {Main}-sequence {Turnoff}.
\newblock \emph{ApJ} 896, 162.
\newblock \doi{10.3847/1538-4357/ab911b}
\bibAnnoteFile{Torres2020}

\bibitem[{Tripathy et~al.(2011)Tripathy, Jain, Salabert, García, Hill, and
  Leibacher}]{Tripathy2011}
Tripathy, S.~C., Jain, K., Salabert, D., García, R.~A., Hill, F., and
  Leibacher, J.~W. (2011).
\newblock Angular-degree dependence of p-mode frequencies during solar cycle
  23.
\newblock \emph{J. Phys. Conf. Ser.} 271, 012055.
\newblock \doi{10.1088/1742-6596/271/1/012055}
\bibAnnoteFile{Tripathy2011}

\bibitem[{Valenti and Fischer(2005)}]{Valenti2005}
Valenti, J.~A. and Fischer, D.~A. (2005).
\newblock Spectroscopic {Properties} of {Cool} {Stars} ({SPOCS}). {I}. 1040
  {F}, {G}, and {K} {Dwarfs} from {Keck}, {Lick}, and {AAT} {Planet} {Search}
  {Programs}.
\newblock \emph{ApJS} 159, 141--166.
\newblock \doi{10.1086/430500}
\bibAnnoteFile{Valenti2005}

\bibitem[{van Driel-Gesztelyi and Green(2015)}]{vanDriel-Gesztelyi2015}
van Driel-Gesztelyi, L. and Green, L.~M. (2015).
\newblock Evolution of {Active} {Regions}.
\newblock \emph{Living Rev. Sol. Phys.} 12, 1.
\newblock \doi{10.1007/lrsp-2015-1}
\bibAnnoteFile{vanDriel-Gesztelyi2015}

\bibitem[{van Saders et~al.(2016)van Saders, Ceillier, Metcalfe, Silva~Aguirre,
  Pinsonneault, García et~al.}]{vanSaders2016}
van Saders, J.~L., Ceillier, T., Metcalfe, T.~S., Silva~Aguirre, V.,
  Pinsonneault, M.~H., García, R.~A., et~al. (2016).
\newblock Weakened magnetic braking as the origin of anomalously rapid rotation
  in old field stars.
\newblock \emph{Nat.} 529, 181--184.
\newblock \doi{10.1038/nature16168}
\bibAnnoteFile{vanSaders2016}

\bibitem[{van Saders and Pinsonneault(2012)}]{vanSaders2012}
van Saders, J.~L. and Pinsonneault, M.~H. (2012).
\newblock The {Sensitivity} of {Convection} {Zone} {Depth} to {Stellar}
  {Abundances}: {An} {Absolute} {Stellar} {Abundance} {Scale} from
  {Asteroseismology}.
\newblock \emph{ApJ} 746, 16.
\newblock \doi{10.1088/0004-637X/746/1/16}
\bibAnnoteFile{vanSaders2012}

\bibitem[{van Saders and Pinsonneault(2013)}]{vanSaders2013}
van Saders, J.~L. and Pinsonneault, M.~H. (2013).
\newblock Fast {Star}, {Slow} {Star}; {Old} {Star}, {Young} {Star}: {Subgiant}
  {Rotation} as a {Population} and {Stellar} {Physics} {Diagnostic}.
\newblock \emph{ApJ} 776, 67.
\newblock \doi{10.1088/0004-637X/776/2/67}
\bibAnnoteFile{vanSaders2013}

\bibitem[{van Saders et~al.(2019)van Saders, Pinsonneault, and
  Barbieri}]{vanSaders2019}
van Saders, J.~L., Pinsonneault, M.~H., and Barbieri, M. (2019).
\newblock Forward {Modeling} of the {Kepler} {Stellar} {Rotation} {Period}
  {Distribution}: {Interpreting} {Periods} from {Mixed} and {Biased} {Stellar}
  {Populations}.
\newblock \emph{ApJ} 872, 128.
\newblock \doi{10.3847/1538-4357/aafafe}
\bibAnnoteFile{vanSaders2019}

\bibitem[{Vaughan(1980)}]{Vaughan1980}
Vaughan, A.~H. (1980).
\newblock Comparison of activity cycles in old and young main-sequence stars.
\newblock \emph{PASP} 92, 392--396.
\newblock \doi{10.1086/130684}
\bibAnnoteFile{Vaughan1980}

\bibitem[{Vaughan and Preston(1980)}]{VaughanPreston1980}
Vaughan, A.~H. and Preston, G.~W. (1980).
\newblock A survey of chromospheric {CA} {II} {H} and {K} emission in field
  stars of the solar neighborhood.
\newblock \emph{PASP} 92, 385--391.
\newblock \doi{10.1086/130683}
\bibAnnoteFile{VaughanPreston1980}

\bibitem[{Vidotto et~al.(2014)Vidotto, Gregory, Jardine, Donati, Petit, Morin
  et~al.}]{Vidotto2014}
Vidotto, A.~A., Gregory, S.~G., Jardine, M., Donati, J.~F., Petit, P., Morin,
  J., et~al. (2014).
\newblock Stellar magnetism: empirical trends with age and rotation.
\newblock \emph{MNRAS} 441, 2361--2374.
\newblock \doi{10.1093/mnras/stu728}
\bibAnnoteFile{Vidotto2014}

\bibitem[{Walter and Bowyer(1981)}]{Walter1981}
Walter, F.~M. and Bowyer, S. (1981).
\newblock On the coronae of rapidly rotating stars. {I}. {The} relation between
  rotation and coronal activity in {RS} {CVn} systems.
\newblock \emph{ApJ} 245, 671--676.
\newblock \doi{10.1086/158842}
\bibAnnoteFile{Walter1981}

\bibitem[{Weber and Davis(1967)}]{Weber1967}
Weber, E.~J. and Davis, L., Jr. (1967).
\newblock The {Angular} {Momentum} of the {Solar} {Wind}.
\newblock \emph{ApJ} 148, 217--227.
\newblock \doi{10.1086/149138}
\bibAnnoteFile{Weber1967}

\bibitem[{Wilson(1963)}]{Wilson1963}
Wilson, O.~C. (1963).
\newblock A {Probable} {Correlation} {Between} {Chromospheric} {Activity} and
  {Age} in {Main}-{Sequence} {Stars}.
\newblock \emph{ApJ} 138, 832.
\newblock \doi{10.1086/147689}
\bibAnnoteFile{Wilson1963}

\bibitem[{Wilson(1968)}]{Wilson1968}
Wilson, O.~C. (1968).
\newblock Flux {Measurements} at the {Centers} of {Stellar} {H}- and
  {K}-{Lines}.
\newblock \emph{ApJ} 153, 221.
\newblock \doi{10.1086/149652}
\bibAnnoteFile{Wilson1968}

\bibitem[{Wilson(1978)}]{Wilson1978}
Wilson, O.~C. (1978).
\newblock Chromospheric variations in main-sequence stars.
\newblock \emph{ApJ} 226, 379--396.
\newblock \doi{10.1086/156618}
\bibAnnoteFile{Wilson1978}

\bibitem[{Wood et~al.(2021)Wood, Müller, Redfield, Konow, Vannier, Linsky
  et~al.}]{Wood2021}
Wood, B.~E., Müller, H.-R., Redfield, S., Konow, F., Vannier, H., Linsky,
  J.~L., et~al. (2021).
\newblock New {Observational} {Constraints} on the {Winds} of {M} dwarf
  {Stars}*.
\newblock \emph{ApJ} 915, 37.
\newblock \doi{10.3847/1538-4357/abfda5}
\bibAnnoteFile{Wood2021}

\bibitem[{Woodard and Noyes(1985)}]{Woodard1985}
Woodard, M.~F. and Noyes, R.~W. (1985).
\newblock Change of solar oscillation eigenfrequencies with the solar cycle.
\newblock \emph{Nature} 318, 449--450.
\newblock \doi{10.1038/318449a0}
\bibAnnoteFile{Woodard1985}

\bibitem[{Wright and Drake(2016)}]{Wright2016}
Wright, N.~J. and Drake, J.~J. (2016).
\newblock Solar-type dynamo behaviour in fully convective stars without a
  tachocline.
\newblock \emph{Nature} 535, 526--528.
\newblock \doi{10.1038/nature18638}
\bibAnnoteFile{Wright2016}

\bibitem[{Wright et~al.(2011)Wright, Drake, Mamajek, and Henry}]{Wright2011}
Wright, N.~J., Drake, J.~J., Mamajek, E.~E., and Henry, G.~W. (2011).
\newblock The {Stellar}-activity-{Rotation} {Relationship} and the {Evolution}
  of {Stellar} {Dynamos}.
\newblock \emph{ApJ} 743, 48.
\newblock \doi{10.1088/0004-637X/743/1/48}
\bibAnnoteFile{Wright2011}

\bibitem[{Wright et~al.(2018)Wright, Newton, Williams, Drake, and
  Yadav}]{Wright2018}
Wright, N.~J., Newton, E.~R., Williams, P. K.~G., Drake, J.~J., and Yadav,
  R.~K. (2018).
\newblock The stellar rotation–activity relationship in fully convective {M}
  dwarfs.
\newblock \emph{MNRAS} 479, 2351--2360.
\newblock \doi{10.1093/mnras/sty1670}
\bibAnnoteFile{Wright2018}

\bibitem[{Yang and Liu(2019)}]{Yang2019}
Yang, H. and Liu, J. (2019).
\newblock The {Flare} {Catalog} and the {Flare} {Activity} in the {Kepler}
  {Mission}.
\newblock \emph{ApJS} 241, 29.
\newblock \doi{10.3847/1538-4365/ab0d28}
\bibAnnoteFile{Yang2019}

\bibitem[{Zhong et~al.(2023)Zhong, Zhang, Yang, and Su}]{Zhong2023}
Zhong, M., Zhang, L., Yang, Z., and Su, T. (2023).
\newblock Magnetic {Activity} of {Different} {Types} of {Variable} {Stars}
  {Observed} by {TESS} {Mission}.
\newblock \emph{Universe} 9, 227.
\newblock \doi{10.3390/universe9050227}
\bibAnnoteFile{Zhong2023}

\bibitem[{Zirin(1970)}]{Zirin1970}
Zirin, H. (1970).
\newblock Active {Regions}. {I}: {The} {Occurrence} of {Solar} {Flares} and the
  {Development} of {Active} {Regions}.
\newblock \emph{Sol. Phys.} 14, 328--341.
\newblock \doi{10.1007/BF00221318}
\bibAnnoteFile{Zirin1970}

\end{thebibliography}

\end{document}